\documentclass[conference,compsoc]{IEEEtran}

\usepackage{tikz}
\usepackage{amsmath}
\usepackage{tabularx}
\usepackage{caption}
\usepackage{multirow}
\usepackage{multicol}
\usepackage{makecell}
\usepackage{adjustbox}
\usepackage{subfigure}
\usepackage{booktabs}

\usepackage{amsmath,amssymb,amsfonts}
\usepackage{filecontents}
\usepackage{tabularx}
\usepackage{listings}
\usepackage{multirow}
\usepackage{multicol}
\usepackage{xcolor}
\usepackage{cleveref}
\usepackage{acronym}
\usepackage{xspace,url}
\usepackage{bbding}
\usepackage{pifont}
\usepackage{wasysym}
\usepackage{tcolorbox}
\usepackage{colortbl}

\usepackage[ruled,linesnumbered]{algorithm2e}

\newcommand{\ie}{\textit{i}.\textit{e}.~}
\newcommand{\eg}{\textit{e}.\textit{g}.~}
\newcommand{\trainbeforepartition}{training-before-partition\xspace}
\newcommand{\partitionbeforetrain}{partition-before-training\xspace}

\acrodef{fl}[FL]{Federated Learning}
\acrodef{gdpr}[GDPR]{the General Data Protection Regulation}
\acrodef{asr}[ASR]{attack success rate}
\acrodef{tsdp}[TSDP]{\textit{TEE-shielded DNN partition}}
\acrodef{mia}[MIA]{\textbf{Membership Inference Attack}}
\acrodef{ms}[MS]{\textbf{Model Stealing}}
\acrodef{dp}[DP]{Differential Privacy}
\acrodef{mpc}[MPC]{Multi-Party Computation}
\acrodef{he}[HE]{Homomorphic Encryption}

\newcommand{\sys}{\textsc{TEESlice}\xspace} 
\newcommand{\tool}{\textsc{TEESlice}\xspace} 
\newcommand{\nettailor}{\textsc{NetTailor}\xspace}

\newcommand{\F}{Fig.}

\newcommand{\T}{Table}
\renewcommand{\S}{Sec.}
\newcommand{\A}{Alg.}
\newcommand{\App}{Appx.}

\newcommand{\parh}[1]{\noindent\textbf{#1}}

\newif\iflongappendix
\longappendixtrue 

\newcommand{\tsdp}{TSDP\xspace}%
\newcommand{\shieldwhole}{shielding-whole-model\xspace}%
\newcommand{\noshield}{white-box\xspace}%

\SetCommentSty{mycommfont}

\newcommand\diagfil[4]{%
  \multicolumn{1}{p{#1}}{\hskip-\tabcolsep
  $\vcenter{\begin{tikzpicture}[baseline=0,anchor=south west,inner sep=0pt,outer sep=0pt]
  \path[use as bounding box] (0,0) rectangle (#1+2\tabcolsep,\baselineskip);
  \node[minimum width={#1+2\tabcolsep},minimum height=\baselineskip+\extrarowheight+\belowrulesep+\aboverulesep,fill=#2] (box)at(0,-\aboverulesep) {};
  \fill [#3] (box.south west)--(box.north east)|- cycle;
  \node[anchor=center] at (box.center) {#4};
  \end{tikzpicture}}$\hskip-\tabcolsep}}

\begin{document}

\title{No Privacy Left Outside: On the (In-)Security of TEE-Shielded DNN Partition for On-Device ML}

\author{\IEEEauthorblockN{
Ziqi Zhang\IEEEauthorrefmark{1}, 
Chen Gong\IEEEauthorrefmark{2}, 
Yifeng Cai\IEEEauthorrefmark{1}, 
Yuanyuan Yuan\IEEEauthorrefmark{3}, \\
Bingyan Liu\IEEEauthorrefmark{4},   
Ding Li\IEEEauthorrefmark{1},
Yao Guo\IEEEauthorrefmark{1},
Xiangqun Chen\IEEEauthorrefmark{1}
}
\IEEEauthorblockA{
    \IEEEauthorrefmark{1}Key Laboratory of High-Confidence Software Technologies (MOE), School of Computer Science, Peking University\\
    \IEEEauthorrefmark{2}School of Computer Science, Peking University\\
    \IEEEauthorrefmark{3}The Hong Kong University of Science and Technology, \\
    \IEEEauthorrefmark{4}School of Computer Science, Beijing University of Posts and Telecommunications\\
\{ziqi\_zhang,gongchen17,caiyifeng,ding\_li,yaoguo,cherry\}@pku.edu.cn, \\ yyuanaq@cse.ust.hk, bingyanliu@bupt.edu.cn}
}

\maketitle

\begin{abstract}
On-device ML introduces new security challenges: DNN models become white-box
accessible to device users. Based on white-box information, adversaries can
conduct effective model stealing (MS) against model weights and membership
inference attack (MIA) against training data privacy. Using Trusted Execution
Environments (TEEs) to shield on-device DNN models aims to downgrade (easy)
white-box attacks to (harder) black-box attacks. However, one major shortcoming
of TEEs is the sharply increased latency (up to 50$\times$).
To accelerate TEE-shield DNN computation with GPUs, researchers proposed several
model partition techniques. These solutions, referred to as TEE-Shielded DNN
Partition (\tsdp), partition a DNN model into two parts, offloading\footnote{In
this paper, ``offload'' refers to executing computation-intensive DNN
operations on insecure devices with strong computation ability (\eg GPUs),
rather than secure devices with weak computation ability (\eg TEEs).} the
privacy-insensitive part to the GPU while shielding the privacy-sensitive part
within the TEE. 
However, the community lacks an in-depth understanding of the seemingly encouraging privacy guarantees
offered by existing \tsdp\ solutions during DNN inference. This paper benchmarks
existing \tsdp\ solutions using both MS and MIA across a variety of DNN models,
datasets, and metrics. We show important findings that existing \tsdp\ solutions
are vulnerable to privacy-stealing attacks and are \textit{not} as safe as
commonly believed. We also unveil the inherent difficulty in deciding optimal
DNN partition configurations (i.e., the highest security with minimal utility
cost) for present \tsdp\ solutions. The experiments show that such ``sweet
spot'' configurations vary across datasets and models.
Based on lessons harvested from the experiments, we present \tool, a novel
\tsdp\ method that defends against MS and MIA during DNN inference. Unlike
existing approaches, \tool follows a \partitionbeforetrain strategy, which
allows for accurate separation between privacy-related weights from public
weights. \tool\ delivers the same security protection as shielding the entire
DNN model inside TEE (the ``upper-bound'' security guarantees) with over
$10\times$ less overhead (in both experimental and real-world environments) than
prior \tsdp\ solutions and no accuracy loss.
We make the code and artifacts publicly available on the Internet.
\end{abstract}

\section{Introduction}

On-device machine learning (ML) has become an important paradigm for latency-
and privacy-sensitive
tasks~\cite{sun2021mind,zhu2021hermes,deng2022understanding,sun2020shadownet}
on mobile and IoT devices. However, on-device learning also introduces new
security threats to the deployed deep neural network (DNN) models: by making
\textit{DNN models white-box accessible to device users}, adversaries can obtain
full model information and easily achieve high attack accuracy with much less
cost for representative attacks like \ac{ms} and
\ac{mia}~\cite{hu2022membership,orekondy2019knockoff,jagielski2020high,papernot2016towards,carlini2019the,leino2020stolen}.
Therefore, one key objective for hardening on-device ML is to \textbf{prevent}
adversaries from accessing the on-device models, thereby \textbf{downgrading}
white-box \ac{ms} and \ac{mia} attacks to black-box (much harder)
settings~\cite{hu2022membership,mo2020darknetz,hou2021model,sun2020shadownet}.

Unfortunately, protecting on-device DNN models is particularly challenging due
to the security-and-utility trade-off. Algorithmic-level protections, such as
\ac{mpc}~\cite{juvekar2018gazelle}, \ac{he}~\cite{gilad2016cryptonets},
Regularization~\cite{nasr2018machine}, and \ac{dp}~\cite{dwork2014DP}, are not
applicable since they are either too computationally expensive for mobile and
IoT devices, or downgrade the accuracy of the protected models
significantly~\cite{hu2022membership,tramer2019slalom}. Employing Trusted
Execution Environments
(TEEs)~\cite{hanzlik2021mlcapsule,lee2019occlumency,kim2020vessels,li2021lasagna}
to directly host DNN models is also not practical, because shielding the whole
DNN model in TEEs (\shieldwhole) leads to about $50\times$ deduction in model
speed due to TEE's limited computation speed~\cite{tramer2019slalom}.

While protecting an entire DNN model with TEEs is infeasible for on-device ML,
recent works have advocated to protect privacy-related portion of a DNN model to
offer high utility and security. In particular, researchers propose \ac{tsdp},
which only puts a subset of the DNN model in TEEs and offloads the rest
computation on
GPUs~\cite{mo2020darknetz,hou2021model,shen2022soter,sun2020shadownet}. Existing
research broadly assumes that the offloaded part is insufficient to expose
critical private information of DNN models, meaning the exposed model parts do
not leak substantially more information than a black-box interface.
However, we argue that this assumption is questionable given a practical threat
model where adversaries have access to abundant public information on the
Internet, such as pre-trained models and public
datasets~\cite{chen2021teacher,chen2022copy,wang2018with}.
To exploit \tsdp, adversaries may use public information to analyze the
offloaded model parts and acquire more privacy than only analyzing black-box
outputs, breaking the promises of \ac{tsdp}. Nevertheless, none of existing
approaches have systematically evaluated their security promise when taking
public information into account. 

This paper conducts the first systematic empirical evaluation on the security of
\ac{tsdp} approaches. We first investigate papers published between 2018--2023
in prestigious venues, including IEEE S\&P, MobiSys, ATC, ASPLOS, RTSS, MICRO,
AAAI, ICLR, MICRO, and TDSC. We then put the reviewed approaches into five
categories based on their technical pipelines, and empirically evaluated each
category with \ac{ms}/\ac{mia} initiated by a practical adversary with public
information on hand.

The experiment shows that existing \ac{tsdp} approaches expose substantial
private information to attackers via the offloaded model weights, enabling
approximately white-box quality attacks toward TEE shielded models.
\ac{ms} attack accuracies toward existing \ac{tsdp} solutions are $3.76\times$
-- $3.92\times$ higher than the black-box (\shieldwhole) baseline. For
comparison, the unprotected \noshield baseline (offloading an entire DNN model
outside TEE) has a $4.24\times$ higher accuracy than the \shieldwhole setting.
The results for \ac{mia} are similar. Existing \ac{tsdp} approaches have
$1.16\times$ -- $1.28\times$ higher \ac{mia} accuracy in comparison to the
\shieldwhole baseline, while the accuracy for the \noshield setting is
$1.36\times$ higher.

Worse still, we encountered high difficulties to augment the security of
existing \ac{tsdp} approaches without making substantial changes to their
methodologies. To illustrate this, we measured the attack accuracy of
\ac{ms}/\ac{mia} using various configurations of existing \ac{tsdp} approaches.
We found it particularly difficult to determine a ``sweet spot'' configuration,
that can maximize a DNN model's utility while still satisfying security
requirements. Specifically, given a
maximally tolerant attack accuracy, existing \ac{tsdp} approaches require
substantially different settings to config the shielded part when protecting
different models and datasets. Therefore, for all existing \ac{tsdp} approaches,
we need to conduct an empirical procedure to identify the ``sweet spot''
configuration for specific models and datasets. However, such empirical
procedures are prohibitively expensive, given the vast number of potential model
and dataset combinations.

The fundamental weakness of existing \tsdp approaches is that they follow a
\textit{\trainbeforepartition} strategy. This involves first training a private
model with a public pre-trained model and private data, and then separating the
model into two parts: a shielded part that runs in TEEs, and an offloaded part
that runs out of TEEs. Since training occurs before model partition,
privacy-related weights may likely pervade the entire model. Thus, it is hard
for existing \ac{tsdp} solutions to accurately \textit{isolate} privacy-related
weights, creating potential attack surfaces. 

To ensure the security of \ac{tsdp} solutions, we propose a slicing-based
approach, called \sys, that accurately \textit{isolates} privacy-related weights
from offloaded weights at the inference stage. \sys adopts a
\textit{\partitionbeforetrain} strategy, which first partitions a DNN model into
a backbone and multiple private slices, then reuses public pre-trained models as
the backbone, and at last trains the slices with private data. Therefore, \sys
accurately separates privacy-related weights from offloaded weights and is able
to shield \textit{all} privacy-related weights in TEEs.

The key challenge to realizing the \partitionbeforetrain strategy is to ensure
that the private slices are small enough to run in TEEs while maintaining a
decent accuracy. To this end, we propose a dynamic pruning algorithm that first
trains the private slices with large sizes that have a sufficient model capacity
for high accuracy, and then optimizes the size of the slices automatically under
a threshold of accuracy loss. In this way, \sys automatically finds the ``sweet
spot'' configuration, which minimizes the number of slices (computation) inside
TEE while keeping the same accuracy as the corresponding unpartitioned
model.

Our evaluation shows that \sys outperforms existing \ac{tsdp} approaches in
terms of security guarantee and utility cost. Attackers can hardly obtain any
private information by analyzing the model architectures, and as a result, \sys\
is shown to achieve the security level of \shieldwhole baseline with $10\times$
less computation cost (in both experimental and real-world
environments) than other \tsdp\ solutions. Besides, \sys reaches a
high-security level with nearly no sacrifice. The statistical evaluation shows
no change between the accuracy of the \sys protected model and the original
unpartitioned model. The offloaded public backbone does not increase the attack
performance as well.  
The contribution of this paper can be summarized as follows:

\begin{itemize}

    \item We systematically evaluate the security guarantee of prior \tsdp
    solutions using two representative attacks, \ac{ms} and \ac{mia}, and reveal
    the security issues of these solutions.
    \item We illustrate the difficulty of improving the security of prior
    \ac{tsdp} approaches without substantially changing their methodologies. 

    \item We propose \sys, a novel \tsdp\ solution for DNN inference that
    isolates privacy from offloaded model parts to provide a strong security
    guarantee using TEEs and cryptographic primitives. Our thorough evaluation
    shows that \sys offers a high security guarantee with moderate overhead and
    no accuracy loss.
\end{itemize}

\noindent \textbf{Availability:} The artifact is available
at~\cite{TEESliceArtifact}. We also provide supplementary
experimental results of this paper at~\cite{TEESliceWebsite}.

\section{Background and Threat Model}
\label{sec:threat_model}

\subsection{Background}

\parh{TEE.}
A Trusted Execution Environment (TEE) is an isolated hardware enclave that
stores and processes sensitive data. Popular TEE implementations include Intel
SGX~\cite{mckeen2013innovative}, AMD SEV~\cite{kaplan2016amd}, and
TrustZone~\cite{alves2004trustzone}. In this paper, we follow prior work and
deem TEE as \textit{a secure area on a potential adversary host device
(including
GPUs)}~\cite{mo2020darknetz,hou2021model,shen2022model,sun2020shadownet}.
It means \textit{the data, code, and the whole computation process} inside TEEs
are secure. Although there are side-channel attacks that may leak
sensitive data from TEE, they are out of our consideration. 

\parh{TSDP Solutions.}
TSDP solutions aim to provide a black-box label-only protection against MS/MIA
by shielding partial DNN models inside TEEs. The motivation is to reduce
inference latency of the straightforward black-box protection that shields the
whole model inside TEEs (increase latency by up to
50$\times$~\cite{tramer2019slalom}).
The security goal of \tsdp solutions is to \textit{downgrade white-box MS/MIA
against on-device models to black-box label-only
attacks}~\cite{mo2020darknetz,hou2021model,sun2020shadownet,shen2022soter}. Such
degeneration is important and practical for deployed DNN models in production
environments. For MS, \tsdp solutions enforce accurate, cheap (usually taking
negligible number of queries) white-box
attacks~\cite{zhu2021hermes,rakin2022deepsteal} to expensive (usually taking
tens of thousands of queries) and inaccurate black-box
attacks~\cite{orekondy2019knockoff}. For MIA, \tsdp solutions provide a
deployment framework to guarantee differential privacy requirements with little
accuracy sacrifice~\cite{hu2022membership}.

\subsection{Threat Model}

\parh{Defender's Goal.}
The goal of this paper (and prior \tsdp solutions) is to degrade white-box
attacks to black-box label-only attacks. We consider the security guarantee of a
black-box baseline, where TEE shields the whole DNN model and only returning
prediction labels, as the \textit{upper bound} security protection offered by
\ac{tsdp}
approaches~\cite{mo2020darknetz,hou2021model,shen2022soter,sun2020shadownet}. We
however do \textit{not} aim to completely mitigate information leakage from TEE
outputs (i.e., prediction labels).

\parh{Model Output.}
Following prior \ac{tsdp} papers and also real-world
productions~\cite{mo2020darknetz,hou2021model,shen2022soter,sun2020shadownet},
we assume that the deployed models only generate labels 
to users (i.e., ``label-only outputs''). In other words, we assume that the confidence of the model
prediction is an intermediate result, therefore, can be protected by TEEs.
This assumption is supported by a comprehensive survey on the output type of
on-device ML systems~\cite{sun2021mind}. Further, we also surveyed the eight
most important on-device ML tasks. For each task, we collect the three most
downloaded Android applications (24 apps in total) over three different
application markets (Google Play, Tencent My App, and 360 Mobile Assistant). We
manually checked the output type of the applications and found that \textit{all}
of the 24 applications only return the prediction label and keep the confidence of models in the intermediate results. 
We list the type of tasks and app names in
\T~\ref{append:tbl:label_only_app}. The survey illustrates that the label-only assumption is realistic for the on-device ML models.

\parh{Defender/Adversary's Capability.}
We assume that both the model owner (defender) and the attacker can use the public
model on the
Internet~\cite{wang2018with,chen2021teacher,pal2019a}
to improve the accuracy of the model or attacks, a realistic setting for modern on-device learning tasks~\cite{pytorchhub,
tensorflowhub,tensorflow_tl,pytorchmodelzoo,zhuang2021a,deng2009imagenet,wang2018with,chen2021teacher}.  The attacker can infer the architecture of the whole protected model, or an equivalent one, based on the
public information, such as the inference results or the unprotected model part,
with existing
techniques~\cite{chen2021teacher,chen2022copy,hou2021model,mo2020darknetz,hashemi2021darknight}.
Besides, we assume that the attacker can query the victim model for limited times
(less than 1\% of the training data), a practical assumption shared by related
work~\cite{rakin2022deepsteal,hua2018reverse,yan2020cache}. For simplicity, we
denote the victim model as $M_{\rm vic}$, the public model as $M_{\rm pub}$, and
the surrogate model produced by model stealing as $M_{\rm sur}$.

\section{Evaluating Existing \tsdp Defenses}
\label{sec:evaluate_existing_solutions}

We first conduct a thorough literature review on recent publications in top
conferences and journals and identify five categories of \ac{tsdp} techniques.
Then, we implement a representative technique from each category and evaluate
their security via \ac{ms} and \ac{mia}. We assess if they were sufficiently
secure against the launched attacks, and we harvest empirical observations to
present lessons from the evaluation.

\begin{table}[!htbp]
    \caption{A taxonomy of existing \tsdp\ solutions. We mark
    \colorbox{blue!30}{representative works} empirically assessed in this study.
    Other works are ignored in our empirical evaluation and are just part of the
    literature review. 
    }
    \centering
    \label{tbl:literature_short}
    \begin{adjustbox}{max width=0.8\linewidth}
    \begin{tabular}{ccc}
    \hline
    Literature                                                 & Conference/Journal & Category                                                                                 \\ \hline
    \colorbox{blue!30}{DarkneTZ}~\cite{mo2020darknetz}            & MobiSys 2020 & \multirow{3}{*}{\begin{tabular}[c]{@{}c@{}}Shielding\\ Deep Layers\end{tabular}}         \\
    PPFL \cite{mo2021ppfl}                    & MobiSys 2021 &                                                                                          \\
    Shredder \cite{mireshghallah2020shredder} & ASPLOS 2020  &                                                                                          \\\hline
    Yerbabuena \cite{gu2018yerbabuena}        & Arxiv 2018   & \multirow{3}{*}{\begin{tabular}[c]{@{}c@{}}Shielding\\ Shallow Layers\end{tabular}}      \\
    \colorbox{blue!30}{Serdab} \cite{elgamal2020serdab}           & CCGRID 2020  &                                                                                          \\
    Origami \cite{narra2019origami}           & Arxiv 2019   &                                                                                          \\\hline
    \colorbox{blue!30}{Magnitude}~\cite{hou2021model}             & TDSC 2022    & \begin{tabular}[c]{@{}c@{}}Shielding\\ Large-Mag. Weights\end{tabular}                   \\\hline
    AegisDNN \cite{xiang2021aegisdnn}         & RTSS 2021    & \multirow{2}{*}{\begin{tabular}[c]{@{}c@{}}Shielding\\ Intermediate Layers\end{tabular}} \\
    \colorbox{blue!30}{SOTER} \cite{shen2022soter}                & ATC 2022     &                                                                                          \\\hline
    \colorbox{blue!30}{ShadowNet} \cite{sun2020shadownet}         & S\&P 2023   & \begin{tabular}[c]{@{}c@{}}Shielding Non-Linear \\ Layers \& Obfuscation\end{tabular}                   \\\hline
    DarKnight \cite{hashemi2021darknight}     & MICRO 2021   & -                                                                                        \\
    Goten \cite{lucien2021goten}              & AAAI 2021    & -                                                                                        \\
    Slalom~\cite{tramer2019slalom}            & ICLR 2018    & -                                                                                        \\
    GINN \cite{asvadishirehjini2022ginn}      & CODASPY 2022 & -                                                                                        \\
    eNNclave \cite{schlogl2020ennclave}       & AISec 2020   & -                                                                                        \\ \hline
    \end{tabular}

    \end{adjustbox}

    \end{table}

\subsection{Literature Summary}
\label{sec:literature_category}

\parh{Defense Taxonomy.}~We identify publications that partition DNNs across
TEEs and GPUs for privacy-preserving ML from leading conferences and journals
from the past five years. \T~\ref{tbl:literature_short} lists 15 important works
from computer security, computer systems, mobile computing, artificial
intelligence, and computer architecture.

\begin{figure*}[!t]
    \centering
    \includegraphics[width=\linewidth]{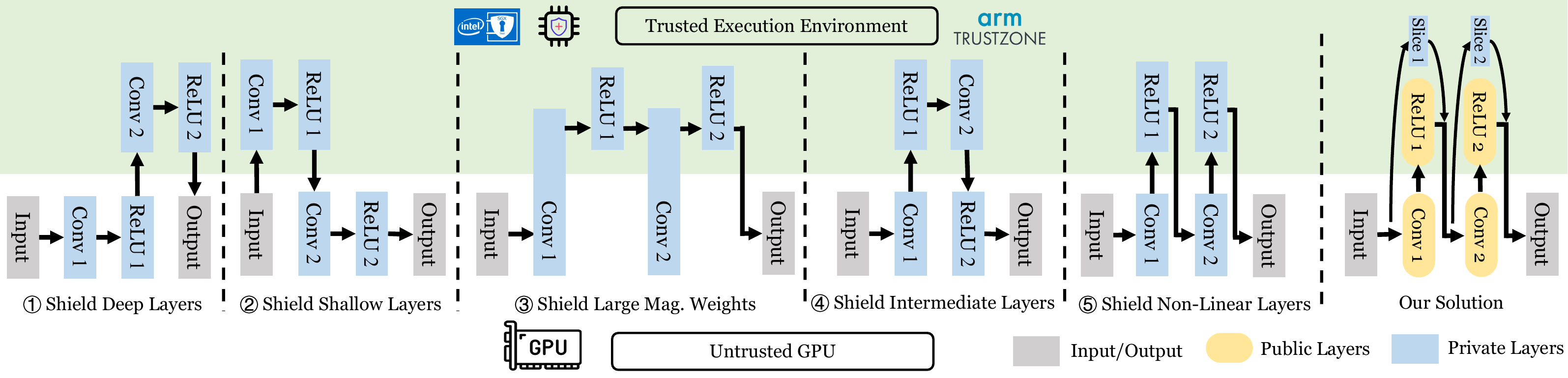}
    \vspace*{-15pt}
    \caption{An illustration of different \tsdp solutions on a four-layer DNN.
    \colorbox{blue!15}{Blue} squares are privacy-related layers, and
    \colorbox{yellow!80}{yellow} rounded squares are privacy-irrelevant (public)
    layers. \ding{172} shields two deep layers (Conv2 and ReLU2) and \ding{173}
    shields two shallow layers (Conv1 and ReLU1). \ding{174} shields the
    large-magnitude weight of each layer. \ding{175} shields two random
    intermediate layers (ReLU1 and Conv2). \ding{176} shields non-linear layers
    (ReLU1 and ReLU2) and obfuscates other layers (Conv1 and Conv2). Our solution
    (introduced in \S~\ref{sec:approach}) shields privacy-related
    slices and non-linear layers of the public backbone model.}
    \label{fig:model_partition_categories}
    \vspace*{-13pt}
    \end{figure*}

Five papers do not meet the requirements under our threat model. Among them
three are unable to defend models against \ac{ms}
(DarKnight~\cite{hashemi2021darknight}, Slalom~\cite{tramer2019slalom}, and
GINN~\cite{asvadishirehjini2022ginn}), one has a stronger assumption that
requires two TEEs to verify each other (Goten~\cite{lucien2021goten}), and the
last one decreases DNN accuracy significantly
(eNNclave~\cite{schlogl2020ennclave}; details in
\S~\ref{sec:experiment:trade-off}). We then divide the remaining ten papers into
five categories depending on the \tsdp schemes.
\F~\ref{fig:model_partition_categories} schematically illustrates each category
using a sample 4-layer DNN model (including two convolution layers and two ReLU
layers). We also include our proposed approach (details in
\S~\ref{sec:approach}) for comparison. The five categories are as follows:

\noindent \ding{172} \underline{Shielding Deep Layers} partitions the DNN
according to the layer depth and places the layers close to the output layer in
the TEE. In \F~\ref{fig:model_partition_categories}, two deepest layers (Conv2
and ReLU2) are shielded.

\noindent \ding{173} \underline{Shielding Shallow Layers} partitions the DNN
according to the layer depth and places the layers close to the input layer in
the TEE. In \F~\ref{fig:model_partition_categories}, two shallowest layers
(Conv1 and ReLU1) are shielded.

\noindent \ding{174} \underline{Shielding Large-Magnitude Weights} partitions
the DNN according to the absolute weight value, and then puts the weights with
large magnitudes and ReLU layers in the TEE.
\F~\ref{fig:model_partition_categories} shields partial convolution layers (to
represent large-magnitude weights) and ReLU layers. 

\noindent \ding{175} \underline{Shielding Intermediate Layers} puts randomly
chosen intermediate layers in the TEE. \F~\ref{fig:model_partition_categories}
shields ReLU1 and Conv2 as the random-selected layers.

\noindent \ding{176} \underline{Shielding Non-Linear Layers and Obfuscation}
partitions the DNN by the layer types and shields non-linear (\eg ReLU) layers
using TEE. The offloaded linear layers (\eg convolution layers) are protected by
lightweight obfuscation algorithms (\eg matrix transformation).
\F~\ref{fig:model_partition_categories} shields the ReLU layers and offloads all
convolution layers.

\subsection{Representative Defenses}
\label{subsec:defense-selection}

\parh{Scheme Selection.}~For each of the five \tsdp\ schemes, we select one
representative solution for evaluation. For shielding deep layers (\ding{172}),
we choose DarkneTZ because according to related
work~\cite{hashemi2021darknight,mo2021ppfl}, it is the state-of-the-art (SOTA)
solution for protecting training data privacy. For shielding shallow layers
(\ding{173}) and shielding intermediate layers (\ding{175}), we select Serdab
and SOTER because they are the most recent papers published in peer-reviewed
conferences. For shielding large-magnitude weights (\ding{174}) and shielding
non-linear layers (\ding{176}), we choose Magnitude~\cite{hou2021model} and
ShadowNet~\cite{sun2020shadownet} since they are the only solutions in their
respective categories. 

\parh{Configuration Setting.}~All these schemes require configurations, e.g.,
for \ding{172}, we need to configure the exact number of ``deep'' layers in the
TEE. Overall, we configure each defense scheme according to their papers. In
particular, for DarkneTZ (\ding{172}), we put the last classification layer into
TEE. For Serdab (\ding{173}), the TEE shields the first four layers. For
Magnitude (\ding{174}), the TEE shields 1\% weights with the largest magnitudes.
For SOTER (\ding{175}), the TEE shields 20\% randomly-selected layers and
multiplies the other offloaded layers with a scalar to conceal the weight
values. For ShadowNet (\ding{176}), the TEE shields all the ReLU layers and
obfuscates all the offloaded convolution layers with matrix transformation and
filter permutation (detailed description in \App~\ref{sec:append:shadownet}).
Furthermore, we note that selecting proper configurations constitutes a key
factor that undermines their security/utility guarantee. We will discuss other
configurations later in \S~\ref{sec:dilemma}.

\subsection{Evaluated Attacks}
\label{sec:evaluated_attacks}

\parh{Attack Selection.}~We consider \ac{ms} and \ac{mia} attacks that can
extract confidential model weights and private training data as the security
benchmark for existing \ac{tsdp} approaches.
For \ac{ms}, we employ standard query-based stealing techniques where the
attacker trains a model from a set of collected data labeled by the
partially-shielded $M_{vic}$. Query-based \ac{ms} has been widely adopted in
literature~\cite{orekondy2019knockoff, jagielski2020high,
shen2022model,orekondy2020prediction}. We leverage the attack implementation
offered by Knockoff Net~\cite{KnockoffNetCode}, a SOTA baseline accepted by
prestigious
publications~\cite{orekondy2020prediction,jagielski2020high,shen2022model}. 
For \ac{mia}, we chose the transfer attack that is designed against the 
label-only scenario~\cite{li2021membership}. Transfer attack builds $M_{\rm
sur}$ to imitate the behavior of $M_{\rm vic}$ and infer the privacy of $M_{\rm
vic}$ from white-box information of $M_{\rm sur}$ (\eg, confidence scores, loss,
and gradients). The intuition is that membership information can transfer from
$M_{\rm vic}$ to $M_{\rm sur}$. We chose the standard confidence-score-based
algorithm to extract membership from $M_{\rm sur}$. In particular, this process
trains a binary classifier, such that given the confidence score of $M_{\rm
sur}$, the classifier predicts if the corresponding DNN input is in the training
data of $M_{\rm vic}$. Confidence-score-based \ac{mia} has been extensively used
in previous attacks~\cite{papernot2016towards, salem2019mlleak,
shokri2017membership,li2021membership,nasr2018machine}, and we reused the attack
implementation from a recent benchmark suite,
\textsc{ML-Doctor}~\cite{MLDoctorCode,liu2022mldoctor}. 
Recent work has consistently employed \textsc{ML-Doctor} for membership
inference~\cite{shen2022model,chen2021when,carlini2022membership}.

\begin{figure}[!t]
\centering
\includegraphics[width=0.85\linewidth]{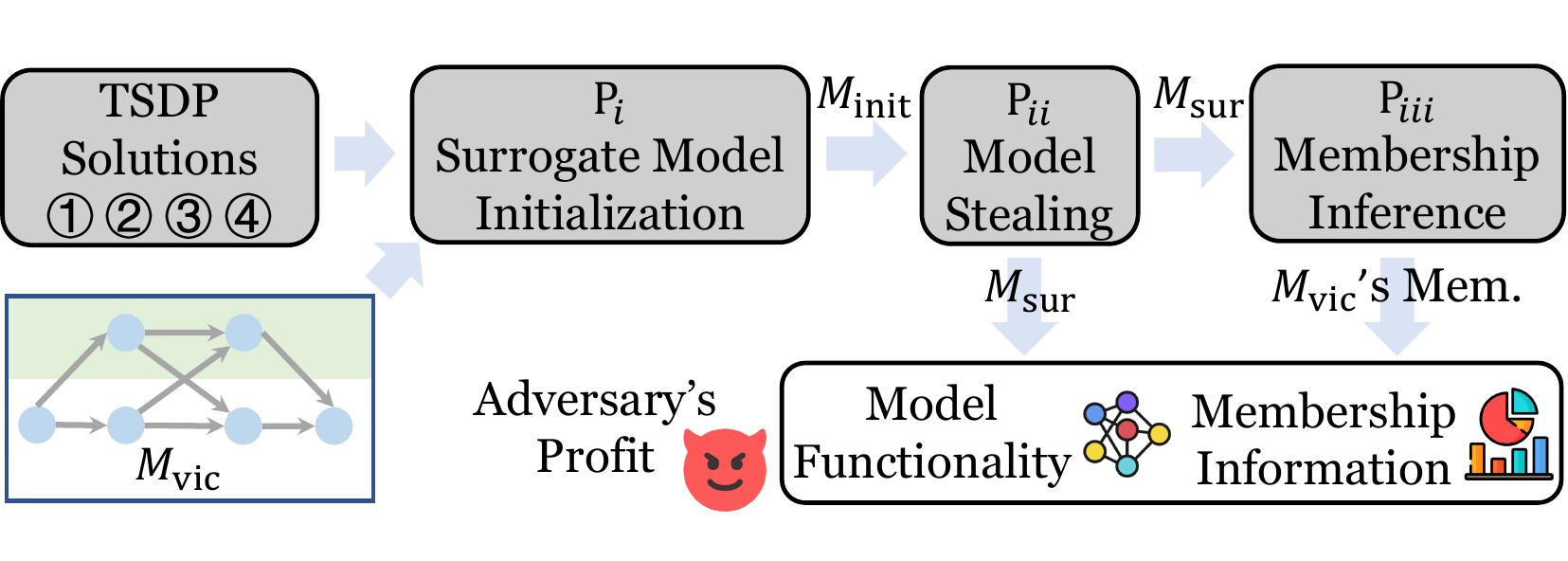}
\vspace{-5pt}
\caption{A three-phase attack pipeline.}
\label{fig:attack_pipeline}
\vspace{-15pt}
\end{figure}

\parh{Attack Pipeline.}~As in \F~\ref{fig:attack_pipeline}, the attack goal is
to get the victim model $M_{vic}$'s functionality and membership information.
The attack pipeline consists of three phases: surrogate model initialization
(P$_{i}$), \ac{ms} (P$_{ii}$), and \ac{mia} (P$_{iii}$). The attacker first
analyzes the target defense scheme and then conducts P$_{i}$ to get an
initialized surrogate model (denoted as $M_{\rm init}$). P$_{ii}$ trains $M_{\rm
init}$ with queried data and outputs the surrogate model $M_{\rm sur}$ (the
recovered victim model). P$_{iii}$ takes $M_{\rm sur}$ as input, uses \ac{mia}
algorithms and outputs $M_{\rm vic}$'s membership privacy. Specifically, in
P$_{iii}$, the adversary first trains a binary classifier. Then, given an input
$i$ and the $M_{\rm sur}$'s prediction confidence score $p$, the classifier
decides if $i$ belongs to the $M_{\rm vic}$'s training data by taking $p$ as its
input~\cite{shokri2017membership,carlini2022membership,yuan2022membership}.

\parh{Surrogate Model Initialization.}~Steps in our attack pipeline are
automated except for the first step, surrogate model initialization (P$_i$). To
run the attacks, attackers first need to construct $M_{\rm init}$ with the
exposed knowledge in the public part of the \ac{tsdp} protected models. The
high-level process to construct $M_{\rm init}$ has three steps. First, the
attacker infers the architecture of the \tsdp\ protected model based on the
offloaded part and the model output with existing
techniques~\cite{chen2021teacher,chen2022copy}. Then, the attacker chooses a
public model $M_{\rm pub}$ (with the same or an equivalent architecture) as
$M_{\rm init}$. Lastly, the attacker transports the model weights in the
offloaded part of $M_{\rm vic}$ to the corresponding parts of $M_{\rm init}$.
For DarkneTZ (\ding{172}), Serdab (\ding{173}), and SOTER (\ding{175}), we use
the offloaded layers to replace the corresponding layers of $M_{\rm init}$. For
Magnitude (\ding{174}), we use the offloaded weights that run on GPUs to replace
the corresponding weights in $M_{\rm init}$. For ShadowNet (\ding{176}), the
attacker uses the public model to decode the obfuscation algorithm (detailed
description in \App~\ref{sec:append:shadownet}) and uses the decoded weights to
initialize $M_{\rm init}$.

\parh{Comparison Baselines.}~To ease the comparison, we also provide baseline
evaluation results. Notably, for \ac{ms}, we consider the
\noshield solution (offloading the whole $M_{\rm vic}$ to GPU, referred to as
``No-Shield'') as the easiest baseline because the adversary can directly use
the offloaded $M_{\rm vic}$ as $M_{\rm sur}$ and does not need to train the
model. We consider a black-box (or \shieldwhole) setting (referred to as
``Black-box'') where attackers can only access the model prediction labels and
identify the $M_{\rm vic}$'s corresponding public models. However, the attacker
cannot leverage the $M_{\rm vic}$'s weights offloaded on GPU to construct
$M_{\rm init}$. This is a challenging setting, and the attacker needs to steal
every layer's weights from $M_{\rm vic}$.
As for \ac{mia}, the \noshield setting denotes that the attackers
can directly use the $M_{\rm vic}$'s output confidence score for membership
inference because the whole $M_{\rm vic}$ is offloaded. Recall, as noted in our
threat model (\S~\ref{sec:threat_model}), \ac{tsdp} solutions often refuse to
return confidence scores and only return predicted labels to mitigate
membership inference. In contrast, the black-box setting denotes that the
attacker directly uses $M_{\rm pub}$ as $M_{\rm init}$ and trains $M_{\rm sur}$
from queried data. Apparently, this is comparable to ``random guess'' (success
rate around 50\%\footnote{In the evaluation setting (\S~\ref{subsec:evaluation-setup}), we follow prior \ac{mia} work to set the size of target training dataset equal the size of target testing dataset.}), given that a $M_{\rm init}$ (\ie, $M_{\rm pub}$) contains no
information on the $M_{\rm vic}$'s private training datasets. 

\subsection{Evaluation Setting}
\label{subsec:evaluation-setup}

\parh{Datasets.}~We use four different datasets and five DNN models to evaluate
different defenses. The dataset and model selection refers to prior \ac{ms} and
\ac{mia} literatures~\cite{liu2022mldoctor,jagielski2020high}. In particular,
the datasets include CIFAR10, CIFAR100~\cite{krizhevsky2009learning},
STL10~\cite{coates2011an}, and UTKFace~\cite{zhang2017age}. CIFAR10 and CIFAR100
are default choices of prior \ac{ms}/\ac{mia}
literatures~\cite{orekondy2020prediction,jagielski2020high,liu2022mldoctor,carlini2022membership,yuan2022membership}.
STL10 and UTKFace are used by \textsc{ML-Doctor} to quantify model
vulnerability~\cite{liu2022mldoctor}. 
We present further details of each dataset in \App~\ref{append:dataset}. 
When evaluating \ac{ms}, we use each dataset's
default train/test split to avoid possible bias. To evaluate \ac{mia}, we follow
the setting of \textsc{ML-Doctor} to partition each dataset into four splits:
the target training dataset, the target testing dataset, the shadow training
dataset, and the shadow testing dataset. The model owner uses the target
training and the target testing datasets to train and evaluate the victim model
$M_{\rm vic}$. The adversary uses the shadow training dataset and the shadow
testing dataset to train the binary classifier. This is a common setting for
evaluating
\ac{mia} in prior work~\cite{carlini2022membership,shokri2017membership,shen2022model,li2021membership}.

\parh{Models.}~The benchmark models include ResNet18~\cite{he2016deep},
VGG16\_BN~\cite{simonyan2015very}, AlexNet~\cite{alex2012imagenet}, and
ResNet34. These models are widely used in prior security
studies~\cite{liu2022mldoctor,chen2021teacher,yuan2022membership,rakin2022deepsteal}.
We mainly report the results on AlexNet, ResNet18, and VGG16\_BN and leave the
results of ResNet34 and VGG19\_BN in
\App~\ref{append:sec:resnet34_vgg19_results}. 
We set the hyper-parameters following the paper and released code of prior
works~\cite{MLDoctorCode,KnockoffNetCode}. As introduced in
\F~\ref{fig:attack_pipeline}, for all cases, we use the public models as
initialization to get a better model
performance~\cite{orekondy2019knockoff,orekondy2020prediction}. For the training
in the \ac{ms} part, we follow the hyper-parameter settings of Knockoff
Nets~\cite{KnockoffNetCode}. Accuracies of the trained models $M_{\rm vic}$ are
reported in \T~\ref{tbl:nettailor_accuracy}. The accuracies are generally
consistent with public results~\cite{STL10ACC,CIFAR10ACC,CIFAR100ACC}. For the
training in the \ac{mia} part, we follow the settings of
\textsc{ML-Doctor}~\cite{MLDoctorCode}. 
Model accuracies are reported in \App~\ref{append:mia_victim_accuracy}. 
All models achieve high accuracy on the
target training dataset, and the accuracies are consistent with prior
works~\cite{liu2022mldoctor}. 
We leave detailed settings of hyper-parameters in
\App~\ref{append:vic_hyperparameters}.

\parh{Metrics.}~Overall, we systematically evaluate the effectiveness of de
facto \tsdp\ solutions using three \ac{ms} metrics and four
\ac{mia}~\cite{jagielski2020high, rakin2022deepsteal, nasr2019comprehensive,
song2020information, yuan2022membership}. Specifically, we record \ac{ms}
accuracy, fidelity, and \ac{asr} according to prior
work~\cite{jagielski2020high,orekondy2019knockoff,zhu2021hermes,rakin2022deepsteal}.
For \ac{mia}, we use confidence-based \ac{mia} accuracy,
gradient-based \ac{mia} accuracy, generalization gap, and confidence gap following prior literature as
well~\cite{nasr2019comprehensive,song2020information,yuan2022membership}.
The definition of these metrics are included in
\App~\ref{append:sec:metric_definition}.
Due to limited space, in the main paper, we only report ``\ac{ms} accuracy''
(denoted as ``Model Stealing'') and ``confidence-based \ac{mia} attack
accuracy'' (denoted as ``Membership Inference'') as the main metrics. 
We report the other metrics in \App~\ref{append:sec:other_metrics:evaluation}
(\T~\ref{tbl:evaluate_solution:fidelity} to
\T~\ref{tbl:evaluate_solution:grad_attack}). 
The findings are consistent with the main paper.
\ac{ms} accuracy denotes the prediction accuracy of the $M_{\rm
sur}$~\cite{zhu2021hermes,orekondy2019knockoff,jagielski2020high,shen2022model}. A higher
accuracy indicates that $M_{\rm sur}$ successfully steals more functionality
from $M_{\rm vic}$. As for the \ac{mia} accuracy, a higher accuracy
denotes that attackers can more accurately decide if a given sample is in
$M_{\rm vic}$'s training dataset.

\subsection{Attack Results}
\label{sec:empirical_evaluation_results}

We aim to answer the following research question primarily:

\vspace{-1pt}
\begin{tcolorbox}[size=small]
    \textbf{RQ1:} How secure are the selected \tsdp\ solutions against
    model and data privacy stealing?
\end{tcolorbox}
\vspace{-3pt}

We report the evaluation results over three models (AlexNet, ResNet18, and
VGG16\_BN) in \T~\ref{tbl:evaluate_solution} (total 12 cases). As
aforementioned, we also report the baseline settings (``No-Shield'' and
``Black-box'') for comparison. 
The results of the other two models are generally consistent with findings in
\T~\ref{tbl:evaluate_solution}. See their results in
\App~\ref{append:sec:resnet34_vgg19_results}
(\T~\ref{tbl:evaluate_solution_append}). 
To compare the performance between different defenses, for each case, we compute
a relative accuracy as the times of the accuracy over the accuracy of the
black-box baseline. We report the average relative accuracy in the last row of
\T~\ref{tbl:evaluate_solution}. A defense scheme is considered more effective if
its corresponding attack accuracy is closer to the black-box baseline (the
relative accuracy is closer to 1.00$\times$). As \T~\ref{tbl:evaluate_solution}
shows, for the \noshield baseline (the whole $M_{\rm vic}$ is offloaded to the
GPU), the relative accuracies are $4.26\times$ for model stealing and
$1.39\times$ for membership inference.

\colorlet{best}{green!30}
\colorlet{high}{red!30}
\colorlet{low}{yellow}

\begin{table*}[!h]
\caption{Attack accuracies regarding representative defense schemes. ``C10'',
``C100'', ``S10'', and ``UTK'' represent CIFAR10, CIFAR100, STL10, and UTKFace,
respectively. The last row reports the average accuracy toward each
defense relative to the baseline black-box solutions. For each setting, we mark
the highest attack accuracy in \colorbox{high}{red} and the lowest accuracy in
\colorbox{low}{yellow}. Attack accuracy toward our approach
(\S~\ref{sec:approach}) is marked with \colorbox{best}{green}.}
\label{tbl:evaluate_solution}
\setlength{\tabcolsep}{1.5pt}
\begin{adjustbox}{max width=1\linewidth}

    \begin{tabular}{cccccccccccccccccccc}
    \hline
                               &      & \multicolumn{8}{c}{Model Stealing $\downarrow$}                                              & & & \multicolumn{8}{c}{Membership Inference $\downarrow$}                              \\ \cline{3-10} \cline{13-20} 
                               &      & No-Shield & \ding{172}DarkneTZ & \ding{173}Serdab  & \ding{174}Magnitude & \ding{175}SOTER & \ding{176}ShadowNet  & Ours     & Black-box & & & No-Shield & \ding{172}DarkneTZ & \ding{173}Serdab  & \ding{174}Magnitude & \ding{175}SOTER & \ding{176}ShadowNet  & Ours     & Black-box \\ \hline
    \multirow{4}{*}{\rotatebox{90}{AlexNet}}   & C10  & 83.72\%   & 77.15\%  & \cellcolor{low}63.58\% & 65.97\%   & 76.90\% & \cellcolor{high}83.57\% & \cellcolor{best}19.04\% & 24.38\%   & & & 67.25\%   & 57.67\%  & 62.96\% & \cellcolor{low}52.67\%   & 62.18\% & \cellcolor{high}69.43\% & \cellcolor{best}50.00\% & 50.00\%   \\
                               & C100 & 56.60\%   & \cellcolor{low}41.57\%  & 46.48\% & 47.86\%   & 50.83\% & \cellcolor{high}56.43\% & \cellcolor{best}8.27\% & 10.68\%   & & & 78.32\%   & \cellcolor{low}63.27\%  & 72.20\% & 71.31\%   & 63.39\% & \cellcolor{high}81.23\% & \cellcolor{best}50.00\% & 50.00\%   \\
                               & S10  & 76.55\%   & \cellcolor{high}75.17\%  & 69.06\% & 73.67\%   & 37.60\% & \cellcolor{low}35.98\% & \cellcolor{best}24.15\% & 15.26\%   & & & 65.12\%   & \cellcolor{low}58.49\%  & 61.51\% & \cellcolor{high}66.26\%   & 59.72\% & 65.57\% & \cellcolor{best}50.00\% & 50.00\%   \\
                               & UTK  & 90.01\%   & \cellcolor{high}88.74\%  & 82.92\% & 86.65\%   & \cellcolor{low}58.86\%  & 73.93\% & \cellcolor{best}52.27\% & 48.62\%   & & & 62.97\%   & 55.84\%  & \cellcolor{low}55.43\% & 56.28\%   & 55.52\% & \cellcolor{high}63.53\% & \cellcolor{best}50.00\% & 50.00\%   \\\hline
    \multirow{4}{*}{\rotatebox{90}{ResNet18}}  & C10  & 95.91\%   & 87.55\%  & \cellcolor{high}93.94\% & \cellcolor{low}89.92\%   & 92.61\% & 91.58\% & \cellcolor{best}31.40\% & 19.88\%   & & & 70.37\%   & 65.01\%  & 66.59\% & 59.12\%   & \cellcolor{low}52.67\% & \cellcolor{high}69.53\% & \cellcolor{best}50.00\% & 50.00\%   \\
                               & C100 & 81.63\%   & \cellcolor{low}70.11\%  & 78.01\% & 74.84\%   & \cellcolor{high}79.28\% & 78.51\% & \cellcolor{best}10.90\% & 15.41\%   & & & 82.75\%   & 81.10\%  & 82.92\% & \cellcolor{low}67.55\%   & 76.31\% & \cellcolor{high}83.73\% & \cellcolor{best}50.00\% & 50.00\%   \\
                               & S10  & 87.45\%   & \cellcolor{high}86.03\%  & 85.05\% & \cellcolor{low}77.08\%   & 80.83\% & 84.38\% & \cellcolor{best}29.19\% & 21.66\%   & & & 76.09\%   & 65.98\%  & \cellcolor{high}74.22\% & 64.29\%   & \cellcolor{low}59.83\% & 74.07\% & \cellcolor{best}50.00\% & 50.00\%   \\
                               & UTK  & 90.78\%   & 85.65\%  & 84.65\% & \cellcolor{low}64.99\%   & 76.43\% & \cellcolor{high}89.42\% & \cellcolor{best}51.95\% & 45.41\%   & & & 62.87\%   & 56.33\%  & 59.25\% & 54.53\%   & \cellcolor{low}51.69\% & \cellcolor{high}63.62\% & \cellcolor{best}50.00\% & 50.00\%   \\\hline
    \multirow{4}{*}{\rotatebox{90}{VGG16\_BN}} & C10  & 92.95\%   & 87.76\%  & \cellcolor{high}91.34\% & 87.35\%   & \cellcolor{low}81.52\% & 90.67\% & \cellcolor{best}30.87\% & 14.62\%   & & & 63.17\%   & \cellcolor{high}64.03\%  & 62.44\% & 58.63\%   & \cellcolor{low}55.20\% & 62.14\% & \cellcolor{best}50.00\% & 50.00\%   \\
                               & C100 & 72.78\%   & \cellcolor{low}63.68\%  & 72.19\% & 68.82\%   & 66.06\% & \cellcolor{high}72.85\% & \cellcolor{best}9.78\% & 10.93\%   & & & 81.22\%   & 78.63\%  & \cellcolor{high}81.34\% & 71.25\%   & \cellcolor{low}50.10\% & 81.13\% & \cellcolor{best}50.00\% & 50.00\%   \\
                               & S10  & 90.03\%   & 89.17\%  & 89.33\% & \cellcolor{low}84.33\%   & \cellcolor{high}89.46\% & 89.43\% & \cellcolor{best}32.92\% & 18.97\%   & & & 66.08\%   & \cellcolor{high}68.20\%  & 66.20\% & 66.97\%   & \cellcolor{low}58.22\%  & 65.85\% & \cellcolor{best}50.00\% & 50.00\%   \\
                               & UTK  & 91.51\%   & 87.60\%  & 89.60\% & 90.28\%   & \cellcolor{low}87.30\% & \cellcolor{high}91.14\% & \cellcolor{best}48.37\% & 45.46\%   & & & 58.73\%   & 52.79\%  & 58.48\% & \cellcolor{high}58.93\%   & \cellcolor{low}51.34\%  & 57.17\% & \cellcolor{best}50.00\% & 50.00\%   \\ \hline
    \multicolumn{2}{c}{Average}       & 4.28$\times$      & 3.92$\times$     & 4.03$\times$    & 3.91$\times$      & 3.76$\times$   & 4.28$\times$  & 1.23$\times$    & 1.00$\times$      & & & 1.39$\times$      & 1.28$\times$     & 1.34$\times$    & 1.25$\times$      & 1.16$\times$   & 1.39$\times$  & 1.00$\times$    & 1.00$\times$      \\ \hline
    \end{tabular}

\end{adjustbox}
\vspace{-10pt}

\end{table*}

From \T~\ref{tbl:evaluate_solution}, we can observe that the defense
effectiveness of all solutions is limited. It is evident that even the lowest
attack accuracy (marked in \colorbox{low}{yellow}), indicating the highest
defense effectiveness, among each setting, is still \textit{much higher} than
the black-box baseline: the lowest attack accuracies are averagely 3.54$\times$
higher than black-box for \ac{ms} and 1.12$\times$ higher for \ac{mia}. Even
worse, the highest attack accuracies (marked in \colorbox{high}{red}) are
similar to that of the \noshield baseline, indicating that the defense schemes
are ineffective.

The relative accuracy (w.r.t.~black-box baselines) of DarkneTZ (\ding{172}) for
\ac{ms} is 3.92$\times$ and 1.28$\times$ for \ac{mia}. For Serdab (\ding{173}),
the relative attack accuracies are 4.03$\times$ and 1.34$\times$. Since the
attack performance toward both Serdab and DarkneTZ is high, we interpret that
shielding a limited number of deep layers (\ding{172}) or shallow layers
(\ding{173}) facilitates limited protection. Magnitude (\ding{174}) achieves a
similar defense effect with DarkneTZ, with 3.91$\times$ higher accuracy for
\ac{ms} and 1.25$\times$ higher accuracy for \ac{mia}. 
Though the DarkneTZ and Magnitude papers empirically demonstrate the defense
effectiveness against naive adversaries (without the surrogate model initialized
by a pre-trained model or public data), we depict that well-designed and
practical attacks can crack such empirical settings.

SOTER (\ding{175}) offers best protection across all solutions for \ac{mia}:
in seven (out of 12) cases, SOTER achieves the lowest \ac{mia} accuracy among
the five solutions. However, its higher security strength does not come for
free. SOTER shields the largest number of layers using TEEs (20\% layers)
compared with other solutions. That is, SOTER has a much higher inference
latency and utility cost. 

Among the five schemes, ShadowNet (\ding{176}) shows the weakest
protection and most ``\colorbox{high}{red cases}'' in
\T~\ref{tbl:evaluate_solution}: five (out of 12) for \ac{ms} and six for
\ac{mia}. The average attack accuracy against ShadowNet is similar to that of
the No-Shield baseline. According to our evaluation~\cite{TEESliceWebsite}, the
adversary can recover 95\% of the obfuscated weights. This indicates that its
lightweight obfuscation (matrix obfuscation and filter permutation) is
insufficient to protect the target model in front of well-designed attacks.

\begin{tcolorbox}[size=small]
\textbf{Answer to RQ1}: Contrary to our expectation, existing \tsdp\ solutions
do not provide a black-box level security guarantee when being exploited by \ac{ms}
and \ac{mia}. That is, their shielded model weights and private training data
are vulnerable to attackers with well-prepared $M_{\rm init}$ on hand.
\end{tcolorbox}

\section{Challenges of Straightforward Mitigations}
\label{sec:dilemma}

Our study and observation in answering \textbf{RQ1} show that straightforward
mitigation to the attacks for existing \ac{tsdp} approaches is to put a larger proportion of a DNN model into TEEs to improve the protection effectiveness.
However, this straightforward solution needs to address the \textit{Security vs.
Utility Trade-off}: putting more portions of a DNN model into TEEs boosts
security but presumably diminishes utility (e.g., increases prediction latency).
Thus, the objective is to find a ``sweet spot'' configuration (i.e., how large
the TEE protected part should be) that satisfies the security requirement while
minimizing the utility overhead. This section will therefore address the
following research question:

\begin{tcolorbox}[size=small]
    \textbf{RQ2:} 
 For each of the five \tsdp\ solutions evaluated in
\S~\ref{sec:evaluate_existing_solutions}, is there a systematic approach to
identify its ``sweet spot'' configuration that simultaneously achieves high
utility and security?
\end{tcolorbox}
\vspace{-5pt}

\subsection{Problem Formalization}
To systematically find the ``sweet spots'' of the optimal size of the TEE
shielded part for the \tsdp\ solutions in
Section~\ref{sec:evaluate_existing_solutions}, we first formalize the problem as
an optimization problem. Formally, let $P$ be a \tsdp\ solution that splits a
DNN model into TEE-shielded and GPU-offloaded portions. Let $C$ denote a
configuration instance of $P$ that specifies to what degree the model is
shielded. We define two evaluation functions, $Security(C)$ and $Utility(C)$,
which quantify the security risk and the utility cost of $C$, respectively. We
also define $Security_{black}$ as the security risk baseline of a black-box setting, which puts the whole DNN model in TEE. As noted in
\S~\ref{sec:evaluated_attacks}, $Security_{black}$ denotes the lower bound of
the security risk (the strongest protection \tsdp\ can offer). Then, given the
security requirement $\Delta$, we formulate the ``sweet spot'' configuration
$C^*$ that satisfies $|\ Security(C) - Security_{black} \ | < \Delta$ with the
minimal $Utility(C)$.  Alternatively, the  ``sweet spot'' is the solution to
Equation~\ref{equ:optimal_config}.
\begin{equation}
    C^* = \mathop{\arg\min}_{ \substack{ | \ Security(C) - Security_{black} \ | < \Delta  }} \ Utility(C)
    \label{equ:optimal_config}
\end{equation}

\subsection{Experimental Settings}
\label{sec:optimal_config_setting}

To answer \textbf{RQ2}, we empirically identify the ``sweet spots'' for the five
\tsdp\ defenses evaluated in \S~\ref{sec:evaluate_existing_solutions}
w.r.t.~different security risk metrics, datasets, and shielded models. We
discuss the details of the experiment setup below.

\parh{Security Risk Metric.}~Consistent with \S~\ref{subsec:evaluation-setup},
we implement $Security(C)$ using seven security metrics. For \ac{ms}, we use
model stealing accuracy, fidelity, and \ac{asr}. For \ac{mia}, we use
confidence-based \ac{mia} accuracy, gradient-based \ac{mia} accuracy,
generalization gap, and confidence gap. Following
\S~\ref{sec:evaluated_attacks}, we mainly report the results of \ac{ms} accuracy
(denoted as ``Model Stealing'') and confidence-based \ac{mia} accuracy (denoted
as ``Membership Inference''). 
We leave the results of other metrics in
\App~\ref{append:sec:other_metrics:optimal_config_qualitative} and
\App~\ref{append:sec:other_metrics:optimal_config_quantitative}; those results
are consistent with the main findings reported in this section.

\parh{Utility Cost Metric.}~As a common setup, we use FLOPs to measure the
utility cost of DNN
models~\cite{hou2021model}. FLOPs is
a platform-irrelevant metric to assess the utility cost $Utility(C)$ by counting
the total number of multiplication and addition operations conducted inside
TEEs. We define $\% FLOPs$ as the ratio of FLOPs in the TEE over the total FLOPs
of the DNN model. According to prior work~\cite{tramer2019slalom}, the
computation speed inside TEE is about 30$\times$ slower than GPUs. Thus, a
larger $\% FLOPs(C)$ indicates fewer computations are offloaded on GPUs, leading
to higher utility costs.

We compute the FLOPs of different layers as follows. For a DNN layer, suppose
the input channel size is $c_{in}$, the output channel size is $c_{out}$,  and
the width and height of the output are $w$ and $h$. The FLOPs of a linear layer
is computed as $2 \times c_{in} \times c_{out}$. The FLOPs of a batch
normalization layer are computed as $2 \times c_{in} \times h \times w$. For a
convolution layer, suppose the kernel size is $k$, the FLOPs are computed as $2
\times c_{in} \times k^2 \times h \times w \times c_{out}$. To validate the
correctness of using $\% FLOPs$ as the utility measurement, we measured the
inference latency of different \tsdp\ solutions over different configurations on
Intel SGX with an industrial-level platform~\cite{shen2020occlum}. Experimental
results show that the inference latency increases monotonously with $\% FLOPs$; details in the Website~\cite{TEESliceWebsite}.

\parh{Datasets and Models.}~The dataset and model selection is the same as
\S~\ref{subsec:evaluation-setup}. Due to space limitations, we will report the
results on AlexNet, ResNet18, and VGG16\_BN. 
The results of ResNet34 and VGG19\_BN are displayed in
\App~\ref{append:sec:resnet34_vgg19_results}.

\parh{Configurations.}~For each \tsdp\ defense benchmarked in
\S~\ref{sec:evaluate_existing_solutions}, we iterate possible configurations to
identify $C^*$. In particular, for the defense that shields deep layers
(\ding{172}), we shield different numbers of consecutive ``deep'' layers
starting from the output layer with TEEs. Similarly, for \ding{173}, which
shields shallow layers, we put different amounts of consecutive layers starting
from the DNN input layer. For ResNet models, we use the residual layers as the
dividing boundaries. For VGG models and AlexNet models, we use convolution
layers as boundaries.

For shielding large-magnitude weights (Magnitude; \ding{174}), the number of protected weights is controlled by a configuration parameter
\texttt{mag\_ratio}. We set the range of \texttt{mag\_ratio} as $\{0, 0.01, 0.1,
0.3, 0.5, 0.7, 0.9, 1\}$, whereas $0.01$ is the recommended setting of
Magnitude. For shielding intermediate layers (SOTER; \ding{175}), the number of
shielded layers is also defined by a configuration parameter,
\texttt{soter\_ratio}. We set the range of \texttt{soter\_ratio} as $\{0, 0.1,
0.2, 0.3, 0.5, 0.7, 0.9, 1\}$ and $0.2$ is the recommended setting of the
original paper. For both \ding{174} and \ding{175}, setting \texttt{mag\_ratio}
(and \texttt{soter\_ratio}) to $0$ represents the \noshield baseline while
setting the parameters to $1$ is the black-box baseline. For shielding
non-linear layers (ShadowNet; \ding{176}), we clarify that ShadowNet does not
need to set any configuration.

\parh{Attack Implementation.}~We re-run the same \ac{ms} and \ac{mia} as in
\S~\ref{sec:evaluated_attacks}. That is, we re-launch the three-phase attack
pipeline, which comprises surrogate model initialization (P$_{i}$), model
stealing (P$_{ii}$), and membership inference (P$_{iii}$).

\begin{figure*}[ht]
    \centering
    \includegraphics[width=0.98\linewidth]{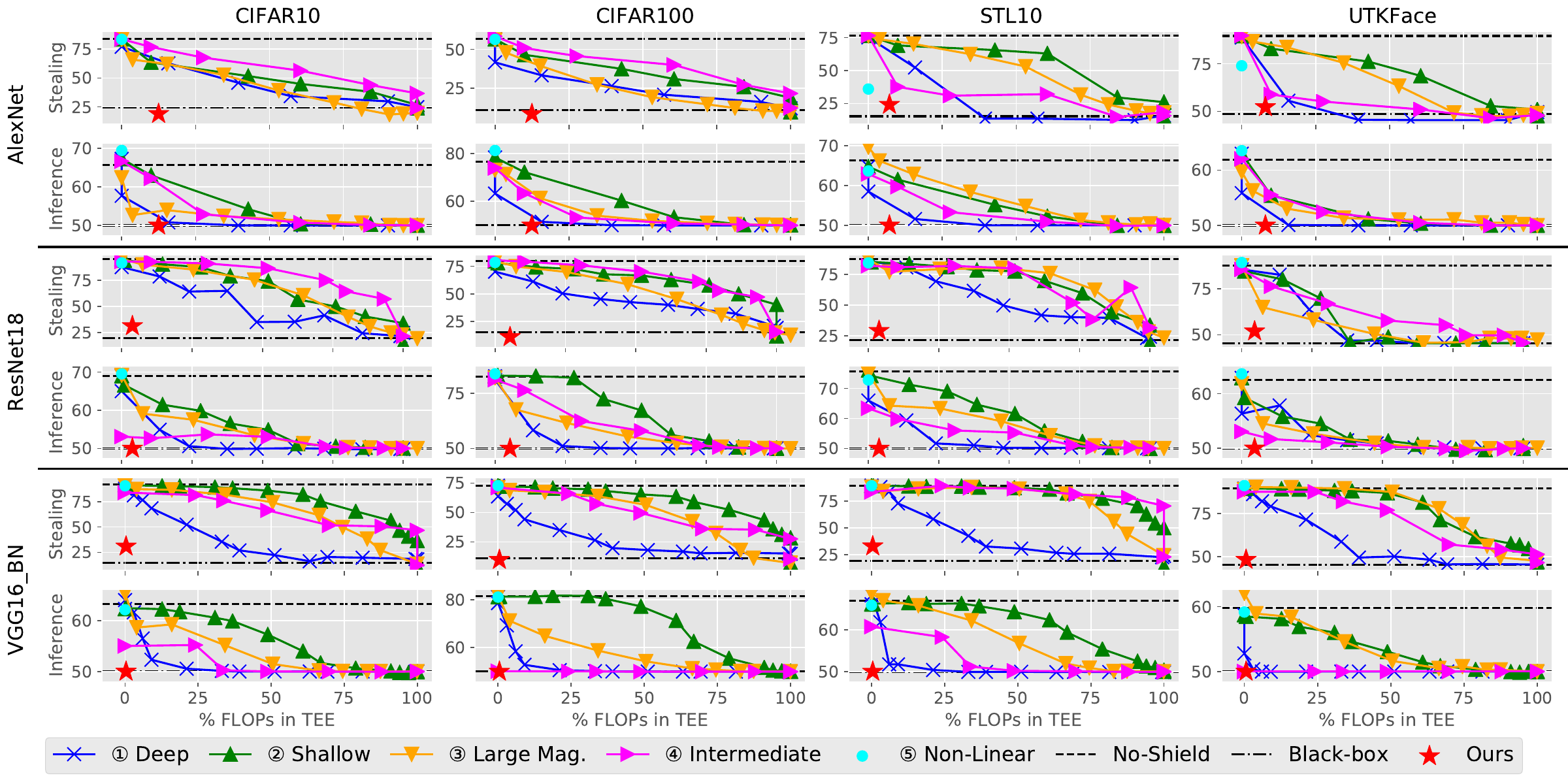}
    \vspace{-5pt}
    \caption{The relationship between $Security$ (y-axis) and $Utility$ (x-axis)
    of various \tsdp\ solutions. The results of \ac{ms} and \ac{mia} for each
    case are shown in two sub-figures. Each curve of `` \ding{172}Deep'',
    ``\ding{173}Shallow'', ``\ding{174}Large Mag.'', ``\ding{175}Intermediate'',
    and ``\ding{176}Non-Linear'' shows the corresponding solution. We
    also plot \noshield and black-box baselines in horizontal lines.}
    \label{fig:acc_mia_flops_one_fig}
    \end{figure*}

\subsection{Qualitative and Quantitative Results}
\label{sec:evaluation_dilemma_results}

We compute both qualitative and quantitative results to explore if the ``sweet
spot'' configuration exists for each scheme and the characteristics of the sweet
spots. For qualitative results, \F~\ref{fig:acc_mia_flops_one_fig} presents the
relationship between $Security$ and $Utility$ of different configurations. Note
that in \F~\ref{fig:acc_mia_flops_one_fig}, we only show the results for \ac{ms}
accuracy (``Model Stealing'') and confidence-based \ac{mia} accuracy
(``Membership Inference'') due to the space limit. In
\F~\ref{fig:acc_mia_flops_one_fig}, the x-axis shows $\%FLOPs$ ($Utility$) of
each partition configuration, and the y-axis shows the accuracy of \ac{ms} and
\ac{mia} ($Security$). For each model, \ac{ms} and \ac{mia} are displayed in two
sub-figures, respectively. In each sub-figure, shielding deep layers
(\ding{172}), shielding shallow layers (\ding{173}), shielding large-magnitude
weights (\ding{174}), shielding intermediate layers (\ding{175}), and shielding
non-linear layers (\ding{176}) are represented by a blue line with crosses, a
green line with up triangles, an orange line with down triangles, a pink line
with right triangles, and an aquamarine circle, respectively. 
We also plot the performance of
the \noshield baseline and the black-box baseline using horizontal black lines. 
We put the results for other
metrics in \App~\ref{append:sec:other_metrics:optimal_config_qualitative}
(\F~\ref{fig:acc_flops_append_one_fig} and
\F~\ref{fig:mia_flops_append_one_fig}).

In a holistic sense, \F~\ref{fig:acc_mia_flops_one_fig} suggests that there is no
systematic and automated approach to identifying the ``sweet spot'' for the five
representative defenses. The shapes of the lines are substantially different
across datasets and models. Given a requirement $\Delta$ for security risk,  it
is tough to set a uniform threshold for $Utility(C^*)$ without a comprehensive
empirical measurement of both $Security$ and $Utility$. For example, for
shielding deep layers (\ding{172}) of AlexNet for model stealing (the first row
in \F~\ref{fig:acc_mia_flops_one_fig}), the shape of the curves for CIFAR10 and
CIFAR100 are very different from the curve for STL10. Given a requirement
$\Delta$ and a model to protect, the locations of ``sweet spots'' are random for
different datasets.

For quantitative results, we measure the values of the $Utility(C^*)$ as defined
in Equation~\ref{equ:optimal_config} (the smallest value of $Utility$ to achieve
$Security_{black}$) for different defenses. \T~\ref{tbl:optimal_config} reports
the ratio of TEE-shielded FLOPs ($\%FLOPs$) to achieve $Security_{black}$ for
each setting for \ac{ms} and \ac{mia}. We omit the approach of shielding non-linear layers
(ShadowNet, \ding{176}) because it does not require configurations. 
The results
of other metrics are displayed in
\App~\ref{append:sec:other_metrics:optimal_config_quantitative}
(\T~\ref{tbl:optimal_config:fidelity} to
\T~\ref{tbl:optimal_config:grad_attack}). 
Overall, \T~\ref{tbl:optimal_config}
implies that \textit{the $Utility$ values to achieve $Security_{black}$ are
distinct across protected models and datasets.} For example, to protect AlexNet
from \ac{ms} with shielding deep layers (\ding{172}), we need to put 100\% of
the protected model in TEE to achieve $Security_{black}$ for CIFAR10 (C1) and
CIFAR100 (C100). However, for STL10 (S10) and UTKFace (UTK), we only need to put
39.44\% of FLOPs in TEE to achieve $Security_{black}$. Further, it is tough, if
not impossible, to predict the actual value of $Utility(C^*)$ before we run the
models empirically because the numbers are irregular. We also observe similar
irregularity for other defense methods when protecting different models and
datasets.

\colorlet{best}{green!30}
\colorlet{high}{red!30}
\colorlet{low}{yellow}

\begin{table*}[!h]
\caption{Different $Utility(C^*)$ ($\% FLOPs(C^*)$) values of ``sweet spot'' in front of \ac{ms} and \ac{mia}. 
A lower value represents a lower utility cost. The $\% FLOPs(C^*)$ for \noshield and black-box baselines are 0\% and 100\%, respectively.
For each \tsdp\ solution (row), we mark the lowest $Utility(C^*)$ with \colorbox{low}{yellow} and the 
highest value with \colorbox{high}{red}.
For each case (model and dataset, column), we mark the lowest $Utility(C^*)$ across all solutions with \colorbox{best}{green}.
The last column is the average utility cost for each solution.
We omit shielding non-linear layers
(ShadowNet, \ding{176}) because it does not require configurations.
}
\label{tbl:optimal_config}
\vspace{-5pt}
\setlength{\tabcolsep}{1.5pt}
\centering
\begin{adjustbox}{max width=0.83\linewidth}

    \begin{tabular}{@{}cccccclcccclcccccc@{}}
    \toprule
                                                                                     &              & \multicolumn{4}{c}{AlexNet}               &  & \multicolumn{4}{c}{ResNet18}              &  & \multicolumn{4}{c}{VGG16\_BN}             &  & \multirow{2}{*}{Average} \\ \cmidrule(lr){3-6} \cmidrule(lr){8-11} \cmidrule(lr){13-16}
                                                                                     &              & C10      & C100     & S10      & UTK      &  & C10      & C100     & S10      & UTK      &  & C10      & C100     & S10      & UTK      &  &                          \\ \midrule
    \multirow{4}{*}{\begin{tabular}[c]{@{}c@{}}Model \\ Stealing\end{tabular}}       & \ding{172}Deep         & \cellcolor{high}100.00\% & \cellcolor{high}100.00\% & 39.44\%  & 39.44\%  &  & \cellcolor{high}100.00\% & \cellcolor{high}100.00\% & \cellcolor{high}100.00\% & \cellcolor{low}37.46\%  &  & \cellcolor{high}100.00\% & \cellcolor{high}100.00\% & \cellcolor{high}100.00\% & 69.23\%  &  & 82.13\%                  \\
                                                                                     & \ding{173}Shallow      & \cellcolor{high}100.00\% & \cellcolor{high}100.00\% & \cellcolor{high}100.00\% & \cellcolor{high}100.00\% &  & \cellcolor{high}100.00\% & \cellcolor{high}100.00\% & \cellcolor{high}100.00\% & \cellcolor{low}38.55\%  &  & \cellcolor{high}100.00\% & \cellcolor{high}100.00\% & \cellcolor{high}100.00\% & \cellcolor{high}100.00\% &  & 94.88\%                  \\
                                                                                     & \ding{174}Large Mag.   & 81.18\%  & 90.58\%  & \cellcolor{high}100.00\% & 71.82\%  &  & \cellcolor{high}100.00\% & 94.71\%  & \cellcolor{high}100.00\% & \cellcolor{low}61.48\%  &  & \cellcolor{high}100.00\% & 87.43\%  & \cellcolor{high}100.00\% & \cellcolor{high}100.00\% &  & 90.60\%                  \\
                                                                                     & \ding{175}Intermediate & \cellcolor{high}100.00\% & \cellcolor{high}100.00\% & 84.31\%  & \cellcolor{low}60.69\%  &  & \cellcolor{high}100.00\% & \cellcolor{high}100.00\% & \cellcolor{high}100.00\% & \cellcolor{high}100.00\% &  & \cellcolor{high}100.00\% & \cellcolor{high}100.00\% & \cellcolor{high}100.00\% & \cellcolor{high}100.00\% &  & 95.42\%                  \\ \midrule
    \multirow{4}{*}{\begin{tabular}[c]{@{}c@{}}Membership \\ Inference\end{tabular}} & \ding{172}Deep         & 15.78\%  & 15.78\%  & 15.78\%  & 15.78\%  &  & \cellcolor{high}23.97\%  & \cellcolor{high}23.97\%  & \cellcolor{high}23.97\%  & \cellcolor{high}23.97\%  &  & 9.03\%   & 21.07\%  & 6.02\%   & \cellcolor{low}3.01\%   &  & 16.51\%                  \\
                                                                                     & \ding{173}Shallow      & 60.56\%  & 84.22\%  & 60.56\%  & 42.81\%  &  & 62.54\%  & 86.52\%  & 76.03\%  & \cellcolor{low}38.55\%  &  & 66.90\%  & \cellcolor{high}90.97\%  & \cellcolor{high}90.97\%  & 60.88\%  &  & 68.46\%                  \\
                                                                                     & \ding{174}Large Mag.   & \cellcolor{low}34.51\%  & 53.17\%  & 71.82\%  & 34.51\%  &  & 61.48\%  & 61.48\%  & \cellcolor{high}76.61\%  & 44.93\%  &  & 50.56\%  & 66.47\%  & 66.47\%  & 50.56\%  &  & 56.05\%                  \\
                                                                                     & \ding{175}Intermediate & 60.69\%  & 60.69\%  & 60.69\%  & 27.95\%  &  & \cellcolor{high}72.80\%  & \cellcolor{high}72.80\%  & \cellcolor{high}72.80\%  & 11.02\%  &  & 33.84\%  & \diagfil{1.2cm}{low}{best}{0.10\%}   & 33.84\%  & \diagfil{1.2cm}{low}{best}{0.10\%}  &  & 42.28\%                  \\ \midrule
    \multicolumn{2}{c}{Ours}                                                                        & \cellcolor{best}12.48\%  & \cellcolor{best}12.48\%  & \cellcolor{best}7.12\%   & \cellcolor{best}8.01\%   &  & \cellcolor{best}3.80\%   & \cellcolor{best}5.33\%   & \cellcolor{best}3.80\%   & \cellcolor{best}4.58\%   &  & \cellcolor{best}0.34\%   & 0.47\%   & \cellcolor{best}0.40\%   & 0.60\%   &  & 4.95\%                   \\ \bottomrule
    \end{tabular}

\end{adjustbox}
\vspace*{-15pt}

\end{table*}

\begin{tcolorbox}[size=small]
\textbf{Answer to RQ2}: It is difficult to systematically
identify the ``sweet spots'' configuration $C^{*}$ for prior TSDP solutions. 
\end{tcolorbox}

\section{Design of \tool}
\label{sec:approach}

Besides systematically benchmarking de facto \tsdp\ solutions (\textbf{RQ1}) and
summarizing their common drawbacks (\textbf{RQ2}), we conclude the root cause
of the \tsdp solutions' vulnerabilities and propose a novel partition scheme,
which alleviates the security vs. utility trade-off and can automatically find
the ``sweet spot'' configuration.

The root cause of \tsdp solutions' weakness is that all the \tsdp approaches
follow a \textit{\trainbeforepartition} strategy, which first trains $M_{vic}$
using private data and then partitions the private model. After training, all
the model weights (including the offloaded part) are updated by the private data
and thus contain private information. During deployment, the private information
in the offloaded weights is exposed to the untrusted environment. As
demonstrated in \textbf{RQ1} and \textbf{RQ2}, attackers could effectively
recover private information of $M_{vic}$ from offloaded privacy-related weights
with the help of public knowledge, \textit{i.e.}, a well-prepared $M_{\rm init}$ on hand.
With this regard, we champion that an ideal \tsdp\ solution should ensure
\textit{the offloaded DNN weights on GPUs are never trained using private data}
and thus, no information is leaked to the untrusted environment.

\begin{figure}[!h]
  \centering
  \includegraphics[width=\linewidth]{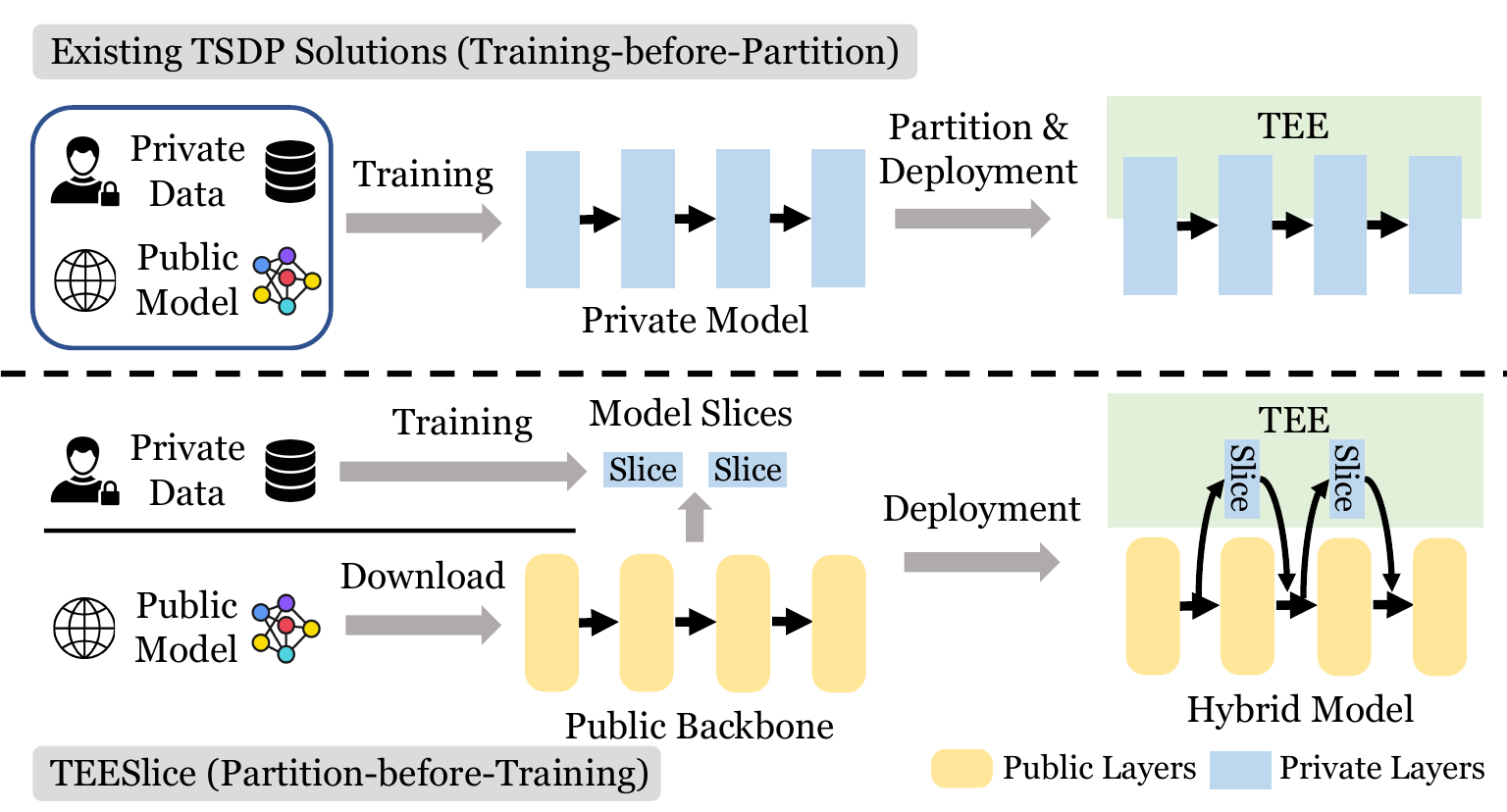}
  \vspace{-10pt}
  \caption{The comparison between \sys and prior \tsdp\ solutions. For \tool, all information generated by private data will be handled in the TEE.}
  \label{fig:teeslice_compare}
  \vspace{-10pt}
\end{figure}

\subsection{Approach Overview}

We propose \sys, a novel partitioning strategy that offloads DNN layers with no
private information to GPUs at the inference stage. Prior solutions use public
pre-trained model and private data to train $M_{\rm vic}$, use heuristic designs
to partition $M_{\rm vic}$, and shield a subset of model parts. On the contrary,
\tool uses a \textit{\partitionbeforetrain} paradigm, which partitions private data from
the pre-trained model and then separately trains privacy-related layers.
\F~\ref{fig:teeslice_compare} shows the comparison between existing \tsdp\
solutions (\trainbeforepartition) and \tool (\partitionbeforetrain). 

We term the public pre-trained model as \textit{backbone}, and the
privacy-related layers as \textit{model slices}~\cite{zhang2020dynamic,zhang2022remos,zhang2022teeslice,zhang2023fedslice}. The model slices are
lightweight and contain the private knowledge of $M_{\rm vic}$. The backbone and
model slices are combined to form a \textit{hybrid model} (denoted as $M_{\rm
hyb}$) that imitates the behavior of $M_{\rm vic}$. Each slice takes the output
of the prior layer (of the backbone) as its input and produces the input for the
next layer (of the backbone). A detailed illustration of \tsdp is shown in
\F~\ref{fig:model_partition_categories}, where the yellow rounded blocks (Layer1
to Layer4) represent layers of the backbone and the blue squares (Slice1 and
Slice2) are privacy-related model slices. The arrows between backbone layers and
model slices indicate the data flow of internal DNN features. For example, the
output of Layer1 is fed to Slice1, and Layer4 takes the outputs of Slice2 and
Layer3 as input.

The key challenge for \sys is to generate small slices that can run in TEEs with
enough little or no accuracy lost. To this end, \sys leverages a two-staged
approach. First, it builds an instance of $M_{\rm hyb}$, a densely sliced model
(denoted as $M_{\rm dense}$), with substantial private slices that can achieve
high accuracy. However, $M_{\rm dense}$ cannot fit into TEEs due to its large
size. Then, \sys prunes $M_{\rm dense}$ with a self-adaptive, iterative pruning
strategy that produces fewer slices without losing the $M_{\rm hyb}$'s
performance. The pruned model is another instance of $M_{\rm hyb}$, which we
call a sparsely sliced model (denoted as $M_{\rm sparse}$). As the pruning is
conducted simultaneously with the training phase, \tool\ can prune slices with
little accuracy drop. Note that the iterative pruning strategy is specifically
designed for the resource-constrained TEE environment.

\sys is partially motivated by \nettailor~\cite{morgado2019nettailor}, which
implements a training framework of $M_{\rm dense}$. However, \nettailor does not
aim to protect model privacy with TEEs and generates a large number of slices. To
meet the constraint of TEEs, \tool generates $M_{\rm sparse}$ with an adaptive
pruning strategy to reduce the computation cost of slices. Besides,
\textsc{NetTailor} lacks cryptographic primitives to securely transmit DNN
intermediate data between GPUs and TEEs. \tool\ employs a one-time pad
(OTP)~\cite{OneTimePad} and Freivalds'
algorithm~\cite{freivalds1977probabilistic} to secure TEE-GPU transmission.

\subsection{Detailed Design}
\label{sec:teeslice_detail}

\sys consists of two stages: model slice extraction (training phase) and hybrid
model deployment (inference phase). The slice extraction stage automatically
finds the ``sweet spot'' by minimizing the utility cost while maintaining
accuracy and security. \sys trains $M_{\rm dense}$ from the public backbone and
then prunes $M_{\rm dense}$ to get $M_{\rm sparse}$. 
In the hybrid model deployment stage, $M_{\rm sparse}$ is deployed across the
TEE and GPU. Model slices and non-linear layers (of the backbone) are deployed
inside TEEs, whereas the other part of the backbone is offloaded on GPUs. 

\subsubsection{Model Slice Extraction}

\parh{Densely Sliced Model Generation.}
Let the $i$-th layer of the public backbone be $L_i$. The private slice is
represented as $ A^i_p $, which connects layer $L_p$ with layer $L_i$. $ A^i_p $
is designed to be lightweight and is $18\times$ smaller than $L_i$. The slices
in $M_{\rm dense}$ connect a layer pair from backbone whenever the distance of
the layers in this pair is less than three. 
During the training stage, \sys assigns one importance scalar $\alpha_p^i$ to
each slice $ A^i_p $ following the same strategy in related
work~\cite{morgado2019nettailor}. The output of $ A^i_p $ is multiplied by
$\alpha_p^i$ and sent to the next layer (of the backbone). A smaller
$\alpha_p^i$ diminishes the influence of $ A^i_p $. The model slices and the
scalars are optimized simultaneously with a loss function that penalizes both
model performance and complexity. The output of this step is $M_{\rm dense}$
with close performance as $M_{\rm vic}$.

\parh{Iterative Slice Pruning.}~The pruning algorithm is guided by the
importance scalar $\alpha_p^i$, which controls the impact of $A^i_p$ on the
model prediction. We iteratively prune the slices with the smallest $\alpha_p^i$
and re-train the hybrid model to maintain the desired accuracy. 
A pre-defined threshold $\delta$ (defined as 1\%) governs the tolerable accuracy
loss during pruning. Let $ACC_{\rm vic}$ represent the accuracy of $M_{\rm
vic}$. The tolerable accuracy is $ACC_{\rm tol} = (1-\delta) \cdot ACC_{\rm
vic}$, and the accuracy of the final $M_{\rm hyb}$ should be greater than
$ACC_{\rm tol}$.

\A~\ref{alg:pruning_pipeline} depicts the pruning pipeline. $\alpha_{\rm setup}$ determines how to prune unimportant slices during the setup phase. The slice layers with $\alpha^l_p < \alpha_{\rm setup}$ are pruned. As a heuristic, we set $\alpha_{\rm init}$ to be $0.05$ according to \textsc{NetTailor}. Iterative pruning requires two hyper-parameters: the number of the pruned slices in each round $n$ and the total number of training rounds $rounds$. Each round begins with an evaluation of the model's accuracy $ACC_r$. If the current model satisfies the performance requirement ($ACC_r > ACC_{\rm tol}$), \sys prunes $n$ model slices with the smallest $\alpha^l_p$ and trains the pruned model. Otherwise, \sys skips the pruning operation and continues training the model. 

\begin{algorithm}
  \footnotesize
    \SetAlgoLined
    \SetKwProg{Fn}{Function}{:}{} \SetKwFunction{FIterativePrune}{Iterative
    Slice Pruning} \KwIn{The densely sliced model $M_{\rm dense}$, pre-defined
    parameters $\alpha_{\rm setup}$, $n$, and $rounds$} \KwOut{The hybrid model
    $M_{\rm hyb}$}
    
    Prune $M_{\rm dense}$ by $\alpha_{\rm setup}$ to get $M_{1}$ \;
    
    \For{$r\leftarrow 1$ \KwTo $rounds$}{
      Compute the accuracy $ACC_r$ of $M_r$ \;
      \If{$ACC_r > ACC_{\rm tol}$}{
          Store the model $M_{\rm hyb} = M_r$ \;
          Select $n$ slices with smallest $\alpha^l_p$ \;
          Prune the selected slices $A^l_p$ \;
      }
      Re-train $M_r$ to get $M_{r+1}$  \;
    }
    \KwRet The hybrid model $M_{\rm hyb}$
    
    \caption{Iterative Slice Pruning.}\label{alg:pruning_pipeline}
  \end{algorithm}

\parh{Automatically Find the Sweet Spot.}~The iterative slice pruning is indeed
an optimization process that automatically finds the ``sweet spot.'' Given the
constraints of model security and accuracy, the iterative slice pruning
optimizes the size of the private model slices to reduce the utility cost.
\A~\ref{alg:pruning_pipeline} only explicitly considers the threshold
$ACC_{tol}$ for accuracy lost because all the private information is in the
slices, which will run in TEEs. Therefore, the confidentiality of $M_{\rm vic}$
remains intact after offloading the backbone. Unlike prior \tsdp solutions,
where it is difficult to find the sweet spot configuration without a
comprehensive evaluation of both security and utility (\S~\ref{sec:dilemma}),
\sys does not have this shortcoming.

\subsubsection{Hybrid Model Deployment}
\label{sec:model_deployment}
When deploying $M_{\rm hyb}$, the private slices and non-linear layers (of the backbone) are shielded by the TEE, while the GPU hosts the backbone's linear layers. Shielding non-linear layers of the backbone is a common practice for prior \tsdp solutions~\cite{tramer2019slalom,hashemi2021darknight,lucien2021goten,hou2021model} because non-linear layers are hard to securely offload to GPUs and only occupy a small fraction (about 1.5\%) of the DNN's computation cost~\cite{tramer2019slalom}. In the illustration figures (\F~\ref{fig:model_partition_categories} and \F~\ref{fig:teeslice_compare}) we omit the non-linears in TEE for simplicity. There are two security challenges to deploy $M_{\rm hyb}$: 1) how to encrypt features transmitted between GPU and TEE, and 2) how to verify the correctness of computations offloaded on GPUs. These two challenges can be solved separately using one-time pad (OTP) and Freivalds' algorithm~\cite{freivalds1977probabilistic}. 

\parh{Feature Encryption.}~For a backbone linear layer $g(\cdot)$, let
$\textbf{h}$ be the plaintext input shielded by TEE. We first quantize
$\textbf{h}$ into a 8-bit representation following prior
literature~\cite{tramer2019slalom, lucien2021goten} and get $\hat{\textbf{h}}$.
Then, we select a large prime value $p$, generate a random mask $\mathbf{r}$ (as
the OTP), and encrypt the feature by
\begin{equation}
    \mathbf{h}_e = (\hat{\mathbf{h}} + \mathbf{r}) \ \% \ p. 
\end{equation}

GPU receives $\mathbf{h}_e$, computes $g(\mathbf{h}_e)$, and returns the result
back to TEE. \sys decrypts the result by computing $g(\hat{\mathbf{h}}) =
g(\mathbf{h}_e) - g(\mathbf{r})$ because
\begin{equation}
  \begin{aligned}
  & \ g( \mathbf{h}_e) - g( \mathbf{r}) 
      =  g( (\hat{\mathbf{h}} + \mathbf{r})\ \% \ p) - g( \mathbf{r}\ \% \ p) \\
      = & \ g( (\hat{\mathbf{h}} + \mathbf{r})\ \% \ p - \mathbf{r}\ \% \ p ) 
      =  \ g( (\hat{\mathbf{h}} + \mathbf{r} - \mathbf{r})\ \% \ p ) \\
      = & \ g( \hat{\mathbf{h}}\ \% \ p ) \ = \ g( \hat{\mathbf{h}} ).
  \end{aligned}
  \end{equation}
The last equation holds as long as $p > 2^8$. Note that following prior
work~\cite{tramer2019slalom}, computing $g(\mathbf{r})$ can be conducted by the
model provider or inside TEE in an offline phase. Both strategies do not
increase the overhead of online inference and do not impede its utility.

\parh{Result Verification.}~Freivalds' algorithm can periodically verify the
computation results on GPUs on all linear layers. Let the weight of the linear
layer $g(\cdot)$ be $\mathbf{W}$ and $g(\mathbf{h}) = \mathbf{h}^\mathsf{T}
\mathbf{W}$, \sys samples a random vector $\mathbf{s}$ that has the same shape
as $g(\mathbf{h})$. \sys then pre-computes $\tilde{\mathbf{s}} = \mathbf{W}
\mathbf{s}$. The verification can be conducted by checking
$g(\mathbf{h})^\mathsf{T} \mathbf{s} = \mathbf{h}^\mathsf{T}
\tilde{\mathbf{s}}$.

\section{Experiments}
\label{sec:teeslice_experiment}

We implement \sys with PyTorch 1.7 and we select ResNet18 as the public backbone
following \textsc{NetTailor}. We have the flexibility to choose the backbone as
commonly-used models without affecting the security
guarantee~\cite{morgado2019nettailor}. We select ResNet18 because it is the
default setting of \textsc{NetTailor}. As a fair setting, we set the
training time for $M_{\rm dense}$ and $M_{\rm sparse}$ to half the time required
to train $M_{\rm vic}$, respectively. Hence, the overall training time for
$M_{\rm sparse}$ and $M_{\rm vic}$ are equivalent. We apply \sys to all the
datasets and victim models in \S~\ref{subsec:evaluation-setup}. The experiments
aim to answer the following RQs:

\begin{tcolorbox}[size=small]
    \textbf{RQ3:} How does \sys compare with representative defenses in
    \S~\ref{sec:evaluated_attacks} w.r.t. security and utility? 
    \textbf{RQ4:} Does \sys sacrifice the accuracy of the original model?
    \textbf{RQ5:} What is performance of \tool on real-world devices? How much can
    \tool speed up compared to the \shieldwhole baseline?
    \textbf{RQ6:} How is \tool's scalability to NLP tasks.
\end{tcolorbox}

\subsection{Security Guarantee and Utility Cost}
\label{sec:experiment:security_utility}

\parh{Security Guarantee.}~We follow the same experiment protocol in
Section~\ref{sec:evaluate_existing_solutions} to compare \sys with five representative \tsdp\ schemes. Specifically, for \sys, we assume the attacker
knows the architecture of $M_{\rm hyb}$ (default assumption), including which
public backbone it uses and the structure of privacy-related model slices. The
attack pipeline is the same as in Section~\ref{sec:evaluate_existing_solutions}.

\T~\ref{tbl:evaluate_solution} reports the results, with attack accuracy against
our approach marked in \colorbox{best}{green}. The results are highly promising:
in all cases, the attack accuracies are comparable with black-box protection and
are better than the best of existing defenses (marked with \colorbox{low}{yellow}).
For \ac{ms}, the relative accuracy of \sys compared to the black-box baseline is
1.24$\times$, while the relative value of the best defense, SOTER (\ding{175}),
is 3.76$\times$. 
For \ac{mia}, the attack accuracy of \sys is similar to the black-box baseline
(random guess). It is because all the feature communications are encrypted, and
the TEE shields all the privacy-related slices. 

\parh{Security Under Other Assumptions of $M_{\rm sur}$.}
\label{sec:experiment:other_assumption}
This section evaluates the security guarantee of \sys with two different
assumptions. The first assumption, \textit{backbone-only}, is a weaker one that
assumes the attacker only knows the public backbone of \sys and does not know the
slice information (slice positions and structures). The second assumption,
\textit{victim-knowing}, is a stronger assumption that assumes the attacker
knows the structure of the original $M_{vic}$. Note that the victim-knowing is
\textit{not} a realistic assumption, and we only evaluate it to show the
performance of \sys under different settings. We also follow the evaluate
protocol and attack pipeline in \S~\ref{sec:evaluate_existing_solutions}
for a fair experiment.

We show the \ac{ms} accuracies of the two additional assumptions with the
default assumption (knowing the structure of $M_{\rm hyb}$) in
\T~\ref{tbl:other_assumption}. 
Note we omit \ac{mia} because the additional assumptions do not introduce new
information of $M_{\rm vic}$'s training data. Thus the results of \ac{mia} are
the same as the default assumption. For each model and dataset
\T~\ref{tbl:other_assumption}, we mark the highest accuracy with
\colorbox{high}{red} and the lowest accuracy with \colorbox{best}{green}. From
\T~\ref{tbl:other_assumption}, we can see that victim-knowing has lower
accuracies than the other two assumptions (seven green cells and two red cells).
The default assumption (Hybrid $M_{\rm hyb}$) and backbone-only perform
similarly (both have five red cells). We suspect that the reason for
victim-knowing's low accuracy is that the $M_{\rm vic}$ in
\T~\ref{tbl:other_assumption} has a smaller model capacity than the backbone. 
In \T~\ref{tbl:other_assumption_append}, a $M_{\rm vic}$ with larger capacity
(ResNet34) has the highest \ac{ms} accuracy for all datasets.

\begin{table}[]
    \caption{Comparison of model stealing accuracy between different attack assumptions of $M_{\rm sur}$.}
    \vspace{-5pt}
    \label{tbl:other_assumption}
    \centering
    \begin{adjustbox}{max width=0.7\linewidth}
    \begin{tabular}{@{}ccccc@{}}
    \toprule
                               &      & Hybrid $M_{\rm hyb}$ & Backbone & Victim $M_{\rm vic}$ \\ \midrule
    \multirow{4}{*}{AlexNet}   & C10  & \cellcolor{best}19.04\% & 19.56\%  & \cellcolor{high}23.71\% \\
                               & C100 & \cellcolor{best}8.27\%  & \cellcolor{high}14.48\%  & 11.9\%  \\
                               & S10  & 24.15\% & \cellcolor{high}32.75\%  & \cellcolor{best}17.14\% \\
                               & UTK  & \cellcolor{high}52.27\% & 51.32\%  & \cellcolor{best}47.0\%  \\ \midrule
    \multirow{4}{*}{ResNet18}  & C10  & \cellcolor{high}31.4\%  & 25.63\%  & \cellcolor{best}17.33\% \\
                               & C100 & 10.9\%  & \cellcolor{high}18.33\%  & \cellcolor{best}7.78\%  \\
                               & S10  & \cellcolor{best}29.19\% & 32.77\%  & \cellcolor{high}32.86\% \\
                               & UTK  & \cellcolor{high}51.95\% & \cellcolor{best}50.86\%  & 51.63\% \\ \midrule
    \multirow{4}{*}{VGG16\_BN} & C10  & \cellcolor{high}30.87\% & 25.65\%  & \cellcolor{best}20.69\% \\
                               & C100 & 9.78\%  & \cellcolor{high}18.44\%  & \cellcolor{best}6.38\%  \\
                               & S10  & \cellcolor{high}32.92\% & 32.51\%  & \cellcolor{best}31.75\% \\
                               & UTK  & \cellcolor{best}48.37\% & \cellcolor{high}52.54\%  & 51.04\% \\ \bottomrule
    \end{tabular}

\end{adjustbox}
\vspace{-15pt}

\end{table}

\parh{Security Under Other Assumptions of Data.}
In \S~\ref{sec:evaluate_existing_solutions} and \S~\ref{sec:dilemma}, we study a
realistic adversary that has a small amount of data. Although our assumption on
the adversary is
realistic~\cite{rakin2022deepsteal,hua2018reverse,yan2020cache}, we still study
the security of \sys with an ideal adversary who has a large amount of data to
verify if \sys ensures the security of DNN models under extreme conditions.
We compare \ac{ms} accuracies on various $M_{\rm sur}$ and a large range
of queried data size between our approach and black-box protection. The goal is
to study if our approach increases \ac{ms} accuracy under the new assumption.
The $M_{\rm sur}$ includes $M_{\rm hyb}$, the backbone (\ie, ResNet18;
backbone-only), and all the $M_{\rm vic}$ in \S~\ref{subsec:evaluation-setup}
(victim-knowing). Following prior work~\cite{orekondy2019knockoff}, we set the
queried data sizes as \{50, 100, 300, 500, 1K, 3K, 5K, 10K, 15K, 20K, 25K,
30K\}. 

\begin{figure}[ht]
    \centering
    \vspace{-5pt}
    \includegraphics[width=\linewidth]{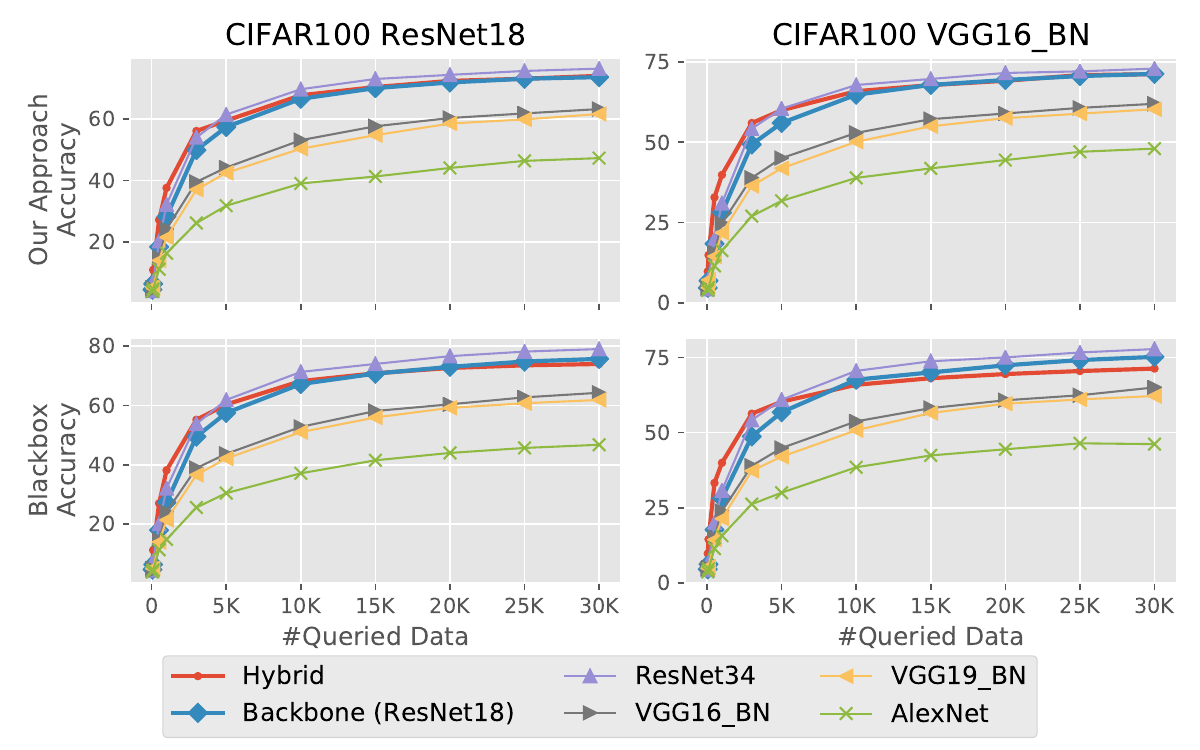}
    \vspace{-15pt}
    \caption{Comparison of \sys and the black-box protection against \ac{ms}
    attacks with different sizes of queried data. We report the accuracy of
    $M_{\rm sur}$, where the first row represents \sys and the second row is for
    the black-box baseline.}
    \label{fig:multi_arch_cifar100_accuracy_res18vgg16}
    \vspace{-5pt}
    \end{figure}

We display \ac{ms} accuracy on CIFAR100 and two models (ResNet18 and VGG16\_BN)
in \F~\ref{fig:multi_arch_cifar100_accuracy_res18vgg16}, the other metrics and
models are displayed in \App~\ref{append:sec:data_assumption}. 
The first column
of \F~\ref{fig:multi_arch_cifar100_accuracy_res18vgg16} means $M_{\rm vic}$ is a
ResNet18 model and the second column means $M_{\rm vic}$ is a VGG16\_BN. The
first row of \F~\ref{fig:multi_arch_cifar100_accuracy_res18vgg16} displays the results of
\tool, and the second row shows the results of the black-box defense. 
We can observe from \F~\ref{fig:multi_arch_cifar100_accuracy_res18vgg16} that,
for all cases, the \ac{ms} accuracy against \tool is similar to that of
black-box baseline. According to the Wilcoxon signed-rank
test~\cite{wilcoxon1992individual}, the null hypothesis is that there is no
difference in accuracy. The p-value is $0.81$, which cannot reject the null
hypothesis. This result indicates that the differences between the accuracy of
\sys and the black-box baseline have no statistical significance. To summarize,
under a different assumption of more queried data, the \ac{ms} accuracy has no
difference between \tool and the black-box baseline.

\parh{Utility Cost.}~We qualitatively compare \tool\ with the security-utility
curves of other defenses in \F~\ref{fig:acc_mia_flops_one_fig}. For both \ac{ms}
and \ac{mia}, \sys achieves a similar level of black-box defense
with a distinguishably smaller $\% FLOPs$ than other defenses.
\textcolor{red}{\ding{72}}, denoting \sys, locates at the bottom left corners for
all cases in \F~\ref{fig:acc_mia_flops_one_fig}. We also quantitatively compare
the $Utility(C^*)$ of \sys with other defenses in \T~\ref{tbl:optimal_config}.
For each case (column), we mark the lowest value of $Utility(C^*)$ with
\colorbox{best}{green}. \sys achieves the lowest utility cost in ten out of 12
cases. The average utility cost of \sys is 4.95\%. On the contrary, the average
utility cost of other defenses ranges from 42.28\% to 95.42\%. That is, \sys
takes $10\times$ less utility cost to achieve the highest (black-box) defense
level.

\begin{tcolorbox}[size=small]
\textbf{Answer to RQ3}: \sys\ features promising, black-box-level security
guarantees under different attack assumptions. The utility cost ($Utility(C^*)$)
of \sys is $10\times$ less than other \tsdp\ solutions. 
\end{tcolorbox}
\vspace{-5pt}

\subsection{Accuracy Loss}
\label{sec:experiment:trade-off}
To answer this research question, we compare the accuracy between $M_{\rm vic}$
and their derived hybrid models $M_{\rm hyb}$ trained by \sys. The result is in
\T~\ref{tbl:nettailor_accuracy}. In general, \sys does not lead to a
considerable loss of accuracy. For AlexNet, \sys achieves a higher accuracy
because the model capacity of the backbone (i.e., ResNet18) is larger than
AlexNet. This phenomenon is consistent with the finding in
\S~\ref{sec:experiment:security_utility} that AlexNet (the lowest model
complexity) always has low accuracy. 
To statistically understand the accuracy loss, we compute the statistical
significance using the Wilcoxon signed-rank test~\cite{wilcoxon1992individual}
across all models except AlexNet. The null hypothesis is that the accuracies
between $M_{\rm vic}$ and $M_{\rm hyb}$ have no difference. The p-value is 0.75
and provides little statistical significance to reject the null hypothesis.
Thus, the pruning processes have little effect on the accuracy of $M_{\rm hyb}$.

We also measure the accuracy loss of eNNclave~\cite{schlogl2020ennclave}, a
recent work sharing similar concepts, by putting $M_{\rm vic}$'s last layer into
TEE while replacing the GPU-offloaded shallow layers with a public backbone. As
clarified in \S~\ref{sec:literature_category}, eNNcalve suffers from low
accuracy. We evaluate the accuracies of eNNclave over all models and datasets
and find that eNNclave has an average downgrade of 34\%, higher than
(about $10\times$) our approach.

\begin{table}[]

\caption{The accuracy comparison between the victim model and the hybrid model
trained by \tool in the form of $M_{\rm vic}$/$M_{\rm hyb}$.
Except for AlexNet where \sys has a higher accuracy due to a larger
model capacity, by average, \sys's relative accuracy loss (the ratio between
the accuracy of $M_{\rm hyb}$ and the accuracy of $M_{\rm vic}$) is 0.34\%.}
\label{tbl:nettailor_accuracy}
\vspace{-5pt}
\setlength{\tabcolsep}{3.0pt}
\begin{adjustbox}{max width=1\linewidth}
    \begin{tabular}{@{}ccccc@{}}
    \toprule
              & CIFAR10         & CIFAR100        & STL10           & UTKFace         \\ \midrule
    AlexNet   & 83.71\%/86.37\% & 56.46\%/61.96\% & 76.54\%/80.17\% & 89.42\%/88.92\% \\
    ResNet18  & 95.47\%/93.65\% & 79.94\%/76.79\% & 87.51\%/86.22\% & 86.97\%/88.24\% \\
    VGG16\_BN & 91.62\%/93.06\% & 73.03\%/73.11\% & 89.67\%/89.42\% & 89.19\%/89.46\% \\
    \bottomrule
    \end{tabular}
\end{adjustbox}

\end{table}

\begin{tcolorbox}[size=small]
\textbf{Answer to RQ4}: Besides achieving a principled security guarantee, \sys
doesn't undermine model accuracy.
\end{tcolorbox}

\subsection{Performance on Real-World Devices}

We created a prototype framework on a Desktop PC with Intel Core i7-8700 3.20GHz
CPU and NVIDIA GeForce GTX 1080 GPU to evaluate \tool's speed-up on real
devices. The framework has two parts: SGX's shielded section and the GPU's
offloaded part. The SGX component is developed in C++ and is compiled using
Intel SGX SDK 2.6 and GCC 7.5. The GPU component is built in PyTorch
1.7 and is supported by CUDA 11.7. We reused code from
Goten~\cite{lucien2021goten} and Slalom~\cite{tramer2019slalom} and implemented
other \tool's operations, such as convolution, the OTP-based feature encryption,
and hybrid model architecture. We emulate production conditions by switching SGX
to hardware mode with all its protection. We ran all experiments ten times and
got the average inference time. We verify that the running time deviates less
than 10\% from the average. We mainly report the throughput (images per second)
as it is more straightforward to evaluate the speed of ML
systems~\cite{tramer2019slalom}. The lowest required throughput for a real-time
on-device ML service is 30 (a latency of 33ms)~\cite{bateni2020neuos}. The
throughput is computed by $1000/average\_latency$.

\begin{table}[]
\caption{The throughput comparison between \shieldwhole, no-shield, and \tool on
a real desktop with SGX and GPU. We switch SGX to the hardware mode to enable
all protections. In parentheses, we present the speedup w.r.t. the \shieldwhole
baseline.}
\label{tbl:realdevice}
\centering
\small
\begin{adjustbox}{max width=0.9\linewidth}

    \begin{tabular}{@{}cccc@{}}
    \toprule
                & AlexNet        & ResNet18       & VGG16 BN       \\ \midrule
    Black-box   & 6.56           & 7.67           & 1.55           \\
    No-Shield   & 495.27 (75.53$\times$) & 288.56 (36.56$\times$) & 103.10 (66.42$\times$) \\ \midrule
    CIFAR10     & 44.67 (6.78$\times$)   & 63.81 (8.32$\times$)   & 72.80 (46.90$\times$)  \\
    CIFAR100    & 47.36 (7.22$\times$)   & 46.63 (6.08$\times$)   & 58.69 (37.81$\times$)  \\
    STL10       & 85.79 (13.08$\times$)  & 65.24 (8.50$\times$)   & 71.35 (45.97$\times$)  \\
    UTKFaceRace & 41.29 (6.30$\times$)   & 58.03 (6.26$\times$)   & 42.34 (27.28$\times$)  \\ \bottomrule
    \end{tabular}

\end{adjustbox}
\vspace{-15pt}

\end{table}





\T~\ref{tbl:realdevice} presents the throughput of \tool on three models
(AlexNet, ResNet18, and VGG16\_BN), as well as two baselines \shieldwhole (in
the SGX) and no-shield (on the GPU). Shielding-whole-model is the throughput
lower bound, and no-shield is the upper bound. For each case, we display the
speedup w.r.t. \shieldwhole baseline in parentheses. For \shieldwhole, the
throughputs on the three models range from $1.55$ to $7.67$, far from the
required throughput of real-time service ($30$). The throughput of no-shield
baseline ranges from $103.10$ to $495.27$, much faster than the real-time
requirement. Besides, the throughput speedup of no-shield compared with
\shieldwhole ranges from $36.56\times$ to $75.53\times$, demonstrating the
efficiency of GPU. The throughputs of \tool range from $41.29$ to $85.79$,
which are, on average, $18.37\times$ faster than the \shieldwhole baseline
and satisfy the real-time requirement of ML services.

To further analyze the performance of \tool, we also logged the latency of
different parts during the inference phase. We break down the inference latency
of \tool into four parts: \texttt{Data Transfer}, \texttt{Slice in TEE},
\texttt{Backbone on GPU}, and \texttt{Non-Linear in TEE}. \texttt{Data Transfer}
is the time to transfer internal results between SGX and GPU. \texttt{Slice in
TEE} is the time to compute the private slices inside SGX. \texttt{Backbone on
GPU} is the time to compute the convolution layers of the backbone on the GPU.
\texttt{Non-Linear in TEE} is the time to compute the non-linear layers (\eg
ReLU) inside SGX (recall \S~\ref{sec:teeslice_detail} that the ReLU layers of
the backbone are computed inside TEE). \T~\ref{tbl:realdevice_breakdown} displays
the percentage of each part over the total inference latency. From the table, we
can see that \texttt{Slice in TEE} occupies 40.49\% of the inference time due to
the constrained computation resources inside SGX. \texttt{Data Transfer} and
\texttt{Non-Linear in TEE} occupy 35.61\% and 20.96\% of the inference time
because all the non-linear layers of the backbone are computed inside SGX.
\texttt{Backbone on GPU} only occupies 2.84\% of the time due to the strong
computation ability of the GPU. Note that although \tool introduces the additional
overhead of \texttt{Data Transfer}, \tool still accelerates the overall
inference time by a large margin.

\begin{table}[]
    \caption{\tool inference time breakdown.}
    \label{tbl:realdevice_breakdown}
    \vspace{-5pt}
    \centering
    \small
    \begin{adjustbox}{max width=1\linewidth}

    \begin{tabular}{@{}cccc@{}}
    \toprule
    \texttt{Data Transfer} & \texttt{Slice in TEE} & \texttt{Backbone on GPU} & \texttt{Non-Linear in TEE} \\ \midrule
    35.61\%       & 40.49\%      & 2.84\%          & 20.96\%           \\ \bottomrule
    \end{tabular}

\end{adjustbox}
\vspace{-5pt}

\end{table}

\begin{tcolorbox}[size=small]
\textbf{Answer to RQ5}: \tool accelerates the throughput by an average of
$18.37\times$ compared with the \shieldwhole baseline and satisfies the
real-time requirement. 
\end{tcolorbox}
\vspace{-5pt}

\subsection{Scalability to NLP Tasks}

The design of \tool is applicable to protect various DNN models. Thus, findings
on protecting computer vision models in \S~\ref{sec:teeslice_experiment} are also
applicable to NLP models. To demonstrate the generalization of \tool, we
evaluate \tool on a representative NLP model, BART~\cite{lewis2019bart}, and
three NLP datasets (SST-2, MRPC, and RTE) from the popular GLUE
dataset~\cite{wang2018glue}. We mainly report \ac{ms} accuracy in
\T~\ref{tbl:nlp}, and omit \ac{mia} accuracy as the partition strategy of \tool
does not leak additional membership information. The comparison baselines are
``No-Shield'' and ``Black-box'', aligned with
\S~\ref{sec:evaluate_existing_solutions}. The results are generally consistent
with \S~\ref{sec:teeslice_experiment}. The attack accuracies of \tool are
comparable with ``Black-box'' and are lower than ``No-Shield''. Besides, the
FLOPs of shielded layers by \tool are significantly lower than ``Black-box''
(over 10$\times$). Thus, we interpret that NLP models can also be effectively
shielded by \tool, and our findings in \textbf{RQ4} are generalizable to NLP
models.

\begin{table}[]
    \caption{\ac{ms} accuracy on NLP tasks against \tool.}
    \label{tbl:nlp}
    \vspace{-5pt}
    \centering
    \small
    \begin{adjustbox}{max width=0.62\linewidth}

    \begin{tabular}{@{}ccccc@{}}
    \toprule
              & SST-2   & MRPC    & RTE     & Average \\ \midrule
    Black-box & 50.92\% & 68.87\% & 48.38\% & 56.05\% \\
    No-Shield & 92.55\% & 85.05\% & 66.79\% & 81.46\% \\
    TEESlice  & 50.92\% & 68.64\% & 46.93\% & 55.49\% \\ \bottomrule
    \end{tabular}

\end{adjustbox}
\vspace{-15pt}

\end{table}

\begin{tcolorbox}[size=small]
\textbf{Answer to RQ6}: \tool is applicable to NLP tasks and manifests
consistently high effectiveness.
\end{tcolorbox}
\vspace{-5pt}

\section{Other Related Work}

Besides the \ac{tsdp} approaches in
Section~\ref{sec:evaluate_existing_solutions}, we notice following areas related
to securing DNN models with TEEs.

\parh{TEE in GPUs.}~Recent work explored implementing trusted architectures
directly inside GPUs to achieve isolation~\cite{volos2018graviton,
hua2020guardnn,NvidiaH100}. Such solutions require customizing hardware
and are designed for server centers. Our solution is primarily for user's end
devices, which requires no change to the hardware or shipped firmware. Thus,
this paper employs commercial GPUs in an ``out-of-the-box'' manner.

\parh{Side Channels.}~TEEs are known as vulnerable toward side-channel
attacks~\cite{nilsson2020a,bulck2018nemesis,kocher2019spectre,chen2020sgxpectre,schaik2019ridl,murdock2020plundervolt}.
While side channels may threaten DNN privacy, various defensive methods have
been proposed to mitigate side channel
breaches~\cite{nilsson2020a,lee2017inferring,chen2018racing,gruss2017strong}.
\sys can be integrated with such defense to reduce side channel
leakages.

\parh{Shielding-Whole-Model by TEE.}~We have reviewed existing \tsdp\ solutions
in \S~\ref{sec:literature_category}. In addition to splitting DNNs and
offloading certain parts of the model on GPUs to speedup model inference, we also notice
existing works explore putting the entire DNN models into
TEEs~\cite{lee2019occlumency, hanzlik2021mlcapsule, li2021lasagna,
kim2020vessels, shen2020occlum}. Nevertheless, these works often notably
sacrifice the utility of the protected DNN models.

\parh{TSDP for DNN Training.}~Researchers have proposed various TSDP solutions
for DNN training to protect the privacy of training data on the cloud
server~\cite{hashemi2021darknight,lucien2021goten,tramer2019slalom}. These
solutions are different from \tool because \tool is designed to
protect the model inference stage.

\section{Discussion}
\label{sec:discussion}

\parh{Other Choices of $Security$ and $Utility$.}~This paper aims to
comprehensively evaluate \ac{tsdp} with seven empirical metrics of $Security$
from well-developed attack toolkits~\cite{liu2022mldoctor,orekondy2019knockoff}.
These seven metrics cover the majority of \ac{ms} and \ac{mia} in literature. We
notice that differential privacy (DP) can also theoretically quantify
$Security$~\cite{dwork2014DP}, but we decide not to use it due to its
prohibitively high computational cost for large models. We leave the evaluation
of other envisioned metrics in future work. 

An intuitive choice of $Utility$ is the model inference latency. However, as
there are various TEE architectures on the market, \textit{e.g.}, Intel
SGX~\cite{mckeen2013innovative}, AMD SEV~\cite{kaplan2016amd}, Intel
TDX~\cite{IntelTDX}, ARM CCA~\cite{ArmCCA}, and
TrustZone~\cite{alves2004trustzone}, evaluating the latency on all DNN models,
TEE architectures and possible configurations is difficult. We leverage FLOP, a
platform-irrelevant function, to form $Utility$, and therefore, our conclusion
should not be affected by the TEE implementation details, and is generally
applicable to the wide range of TEE architectures.

\parh{Attacks to Black-box Models.}~Attackers can still compromise \sys with
black-box attacks~\cite{juuti2019prada, orekondy2019knockoff,
orekondy2020prediction, papernot2017practical,
tramer2016stealing,li2021membership,mehnaz2022are}. However, the black-box
attacks are \textit{much less effective} due to the lack of information about
model architectures and model weights. That is, we deem black-box attacks as the
upper bound security guarantee that can be offered by TEEs. Several methods are
proposed to mitigate black-box
attacks~\cite{juuti2019prada,orekondy2020prediction};
we view those defenses are orthogonal to TEE-based defenses.

\parh{Hyper-Parameters of \sys.}
We clarify that although \sys has involved hyper-parameters in
the training phase, those parameters are merely used for reducing the computation cost
inside TEE (amount of privacy-related slices) instead of influencing privacy
leakage. The training phase of \sys relies on several hyper-parameters,
including $\delta$, $\alpha_{\rm setup}$, $n$, and $rounds$. They are mundane in
training our model setup. Thus, we clarify that it is unnecessary to tune those
parameters and benchmark if their different values may influence privacy leakage.

\parh{Scalability to Large Language Models (LLMs).}
Over the last few months, LLMs (such as ChatGPT~\cite{ChatGPT} and
LLaMA~\cite{LLaMA}) have achieved great advances. The sizes of LLMs (containing
up to hundreds of billions of parameters~\cite{Awesome_LLM}) are larger than
traditional CNNs (only hundreds of millions of
parameters~\cite{Keras_application}) and thus introduce greater challenge to
TEEs-shielded model protection solutions. However, we note that \tool is also
applicable to LLMs to protect the sensitive model privacy with TEEs. The
\partitionbeforetrain strategy can be integrated with recent LLMs'
parameter-efficient training techniques (\eg LoRA~\cite{hu2022lora}) to
efficiently shield LLMs' critical privacy-related slices in the TEEs. The size
of shielded slices is only a small fraction of the entire model (up to 10,000 times
smaller~\cite{hu2022lora}) and thus can significantly improve the inference
latency. We believe \tool can be a promising solution to protect the privacy of
LLMs in the future.

\parh{New TEE Architectures.}
Desipte the traditional TEE architectures (\eg Intel
SGX~\cite{mckeen2013innovative} and ARM TrustZone~\cite{alves2004trustzone})
that have been widely used in the industry, new TEE architectures are still
emerging (\eg Intel TDX~\cite{IntelTDX} and ARM CCA~\cite{ArmCCA}). Such new
architectures may have stronger computation abilities. For example, Intel TDX
has a larger encrypted memory of 1 TB~\cite{IntelTDX}. Although such new TEEs
may mitigate the performance overhead of \shieldwhole solutions, they do not
harm the practicality of \tool because the computation speed of such new TEEs is
still not comparable with GPUs, not to say the GPU architectures are also
evolving. We believe \tool can be a promising solution to bridge the gap between
the new TEEs and the evolving GPUs.

\parh{Application Scope of \tool.}
This paper focuses on an important application scope: protecting DNN privacy on the user's
end devices with TEEs. With the development of hardware architectures, many mobile/IoT devices are already equipped with TEEs by default, such as TrustZone in Raspberry Pi~\cite{TrustZoneRP3} and Android 7~\cite{TrustZoneAndroid}. With the increasing awareness on the user privacy and companies' intellectual property embedded in the DNN model, we believe this topic will attract more attention in the future.

\parh{Threats to Validity.}
In \S~\ref{sec:evaluate_existing_solutions} and \S~\ref{sec:dilemma}, we use a practical adversary to evaluate existing solutions. The adversary has access to public data/models and constructs a surrogate model to perform \ac{ms} and \ac{mia}. We argue that this adversary is realistic and reasonable and we do not make any abnormal assumptions. The usage of public data/models is consistent with the assumption of existing works in this line of research~\cite{orekondy2019knockoff,wang2018with,chen2021teacher,orekondy2020prediction,liu2022mldoctor}.

\parh{Differential Privacy (DP).}
DP is a promising technique to theoretically quantify the privacy leakage of DNN
training data to defense against \ac{mia}~\cite{dwork2014DP}. Nevertheless, DP
is not designed to defense \ac{ms}. Besides, recent works show that DP may
provide insufficient privacy~\cite{mo2021ppfl}, harm
utility/fairness~\cite{bagdasaryan2019differential}, or degrade
performance~\cite{subramani2021enabling}.

\section{Conclusion}

We have systematically examined existing \tsdp\ solutions and uncovered their
defects in front of privacy stealing attacks. Further, we illustrate
the hurdles of identifying ``sweet spot'' DNN partition configurations, which
frequently vary between models and datasets. With lessons harvested from
attacking prior \tsdp\ solutions, we present \tool, a novel
\tsdp\ method that leverages the \partitionbeforetrain strategy. It achieves high-accuracy, high-security protection (comparable
with \shieldwhole baseline), and with much less (about $10\times$) computation overhead.

\section{Acknowledgments}
We would like to thank the anonymous reviewers for their valuable
feedback of this paper. Ding Li and Yao Guo are corresponding authors. This work was partly supported by the National
Natural Science Foundation of China (62172009,62141208). The HKUST authors were supported in part by the research fund provided by HSBC and the HKUST-VPRDO 30 for 30 Research Initiative Scheme under the contract Z1283.

\bibliographystyle{abbrv}
\bibliography{reference}

\newpage
\appendices
\section{Meta-Review}

\subsection{Summary}
This paper aims to mitigate the model-stealing and membership-inference attacks against TEE-shielded DNN models. The authors study the weaknesses of existing solutions and, based on their observations, propose a "partition-before-training" strategy that partitions private data from the pre-trained model and then separately trains privacy-related layers. The evaluation results show that their new solution improves privacy with little accuracy loss.

\subsection{Scientific Contributions}
\begin{itemize}
\item Identifies an Impactful Vulnerability
\item Provides a Valuable Step Forward in an Established Field
\end{itemize}

\subsection{Reasons for Acceptance}
\begin{enumerate}
\item This paper identifies an impactful vulnerability. Specifically, the authors demonstrate via a systematic evaluation that existing solutions that leverage TEEs to protect DNN model data and weights still have exploitable shortcomings when it comes to protecting privacy.
\item This paper provides a valuable step forward in an establish field by contributing a comprehensive and rigorous assessment of the current state-of-the-art techniques under various settings and criteria. They have selected two well-established attacks that target on-device ML models and demonstrated how the current techniques perform in mitigating them. They have also extended the assessment to identify an optimal setting for the current techniques that yields the best outcome, and shown that it is not a simple task to find a universal optimal setting.
\end{enumerate}

\subsection{Noteworthy Concerns} 
Several reviewers are concerned about the security and practicality of using OTP in the proposed solution. There has been extensive technical discussion regarding the security of OTP-based feature encryption schemes, starting with Slalom published in ICLR 2019. In a similar vein, reviewers expressed concerns about ensuring that the TEE does not exhaust the OTP during heavy long-term use, which might introduce an overhead not captured in the paper's evaluation. However, since OTP-based feature encryption schemes have been used in several prior solutions and has not been decisively proven insecure or impractical, the reviewers concluded that there is value in publishing this technique.

\section{Response to the Meta-Review} 

The meta-review notes the security and practicality of using OTP in TEESlice. For the security concern, the generated mask is never reused and the ciphertext lies in the finite field $\mathbb{Z}_p$ of integers modulo a prime $p$, thus it is provably secure against any cryptanalysis~\cite{tramer2019slalom}. For the practicallity concern, we note that the size of random pads does not need to be long or infinite because we can send a small amount of pads at the setup phase and periodically update the pads. A straightforward solution is that TEESlice receives a proporate number of pads at the setup phase. When the pads are about to run up (e.g. less than 5\%), TEESlice can request the model vendor for a batch of new pads. The OTP-based scheme has been adopted by prior work~\cite{tramer2019slalom,hou2021model,sun2020shadownet}, thus we believe OTP will not harm the security and utility of TEESlice.
\section{Artifact}
\label{append:sec:artifact}

We submitted the artifact in the supplementary. The artifact includes the code
to attack the \tsdp\ solutions, the code to train the models, and the script to
run the results in the paper. The structure of the submitted artifact is as
follows:

\begin{itemize}
    \item \texttt{readme.md} describes the functionality of code files and
    directories in the artifact.

    \item \texttt{plot/} contains plotting scripts to generate figures and
    tables from the experiment results.

    \item \texttt{model-stealing/} includes the code to perform model stealing
    attacks against \tsdp solutions

    \item \texttt{membership-inference/} includes the code to perform membership
    inference attacks against \tsdp solutions.
    \item \texttt{soter-attack/} includes the code to attack SOTER.
    \item \texttt{real-device} includes the prototype of the system on real devices.
\end{itemize}

\section{Hyper-parameters of Training $M_{vic}$}
\label{append:vic_hyperparameters}

For the training in \ac{ms} part, we follow the hyper-parameter settings of
Knockoff Nets~\cite{KnockoffNetCode}: we use a mini-batch size of 64, select
cross-entropy loss, use SGD with weight decay of 5e-4, a momentum of 0.5, and
train the victim models for 100 epochs. The learning rate is originally set to
0.1 and decays by 0.1 every 60 epochs. Accuracies of the trained models $M_{\rm
vic}$ are reported in \T~\ref{tbl:nettailor_accuracy}. The accuracies are
generally consistent with public results~\cite{STL10ACC,CIFAR10ACC,CIFAR100ACC}.

For the training in \ac{mia} part, we follow the hyper-parameter settings of
\textsc{ML-Doctor}~\cite{MLDoctorCode}: we set mini-batch size as 64, select
cross-entropy loss, use SGD optimizer with the weight decay as 5e-4, the
momentum as 0.9, the learning rate as 1e-2, and the training epoch as 100. The
hyper-parameters to train shadow models are the same as victim models. Model
accuracies are reported in \App~\ref{append:mia_victim_accuracy}; in short, all
models achieve high accuracy on the target training dataset and the accuracies
are consistent with prior works~\cite{liu2022mldoctor}.

\section{Model Accuracy for Privacy Inference}
\label{append:mia_victim_accuracy}

We clarify that in \T~\ref{append:tbl:mia_accuracy}, we display the
training/testing accuracy of each setting in the membership inference experiments.
The accuracies of certain datasets are higher than the number reported in~\cite{liu2022mldoctor}
because we use a publicly available model as initialization to get a higher
performance.

\begin{table}[]
    \caption{Performance of $M_{\rm vic}$ used in the experiments of
    membership inference attacks. Each box displays the test/training accuracy
    for each case.}
    \label{append:tbl:mia_accuracy}
    \begin{adjustbox}{max width=1\linewidth}
    \begin{tabular}{@{}ccccc@{}}
    \toprule
              & CIFAR10          & CIFAR100        & STL10            & UTKFace          \\ \midrule
    AlexNet   & 77.92\%/100.00\% & 47.91\%/99.99\% & 72.09\%/100.00\% & 83.94\%/99.96\%  \\
    ResNet18  & 85.05\%/100.00\% & 61.69\%/99.99\% & 77.57\%/99.97\%  & 87.15\%/100.00\% \\
    ResNet34  & 85.29\%/100.00\% & 63.00\%/99.97\% & 78.71\%/100.00\% & 87.91\%/99.96\%  \\
    VGG16\_BN & 92.95\%/100.00\% & 71.09\%/99.99\% & 88.89\%/100.00\% & 89.33\%/99.96\%  \\
    VGG19\_BN & 91.31\%/100.00\% & 70.73\%/99.99\% & 86.37\%/100.00\% & 88.01\%/99.93\%  \\ 
    \bottomrule
    \end{tabular}
\end{adjustbox}
\end{table}

\section{Attack ShadowNet}
\label{sec:append:shadownet}

In this section, we will analyze the defense mechanism of ShadowNet in front of
the proactive adversary described in \S~\ref{sec:evaluated_attacks}. We will
first briefly summarize the design of ShadowNet
(\App~\ref{sec:append:shadownet}.\ref{sec:append:shadownet:description}), then describe the attack pipeline
(\App~\ref{sec:append:shadownet}.\ref{sec:append:shadownet:attack}).

\subsection{ShadowNet Defense Description}
\label{sec:append:shadownet:description}

In this part we describe the defense mechanism of
ShadowNet~\cite{sun2020shadownet} in detail. ShadowNet uses two techniques to
protect the outsourced weights: additive mask and filter permutation. Let
$\mathbf{w} = [\mathbf{w}_1, \mathbf{w}_2, \cdots, \mathbf{w}_n]$ be the weight
matrix of a linear layer, where $n$ is the number of filters. We describe each
technique as follow.

For additive mask, the defender selects a set of random masks $\mathbf{F} =
[\mathbf{f}_1, \cdots, \mathbf{f}_{m-n}]$, where $m = \lceil r \cdot n
\rceil$. $r$ is the pre-defined expansion ratio and we follow the original
paper to set $r=1.2$. For each filter weight $\mathbf{w}_i$, defender
randomly selects a mask $\mathbf{f}'_i \in \mathbf{F}$ to add and combine
all the masked weights and random masks:

\begin{equation}
    \mathbf{w}' = [\mathbf{w}_1 + \mathbf{f}'_1, \cdots, \mathbf{w}_n + \mathbf{f}'_n, \mathbf{f}_1, \cdots, \mathbf{f}_{m-n} ],
\end{equation}
For filter permutation, the defender selects a random permutation and permute
the masked weight $\mathbf{w}'$. The permuted weight $\hat{\mathbf{w}}$ is sent
to the GPU.

\subsection{Attack Pipeline}
\label{sec:append:shadownet:attack}

\begin{figure}[!t]
    \centering
    \includegraphics[width=\linewidth,trim={0 490 0 0},clip]{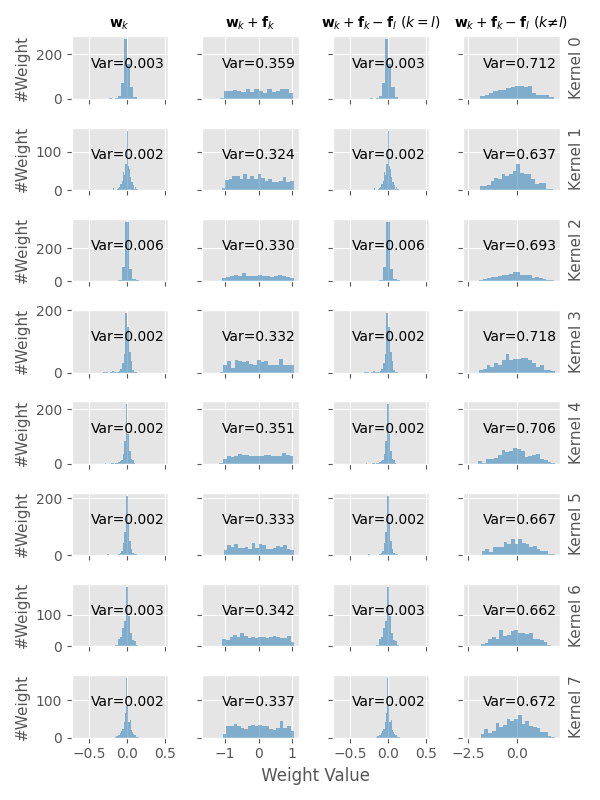}
    \caption{Weight distribution and variances of the first eight filters for the first convolution layer in ResNet18.}
    \label{fig:shadownet:attack_multi_kernel}
    \end{figure}

In this part, we will illustrate the detailed pipeline of attacking ShadowNet.
The attacker first removes the additive mask from the obfuscated weights, then
use the public model to recover permutation. The removal of random masks is
motivated by the observation that the value distribution of weights is different
from the distribution of added noise. In
\F~\ref{fig:shadownet:attack_multi_kernel}, we plot the first eight filters of
the first convolution layer in ResNet18.  
The first column of the figure shows the distribution of original weights and the second column shows the weight distribution after adding random noise. The variance after adding random noise is about $100\times$ larger than the variance of the original weights.
In our implementation, we set a threshold of $0.01$ to distinguish variance of different distribution.

We then describe the pipeline of pairwise filter differentiation.
Let $\hat{\mathbf{w}} = [ \hat{\mathbf{w}}_1, \hat{\mathbf{w}}_2, \cdots, \hat{\mathbf{w}}_m ]$ be the obfuscated weight of one linear layer. The attacker iterates all possible filter pair $(\hat{\mathbf{w}}_i, \hat{\mathbf{w}}_j)$ and computes the differentiation $\hat{\mathbf{w}}_{(i,j)} = \hat{\mathbf{w}}_i - \hat{\mathbf{w}}_j$. There are different conditions for $(\hat{\mathbf{w}}_i, \hat{\mathbf{w}}_j)$:
\begin{enumerate}
    \item If $\hat{\mathbf{w}}_i = \mathbf{w}_k + \mathbf{f}'_k$ and $\hat{\mathbf{w}}_j = \mathbf{f}'_k$, then we have $\hat{\mathbf{w}}_{(i,j)} = \mathbf{w}_k$ and the variance of $\hat{\mathbf{w}}_{(i,j)}$ should be smaller than $0.01$. It means we fully recover the weight of one filter.
    \item If $\hat{\mathbf{w}}_i = \mathbf{w}_k + \mathbf{f}'_k$ and $\hat{\mathbf{w}}_j = \mathbf{w}_l + \mathbf{f}'_k$, then we have $\hat{\mathbf{w}}_{(i,j)} =  \mathbf{w}_k - \mathbf{w}_l$. The variance of $\hat{\mathbf{w}}_{(i,j)}$ is expected to be smaller than $0.01$. Although we don't recover the original weight, we remove the additive masks.
    \item If $\hat{\mathbf{w}}_i = \mathbf{w}_k + \mathbf{f}'_k$ and $\hat{\mathbf{w}}_j = \mathbf{f}'_l$, where $l \neq k$, then we have $\hat{\mathbf{w}}_{(i,j)} = \mathbf{w}_k + \mathbf{f}'_k - \mathbf{f}'_l$. The distribution of $\hat{\mathbf{w}}_{(i,j)}$ is similar to the distribution of $\mathbf{f}'_k - \mathbf{f}'_l$ and the variance is larger than $0.01$. 
    \item If $\hat{\mathbf{w}}_i = \mathbf{w}_k + \mathbf{f}'_k$ and $\hat{\mathbf{w}}_j = \mathbf{w}_l + \mathbf{f}'_l$, the situation is similar to previous one. 
    \item If $\hat{\mathbf{w}}_i = \mathbf{f}'_k$ and $\hat{\mathbf{w}}_j = \mathbf{f}'_l$, then $\hat{\mathbf{w}}_{(i,j)} = \mathbf{f}'_k - \mathbf{f}'_l$. The distribution of $\hat{\mathbf{w}}_{(i,j)}$ is random distribution and the variance is larger than $0.01$.
\end{enumerate}
As discussed above, by identifying if the variance of $\hat{\mathbf{w}}_{(i,j)}$
is smaller than $0.01$, we can remove all conditions of random mask. The
recovered weight matrixes include the weight for all filters $\mathbf{w}_k$ and
possibly some difference between filter weights $\mathbf{w}_k - \mathbf{w}_l$.
The redundant weight difference can be easily removed. Thus we can recover all
the filter weights despite their position in the layer.

After recovering weight values, we recover filter positions with the information
of public models. For each recovered filter, we compute the $L_2$ distance with
all the public model's filters on the same layer. We use the position of public
model's filter that has the smallest distance with the recovered filter as the
recovered position. Experimental results show that in this way, we can precisely
recover more than 95\% of the filter positions. At last, we train the recovered
model with the queried data to fully recover the model functionality. Note that
the attack against ShadowNet does not require any additional data or
information. 

\section{Validation of $\% FLOPs$}
\label{append:sec:flops_validation}

\begin{figure}[!t]
\centering
\includegraphics[width=\linewidth]{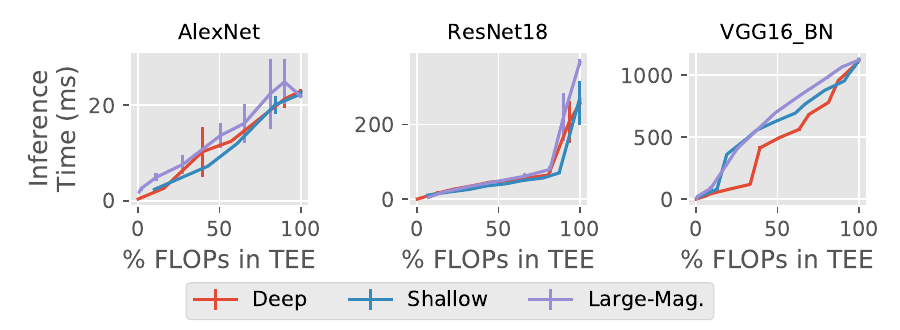}
\caption{Correlation between $\% FLOPs$ and inference latency on
Occlum~\cite{shen2020occlum}. Inference time is averaged over 10 runs.}
\label{fig:occlum_time_flops}
\end{figure}

To validate the correlation between $\% FLOPs$ and utility cost, we did a toy
experiment on the real-world commercial TEE platform, Occlum~\cite{shen2020occlum}, based on Intel SGX.
We evaluate the model inference time on three defense mechanisms: shielding deep
layers (\ding{172}), shielding shallow layers (\ding{173}), and shielding large-magnitude layers (\ding{174}). We repeat each
experiment ten times and plot the average inference time and standard
error on three models in \F~\ref{fig:occlum_time_flops}. As the figure shows,
for all TSDP solutions, inference time monotonically increases as $\% FLOPs$
increases. Thus using $\% FLOPs$ to represent inference time and utility cost is
reasonable.

\section{Label-Only Applications}
\label{append:sec:label_only}

\begin{table}[]

    \caption{We surveyed 24 different applications that contain on-device ML
    models from eight important on-device tasks according to prior
    work~\cite{sun2021mind}. \textit{ALL} of these applications only return
    prediction labels, rather confidences or probabilities. }
    \label{append:tbl:label_only_app}
    \begin{adjustbox}{max width=1\linewidth}

    \begin{tabular}{cc}
    \hline
    Task                                     & App Name                             \\ \hline
    \multirow{3}{*}{OCR}                     & Scanner Almighty                     \\
                                             & Universal Scanner King               \\
                                             & King Of Universal Scanning Treasure  \\ \hline
    \multirow{3}{*}{Handwriting Recognition} & Photo To Text Recognition Extraction \\
                                             & Universal Scan Recognition           \\
                                             & Scanning Almighty                    \\ \hline
    \multirow{3}{*}{Speech Recognition}      & Speech To Text                       \\
                                             & Voice-To-Text Assistant              \\
                                             & Text-To-Speech Assistant             \\ \hline
    \multirow{3}{*}{Card Recognition}        & Identity Verification Assistant      \\
                                             & Identity Scanning Recognition        \\
                                             & Id Card Scanning                     \\ \hline
    \multirow{3}{*}{Face Tracking}           & Face Recognition Realtime            \\
                                             & PopFaces                             \\
                                             & Luxand Face Recognition              \\ \hline
    \multirow{3}{*}{Hand Detection}          & Live2DViewerEX Motion Tracker        \\
                                             & ManoMotion Tech PRO                  \\
                                             & ManoMotion Tech Lite                 \\ \hline
    \multirow{3}{*}{Liveness Detection}      & X facial liveness detection          \\
                                             & Luxand Face Recognition              \\
                                             & Oz Liveness Demo Application         \\ \hline
    \multirow{3}{*}{Face Recognition}        & PopFaces                             \\
                                             & Face Recognition Realtime            \\
                                             & Luxand Face Recognition              \\ \hline
    \end{tabular}

\end{adjustbox}
\end{table}

In this section, we argue that label-only ML service is popular and important
for on-device ML systems. We conduct a survey on the percentage of how much
on-device ML systems in Android apps only return labels, instead of confidence
scores or top-k confidences. Following prior work on the security of on-device
ML models~\cite{sun2021mind}, our survey covers eight important tasks (as listed
in \T~\ref{append:tbl:label_only_app}), three different application markets
(Google Play, Tencent My App, and 360 Mobile Assistant), and 24 most downloaded
applications (three apps for each task). We found that \textit{all} of the
surveyed applications only return the prediction label, instead of top-k
confidences. We list the name of the applications in \T~\ref{append:tbl:label_only_app}.

\section{Dataset Description}
\label{append:dataset}
In this section, we describe the detailed information of the datasets:
\begin{itemize}
    \item \textbf{CIFAR10}~\cite{krizhevsky2009learning} is a widely-used image
    classification dataset. The data is sampled from the large-scale ImageNet
    dataset. CIFAR10 contains 60K 32x32 images in 10 classes. The images are
    split into 50K training samples and 10K test samples.

    \item \textbf{CIFAR100}~\cite{krizhevsky2009learning} is similar to CIFAR10
    and is drawn from the same data distribution. CIFAR100 has 100 classes and
    is more difficult to classify than CIFAR10. Similar to CIFAR10, CIFAR100 has
    50K training images and 10K test images. Averagely, each class has 500
    images to train.

    \item \textbf{STL10}~\cite{coates2011an} is a 10-classes dataset but is much
    smaller than CIFAR10. Each class of STL10 has 13K training images. The
    classes include ship, truck, airplane, car, cat, bird, deer, dog, horse, and
    monkey.

    \item \textbf{UTKFace}~\cite{zhang2017age} is a face dataset labeled with
    age, gender, and race. Following~\cite{liu2022mldoctor}, we use the races
    (White, Black, Asian, and Indian) as classification labels, and there are
    22K valid images. We follow the common dataset split ratio $9:1$ to split
    the dataset into a training split (19.8K images) and a test split (2.2K
    images). 
\end{itemize}

\section{Results on ResNet34 and VGG19\_BN}
\label{append:sec:resnet34_vgg19_results}

\begin{figure*}[!t]
\centering
\includegraphics[width=\linewidth]{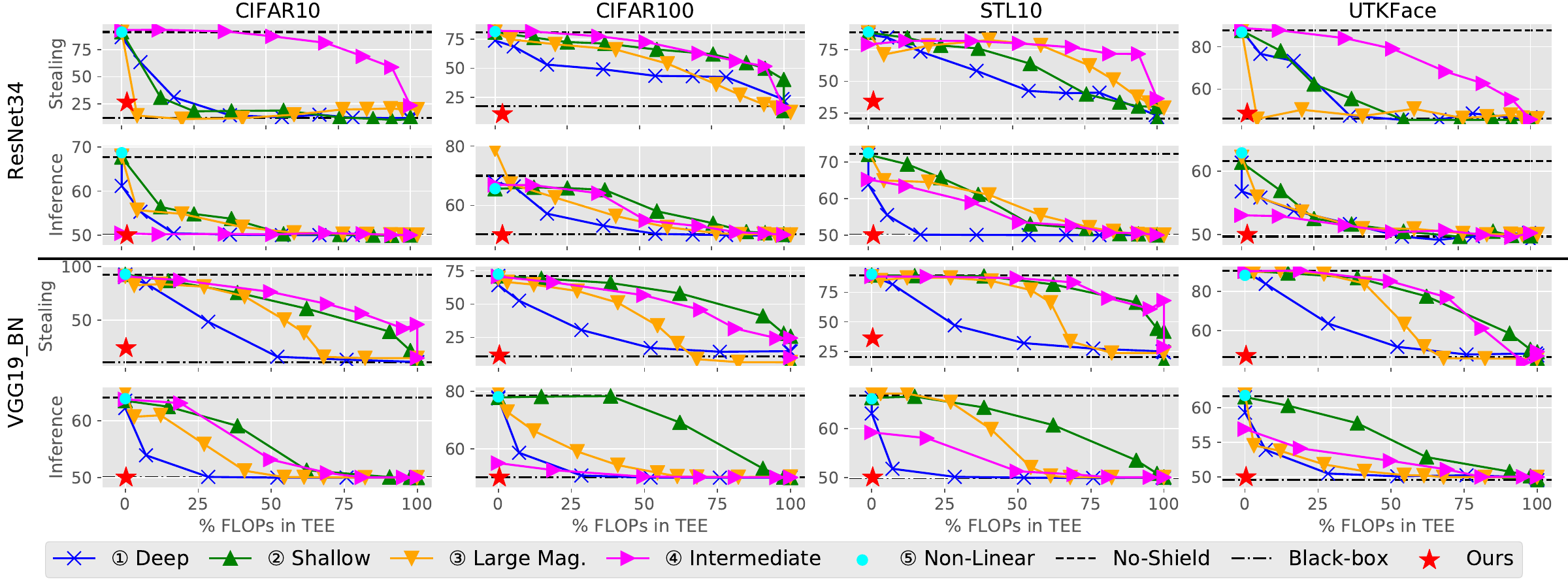}
\caption{Model stealing and membership inference on ResNet34 and VGG19\_BN.}
\label{fig:acc_mia_flops_one_fig_append}
\end{figure*}

In this section, we report evaluation results of the model stealing attack and membership inference attack on ResNet34 and VGG19\_BN. The results of de facto attack (\S~\ref{sec:empirical_evaluation_results}) is displayed in \T~\ref{tbl:evaluate_solution_append}. 
The relationship between $Security$ and $Utility$ (\S~\ref{sec:evaluation_dilemma_results}) is displayed in \F~\ref{fig:acc_mia_flops_one_fig_append}.
The results of $Utility(C^*)$ ($\% FLOPs(C^*)$) values of ``sweet spot'' configuration (\S~\ref{sec:evaluation_dilemma_results}) is displayed in \T~\ref{tbl:optimal_config_append}.
The model stealing accuracies of other attack assumptions against \tool (\S~\ref{sec:experiment:other_assumption}) is displayed in \T~\ref{tbl:other_assumption_append}.
The results are generally consistent with the findings in the main paper. 

\begin{table*}[h]
    \caption{Attack accuracies regarding representative defense schemes. ``C10'',
    ``C100'', ``S10'', and ``UTK'' represent CIFAR10, CIFAR100, STL10, and UTKFace,
    respectively. The last row reports the average accuracy toward each
    defense relative to the baseline black-box solutions. For each setting, we mark
    the highest attack accuracy in \colorbox{high}{red} and the lowest accuracy in
    \colorbox{low}{yellow}. Attack accuracy toward our approach
    (\S~\ref{sec:approach}) is marked with \colorbox{best}{green}.}
    \label{tbl:evaluate_solution_append}
    \begin{adjustbox}{max width=1\linewidth}
    
        \begin{tabular}{ccccccccccccccccccc}
        \hline
                                   &      & \multicolumn{8}{c}{Stealing Accuracy}                                                  &  & \multicolumn{8}{c}{Conf. Attack Accuracy}                                              \\ \cline{3-10} \cline{12-19} 
                                   &      & White-box & DarkneTZ & Serdab  & Magnitude & SOTER  & ShadowNet & Our     & Black-box &  & White-box & DarkneTZ & Serdab  & Magnitude & SOTER & ShadowNet & Our     & Black-box \\ \hline
        \multirow{4}{*}{\rotatebox{90}{ResNet34}}  & C10  & 91.05\%   & 86.28\%  & 30.88   & \cellcolor{low}14.31\%   & \cellcolor{high}93.10\% & 91.03\% & \cellcolor{best}26.43\% & 12.07\%   &  & 67.72\%   & 61.19\%  & 56.41\% & 55.68\%   & \cellcolor{low}50.16\% & \cellcolor{high}66.74\% & \cellcolor{best}50.00\% & 50.00\%   \\
                                   & C100 & 80.85\%   & \cellcolor{low}73.64\%  & 76.22\% & 74.68\%   & \cellcolor{high}81.47\% & 81.34\% & \cellcolor{best}10.67\%   & 17.22\%   &  & 65.56\%   & \cellcolor{high}67.78\%  & 66.06\% & 67.29\%   & 66.84\%  & 65.63\%  & \cellcolor{best}50.00\% & 50.08\%   \\
                                   & S10  & 88.29\%   & 86.74\%  & 84.09\% & \cellcolor{low}70.94\%   & 81.81\% & \cellcolor{high}88.79\%  & \cellcolor{best}34.19\% & 20.77\%   &  & 72.03\%   & 63.85\%  & \cellcolor{high}69.40\% & 64.95\%   & \cellcolor{low}63.42\% & 67.52\& & \cellcolor{best}50.00\% & 50.00\%   \\
                                   & UTK  & 87.65\%   & \cellcolor{high}87.83\%  & 78.07\% & \cellcolor{low}45.87\%   & \cellcolor{high}87.83\%  & 86.97\% & \cellcolor{best}48.50\% & 46.16\%   &  & 61.33\%   & 56.80\%  & 56.91\% & 55.93\%   & \cellcolor{low}52.92\%  & \cellcolor{high}62.85\% & \cellcolor{best}50.00\% & 50.00\%   \\ \midrule
        \multirow{4}{*}{\rotatebox{90}{VGG19\_BN}} & C10  & 92.59\%   & 90.34\%  & 86.87\% & \cellcolor{low}82.66\%   & 87.34\% & \cellcolor{high}92.89\%  & \cellcolor{best}24.08\% & 10.90\%   &  & 63.57\%   & 62.28\%  & 62.52\% & \cellcolor{low}60.74\%   & \cellcolor{high}63.15\% & 62.94\%  & \cellcolor{best}50.00\% & 50.00\%   \\
                                   & C100 & 71.17\%   & \cellcolor{low}64.30\%  & 69.23\% & 66.55\%   & 66.32\%  & \cellcolor{high}72.48\%  & \cellcolor{best}11.47\% & 10.54\%   &  & 77.94\%   & 77.58\%  & \cellcolor{high}78.22\% & 72.91\%   & \cellcolor{low}52.53\%  & 78.16\% & \cellcolor{best}50.00\% & 50.01\%   \\
                                   & S10  & 89.62\%   & 89.31\%  & 88.99\% & \cellcolor{low}85.26\%   & \cellcolor{high}88.41\%  & 87.36\%  & \cellcolor{best}36.11\% & 19.93\%   &  & 66.35\%   & 63.14\%  & 66.62\% & \cellcolor{high}67.25\%   & \cellcolor{low}58.06\%  & 66.17\% & \cellcolor{best}50.00\% & 49.91\%   \\
                                   & UTK  & 90.55\%   & 90.51\%  & 89.33\% & 89.83\%   & \cellcolor{high}90.74\% & 88.26\%  & \cellcolor{best}47.09\% & 46.44\%   &  & 61.50\%   & 59.27\%  & 60.28\% & 54.55\%   & \cellcolor{low}54.13\%  & \cellcolor{high}61.74\% & \cellcolor{best}50.00\% & 50.00\%   \\ \midrule
        \multicolumn{2}{c}{Average}       & 5.01$\times$      & 4.79$\times$     & 4.21$\times$    & 3.76$\times$     & 4.87$\times$ & 4.98$\times$ &  1.45$\times$    & 1.00$\times$      & & 1.34$\times$      & 1.28$\times$     & 1.29$\times$    & 1.25$\times$      & 1.16$\times$  & 1.32$\times$   & 1.00$\times$    & 1.00$\times$      \\ \hline
    
                                \end{tabular}
    
    \end{adjustbox}
    \end{table*}

\colorlet{best}{green!30}
\colorlet{high}{red!30}
\colorlet{low}{yellow}

\begin{table*}[!h]
\caption{Different $Utility(C^*)$ ($\% FLOPs(C^*)$) values of ``sweet spot'' in front of model stealing attack and membership inference attacks. 
A lower value represents a lower utility cost. The $\% FLOPs(C^*)$ for \noshield and black-box baselines are 0\% and 100\%, respectively.
For each \tsdp\ solution (row), we mark the lowest $Utility(C^*)$ with \colorbox{low}{yellow} and the 
highest value with \colorbox{high}{red}.
For each case (model and dataset, column), we mark the lowest $Utility(C^*)$ across all solutions with \colorbox{best}{green}.
The last column is the average utility cost for each solution.
}
\label{tbl:optimal_config_append}
\vspace{-5pt}
\setlength{\tabcolsep}{1.5pt}
\centering
\begin{adjustbox}{max width=0.6\linewidth}

    \begin{tabular}{@{}cccccclcccccc@{}}
    \toprule
                                                                                     &              & \multicolumn{4}{c}{ResNet34}              &  & \multicolumn{4}{c}{VGG19\_BN}             &  & \multirow{2}{*}{Average} \\ \cmidrule(lr){3-6} \cmidrule(lr){8-12}
                                                                                     &              & C10      & C100     & S10      & UTK      &  & C10      & C100     & S10      & UTK      &  &                          \\ \midrule
    \multirow{4}{*}{\begin{tabular}[c]{@{}c@{}}Model \\ Stealing\end{tabular}}       & Deep         & 80.01\%  & \cellcolor{high}100.00\% & \cellcolor{high}100.00\% & \cellcolor{low}37.47\%  &  & 75.79\%  & \cellcolor{high}100.00\% & \cellcolor{high}100.00\% & 75.79\%  &  & 83.63\%                  \\
                                                                                     & Shallow      & 75.50\%  & \cellcolor{high}100.00\% & \cellcolor{high}100.00\% & \cellcolor{low}56.05\%  &  & \cellcolor{high}100.00\% & \cellcolor{high}100.00\% & \cellcolor{high}100.00\% & \cellcolor{high}100.00\% &  & 91.44\%                  \\
                                                                                     & Large Mag.   & 20.35\%  & 94.96\%  & \cellcolor{high}100.00\% &  \cellcolor{low}5.22\%   &  & \cellcolor{high}100.00\% & 67.68\%  & \cellcolor{high}100.00\% & 67.98\%  &  & 69.56\%                  \\
                                                                                     & Intermediate & \cellcolor{high}100.00\% & \cellcolor{high}100.00\% & \cellcolor{high}100.00\% & \cellcolor{high}100.00\% &  & \cellcolor{high}100.00\% & 100.00   & \cellcolor{high}100.00\% &  \cellcolor{low}95.30\%  &  & 99.41\%                  \\ \midrule
    \multirow{4}{*}{\begin{tabular}[c]{@{}c@{}}Membership \\ Inference\end{tabular}} & Deep         & 18.01\%  &  \cellcolor{high}55.49\%  & 18.01\%  & 37.47\%  &  & 28.42\%  & 28.42\%  &  \cellcolor{low}7.11\%   & 28.42\%  &  & 27.67\%                  \\
                                                                                     & Shallow      &  \cellcolor{low}56.05\%  & 87.03\%  & 75.50\%  & 38.02\%  &  & 62.11\%  &  \cellcolor{high}97.63\%  &  \cellcolor{high}97.63\%  & 90.53\%  &  & 75.56\%                  \\
                                                                                     & Large Mag.   & 40.84\%  & 58.31\%  &  \cellcolor{high}74.82\%  & 40.84\%  &  & 40.90\%  & 54.49\%  & 54.49\%  &  \cellcolor{low}27.02\%  &  & 48.96\%                  \\
                                                                                     & Intermediate &  \diagfil{1.2cm}{low}{best}{0.60}\%   &  \cellcolor{high}83.89\%  &  \cellcolor{high}83.89\%  & 36.36\%  &  & 69.26\%  & 19.05\%  & 49.82\%  & 69.26\%  &  & 51.52\%                  \\ \midrule
    \multicolumn{2}{c}{Ours}                                                                        & 1.83\%   &  \cellcolor{best}2.56\%   & \cellcolor{best}1.73\%   & \cellcolor{best}1.92\%   &  & \cellcolor{best}0.27\%   & \cellcolor{best}0.37\%   & \cellcolor{best}0.35\%   & \cellcolor{best}0.37\%   &  & 1.17\%                   \\ \bottomrule
    \end{tabular}

\end{adjustbox}
\end{table*}

\begin{table}[]
    \caption{Comparison of model stealing accuracy between different attack assumptions on ResNet34 and VGG19\_BN.}
    \label{tbl:other_assumption_append}
    \centering
    \begin{adjustbox}{max width=0.8\linewidth}

    \begin{tabular}{@{}ccccc@{}}
    \toprule
                               &      & Hybrid  & Backbone & Victim  \\ \midrule
    \multirow{4}{*}{ResNet34}  & C10  & 26.43\% & \cellcolor{best}24.01\%  & \cellcolor{high}26.45\% \\
                               & C100 & \cellcolor{best}10.67\% & 16.46\%  & \cellcolor{high}20.54\% \\
                               & S10  & 34.19\% & \cellcolor{best}32.71\%  & \cellcolor{high}42.02\% \\
                               & UTK  & \cellcolor{best}48.5\%  & 51.63\%  & \cellcolor{high}52.18\% \\ \midrule
    \multirow{4}{*}{VGG19\_BN} & C10  & 24.08\% & \cellcolor{high}25.73\%  & \cellcolor{best}22.09\% \\
                               & C100 & \cellcolor{best}11.47\% & \cellcolor{high}18.19\%  & 14.69\% \\
                               & S10  & \cellcolor{high}36.11\% & \cellcolor{best}32.74\%  & 37.67\% \\
                               & UTK  & \cellcolor{best}47.09\% & \cellcolor{high}52.72\%  & 51.54\% \\ \bottomrule
    \end{tabular}

\end{adjustbox}
\end{table}

\section{Results on Other Metrics}
\label{append:sec:other_metrics}

As noted in our main paper, when assessing the security of existing \tsdp\
solutions, we consider seven metrics. This Appendix section will
re-visit the employed $Security$ metrics and then display the results of
the other $Security$ metrics. 

\subsection{Metric Definitions}
\label{append:sec:metric_definition}

\noindent \textit{Accuracy}~\cite{rakin2022deepsteal} measures how much test
samples can be correctly classified by the attacker's surrogate model. Achieving
high accuracy is a primary goal of model stealing attacks. 

\noindent \textit{Fidelity}~\cite{rakin2022deepsteal} is the percentage of
test samples with identical prediction between $M_{\rm sur}$ and $M_{\rm vic}$,
including the samples that are misclassified by $M_{\rm
vic}$.
\noindent \textit{Attack Success Rate (ASR)}~\cite{rakin2022deepsteal} is the
percentage that the adversarial samples generated by $M_{\rm sur}$ can
successfully mislead the output of $M_{\rm vic}$. It measures the
transferability of adversarial
samples~\cite{papernot2016transferability,rakin2022deepsteal}. We use the
popular PGD attack~\cite{madry2018towards} to generate adversarial samples.
Following~\cite{rakin2022deepsteal}, we use $L_{\inf}$ norm, $\epsilon=0.03$,
and iteration step of 7 for add datasets. We adopt the PGD implementation public tools~\cite{ding2019advertorch}. 

\noindent \textit{Generalization Gap}~\cite{yuan2022membership} is the
difference between the average accuracies on $M_{\rm vic}$'s training dataset
and the test dataset. The more a model remembers the privacy information of the
training dataset, the larger the generalization gap is. The generalization gap
strongly connects with membership inference attacks~\cite{yeom2018privacy}. 

\noindent \textit{Confidence Gap}~\cite{yuan2022membership} calculates the
difference in average confidence between $M_{\rm vic}$'s training dataset and
test dataset. Similar to the generalization gap, the confidence gap positively
relates to the extent that $M_{\rm sur}$ remembers training data. 

\noindent \textit{Confidence-Attack
Accuracy}~\cite{nasr2018machine,liu2022mldoctor} represents the membership
classification accuracy based on the output confidence. The attack algorithm
uses model posterior as input to infer data membership information. We use the
``Black-Box/Shadow'' implementation of
\textsc{ML-Doctor}~\cite{MLDoctorCode,liu2022mldoctor}.

\noindent \textit{Gradient-Attack Accuracy}~\cite{nasr2019comprehensive}
represents the white-box membership attack accuracy based on the internal
gradients. This attack uses gradient information and loss value to predict data
membership. We use the ``White-Box/Shadow'' attack implementation of
\textsc{ML-Doctor}~\cite{MLDoctorCode,liu2022mldoctor}.

\subsection{\tsdp Solution Evaluation}
\label{append:sec:other_metrics:evaluation}

This part includes the evaluation results of representative defense schemes in
\S~\ref{sec:empirical_evaluation_results}.
\T~\ref{tbl:evaluate_solution:fidelity} to
\T~\ref{tbl:evaluate_solution:grad_attack} reports the results of the other five
metrics that are not reported in the main paper; our overall findings at
this step are consistent with the findings and lessons summarized in the main
paper. Note that the confidence gap and generalization gap of our approach and
random-guess are 0\% because, for a $M_{\rm sur}$ that never sees $M_{\rm vic}$'s
training data, the scores are the same between the training and
testing dataset.

\begin{table}[!htbp]
    \caption{\textit{Fidelity} regarding representative defense schemes.}
    \centering
    \label{tbl:evaluate_solution:fidelity}
    \begin{adjustbox}{max width=\linewidth}

    \begin{tabular}{cccccccccc}
    \hline
                               &      & White-box & DarkneTZ & Serdab  & Magnitude & SOTER   & ShadowNet & Ours    & Black-box \\ \hline
    \multirow{4}{*}{\rotatebox{90}{AlexNet}}   & C10  & 99.95\%   & 83.62\%  & 66.21\% & 69.34\%   & 83.94\% & 100.00    & 19.84\% & 24.81\%   \\
                               & C100 & 96.69\%   & 52.47\%  & 59.61\% & 61.39\%   & 66.98\% & 100.00    & 9.47\%  & 11.72\%   \\
                               & S10  & 99.16\%   & 89.84\%  & 81.06\% & 88.61\%   & 39.91\% & 37.71     & 23.50\% & 15.99\%   \\
                               & UTK  & 99.82\%   & 95.64\%  & 87.97\% & 92.78\%   & 60.17\% & 77.38     & 53.45\% & 49.36\%   \\ \hline
    \multirow{4}{*}{\rotatebox{90}{ResNet18}}  & C10  & 99.72\%   & 88.97\%  & 96.14\% & 91.14\%   & 93.93\% & 98.24     & 31.72\% & 20.16\%   \\
                               & C100 & 95.40\%   & 75.72\%  & 88.49\% & 81.29\%   & 86.92\% & 92.42     & 11.21\% & 15.49\%   \\
                               & S10  & 99.51\%   & 93.10\%  & 91.60\% & 80.06\%   & 84.04\% & 97.20     & 29.27\% & 21.46\%   \\
                               & UTK  & 96.96\%   & 92.51\%  & 91.78\% & 67.30\%   & 77.79\% & 95.73     & 53.45\% & 46.55\%   \\ \hline
    \multirow{4}{*}{\rotatebox{90}{ResNet34}}  & C10  & 99.57\%   & 89.17\%  & 31.50\% & 14.18\%   & 93.72\% & 97.26     & 26.54\% & 11.78\%   \\
                               & C100 & 95.23\%   & 81.16\%  & 84.01\% & 80.66\%   & 89.59\% & 91.59     & 10.97\% & 17.70\%   \\
                               & S10  & 98.61\%   & 93.83\%  & 90.08\% & 72.88\%   & 85.49\% & 98.36     & 35.46\% & 20.99\%   \\
                               & UTK  & 98.82\%   & 95.23\%  & 80.84\% & 44.96\%   & 92.19\% & 96.39     & 47.09\% & 45.23\%   \\ \hline
    \multirow{4}{*}{\rotatebox{90}{VGG16\_BN}} & C10  & 98.34\%   & 91.62\%  & 95.70\% & 89.87\%   & 83.16\% & 97.18     & 31.11\% & 14.93\%   \\
                               & C100 & 93.92\%   & 73.45\%  & 88.96\% & 78.40\%   & 73.52\% & 89.50     & 10.13\% & 11.38\%   \\
                               & S10  & 99.49\%   & 96.40\%  & 97.22\% & 86.81\%   & 92.97\% & 97.29     & 33.58\% & 19.31\%   \\
                               & UTK  & 98.50\%   & 94.96\%  & 96.55\% & 92.73\%   & 91.01\% & 96.82     & 48.86\% & 46.50\%   \\ \hline
    \multirow{4}{*}{\rotatebox{90}{VGG19\_BN}} & C10  & 99.14\%   & 93.99\%  & 89.55\% & 83.51\%   & 89.60\% & 96.38     & 24.08\% & 10.75\%   \\
                               & C100 & 93.43\%   & 76.06\%  & 85.46\% & 77.17\%   & 75.64\% & 96.27     & 12.02\% & 10.66\%   \\
                               & S10  & 99.50\%   & 97.39\%  & 94.96\% & 88.61\%   & 93.11\% & 98.86     & 35.40\% & 19.55\%   \\
                               & UTK  & 97.68\%   & 96.19\%  & 94.01\% & 95.00\%   & 94.91\% & 95.84     & 44.87\% & 43.98\%   \\ \hline
    \end{tabular}

    \end{adjustbox}
    \end{table}

\begin{table}[!htbp]
    \caption{\textit{ASR} regarding representative defense schemes.}
    \centering
    \label{tbl:evaluate_solution:asr}
    \begin{adjustbox}{max width=\linewidth}

    \begin{tabular}{cccccccccc}
    \hline
                               &      & White-box & DarkneTZ & Serdab  & Magnitude & SOTER    & ShadowNet & Ours    & Black-box \\ \hline
    \multirow{4}{*}{\rotatebox{90}{AlexNet}}   & C10  & 98.77\%   & 99.08\%  & 66.21\% & 83.38\%   & 99.05\%  & 100.00    & 8.62\%  & 10.00\%   \\
                               & C100 & 99.12\%   & 95.15\%  & 59.61\% & 92.95\%   & 99.16\%  & 100.00    & 18.94\% & 24.89\%   \\
                               & S10  & 100.00\%  & 100.00\% & 81.06\% & 100.00\%  & 58.45\%  & 64.08     & 6.29\%  & 13.12\%   \\
                               & UTK  & 100.00\%  & 100.00\% & 87.97\% & 100.00\%  & 92.71\%  & 100.00    & 2.62\%  & 9.33\%    \\ \hline
    \multirow{4}{*}{\rotatebox{90}{ResNet18}}  & C10  & 100.00\%  & 100.00\% & 96.14\% & 99.73\%   & 99.73\%  & 100.00    & 14.99\% & 9.12\%    \\
                               & C100 & 100.00\%  & 99.68\%  & 88.49\% & 100.00\%  & 100.00\% & 100.00    & 25.97\% & 25.98\%   \\
                               & S10  & 100.00\%  & 100.00\% & 91.60\% & 99.70\%   & 100.00\% & 100.00    & 10.75\% & 10.45\%   \\
                               & UTK  & 100.00\%  & 99.10\%  & 91.78\% & 81.79\%   & 100.00\% & 100.00    & 1.49\%  & 11.04\%   \\ \hline
    \multirow{4}{*}{\rotatebox{90}{ResNet34}}  & C10  & 100.00\%  & 100.00\% & 31.50\% & 6.27\%    & 100.00\% & 100.00    & 9.12\%  & 10.68\%   \\
                               & C100 & 100.00\%  & 100.00\% & 84.01\% & 98.43\%   & 100.00\% & 100.00    & 26.73\% & 23.74\%   \\
                               & S10  & 100.00\%  & 100.00\% & 90.08\% & 98.48\%   & 100.00\% & 100.00    & 7.88\%  & 8.48\%    \\
                               & UTK  & 100.00\%  & 100.00\% & 80.84\% & 8.90\%    & 100.00\% & 100.00    & 2.08\%  & 4.15\%    \\ \hline
    \multirow{4}{*}{\rotatebox{90}{VGG16\_BN}} & C10  & 100.00\%  & 100.00\% & 95.70\% & 100.00\%  & 100.00\% & 100.00    & 10.48\% & 16.57\%   \\
                               & C100 & 100.00\%  & 100.00\% & 88.96\% & 97.88\%   & 98.98\%  & 99.66     & 21.20\% & 25.80\%   \\
                               & S10  & 100.00\%  & 100.00\% & 97.22\% & 100.00\%  & 100.00\% & 100.00    & 14.91\% & 22.22\%   \\
                               & UTK  & 100.00\%  & 100.00\% & 96.55\% & 100.00\%  & 100.00\% & 100.00    & 8.05\%  & 7.18\%    \\ \hline
    \multirow{4}{*}{\rotatebox{90}{VGG19\_BN}} & C10  & 100.00\%  & 100.00\% & 89.55\% & 100.00\%  & 100.00\% & 100.00    & 7.78\%  & 12.08\%   \\
                               & C100 & 100.00\%  & 100.00\% & 85.46\% & 99.29\%   & 97.59\%  & 100.00    & 21.79\% & 26.96\%   \\
                               & S10  & 100.00\%  & 100.00\% & 94.96\% & 100.00\%  & 100.00\% & 100.00    & 12.72\% & 27.46\%   \\
                               & UTK  & 100.00\%  & 100.00\% & 94.01\% & 100.00\%  & 100.00\% & 100.00    & 7.14\%  & 7.86\%    \\ \hline
    \end{tabular}

    \end{adjustbox}
    \end{table}

\begin{table}[!htbp]
    \caption{\textit{Generalization Gap} regarding representative defense schemes.}
    \centering
    \label{tbl:evaluate_solution:gen_gap}
    \begin{adjustbox}{max width=\linewidth}

    \begin{tabular}{cccccccccc}
    \hline
                               &      & White-box & DarkneTZ & Serdab  & Magnitude & SOTER   & ShadowNet & Ours & Black-box \\ \hline
    \multirow{4}{*}{\rotatebox{90}{AlexNet}}   & C10  & 22.53\%   & 24.25\%  & 22.77\% & 14.5\%    & 22.67\% & 22.0\%    & 0\%  & 0\%       \\
                               & C100 & 52.75\%   & 46.69\%  & 52.92\% & 53.65\%   & 50.42\% & 52.0\%    & 0\%  & 0\%       \\
                               & S10  & 28.65\%   & 28.86\%  & 29.35\% & 30.8\%    & 29.08\% & 28.0\%    & 0\%  & 0\%       \\
                               & UTK  & 16.17\%   & 16.5\%   & 14.65\% & 15.05\%   & 14.26\% & 16.0\%    & 0\%  & 0\%       \\ \hline
    \multirow{4}{*}{\rotatebox{90}{ResNet18}}  & C10  & 15.43\%   & 15.93\%  & 16.51\% & 15.98\%   & 14.58\% & 15.0\%    & 0\%  & 0\%       \\
                               & C100 & 38.87\%   & 41.29\%  & 39.26\% & 39.61\%   & 40.52\% & 39.0\%    & 0\%  & 0\%       \\
                               & S10  & 23.46\%   & 24.46\%  & 23.69\% & 25.69\%   & 23.72\% & 24.0\%    & 0\%  & 0\%       \\
                               & UTK  & 12.73\%   & 13.27\%  & 14.65\% & 12.61\%   & 7.96\%  & 13.0\%    & 0\%  & 0\%       \\ \hline
    \multirow{4}{*}{\rotatebox{90}{ResNet34}}  & C10  & 15.39\%   & 15.03\%  & 16.93\% & 15.89\%   & 4.17\%  & 15.0\%    & 0\%  & 0\%       \\
                               & C100 & 36.9\%    & 39.24\%  & 37.58\% & 38.77\%   & 44.15\% & 37.0\%    & 0\%  & 0\%       \\
                               & S10  & 21.57\%   & 23.05\%  & 23.32\% & 23.26\%   & 26.92\% & 21.0\%    & 0\%  & 0\%       \\
                               & UTK  & 12.52\%   & 13.77\%  & 13.19\% & 12.52\%   & 11.01\% & 13.0\%    & 0\%  & 0\%       \\ \hline
    \multirow{4}{*}{\rotatebox{90}{VGG16\_BN}} & C10  & 7.35\%    & 7.72\%   & 7.55\%  & 11.52\%   & 11.93\% & 7.0\%     & 0\%  & 0\%       \\
                               & C100 & 29.1\%    & 32.02\%  & 29.22\% & 31.99\%   & 41.21\% & 29.0\%    & 0\%  & 0\%       \\
                               & S10  & 11.08\%   & 12.12\%  & 11.2\%  & 13.05\%   & 15.26\% & 11.0\%    & 0\%  & 0\%       \\
                               & UTK  & 10.73\%   & 12.18\%  & 10.56\% & 12.41\%   & 13.47\% & 11.0\%    & 0\%  & 0\%       \\ \hline
    \multirow{4}{*}{\rotatebox{90}{VGG19\_BN}} & C10  & 9.4\%     & 8.69\%   & 10.55\% & 10.61\%   & 8.91\%  & 9.0\%     & 0\%  & 0\%       \\
                               & C100 & 29.58\%   & 30.83\%  & 29.91\% & 32.76\%   & 33.23\% & 29.0\%    & 0\%  & 0\%       \\
                               & S10  & 14.28\%   & 15.69\%  & 15.17\% & 16.34\%   & 14.34\% & 14.0\%    & 0\%  & 0\%       \\
                               & UTK  & 11.38\%   & 11.78\%  & 12.57\% & 13.74\%   & 12.87\% & 12.0\%    & 0\%  & 0\%       \\ \hline
    \end{tabular}

    \end{adjustbox}
    \end{table}

\begin{table}[!htbp]
    \caption{\textit{Confidence Gap} regarding representative defense schemes.}
    \centering
    \label{tbl:evaluate_solution:conf_gap}
    \begin{adjustbox}{max width=\linewidth}

    \begin{tabular}{cccccccccc}
        \hline
                                   &      & White-box & DarkneTZ & Serdab  & Magnitude & SOTER   & ShadowNet & Ours & Black-box \\ \hline
        \multirow{4}{*}{\rotatebox{90}{AlexNet}}   & C10  & 23.76\%   & 20.46\%  & 22.97\% & 13.49\%   & 22.88\% & 23.0\%    & 0\%  & 0\%       \\
                                   & C100 & 54.04\%   & 39.52\%  & 52.76\% & 53.31\%   & 39.53\% & 54.0\%    & 0\%  & 0\%       \\
                                   & S10  & 29.2\%    & 29.1\%   & 29.7\%  & 30.69\%   & 29.68\% & 29.0\%    & 0\%  & 0\%       \\
                                   & UTK  & 16.4\%    & 16.5\%   & 14.68\% & 14.93\%   & 14.24\% & 16.0\%    & 0\%  & 0\%       \\ \hline
        \multirow{4}{*}{\rotatebox{90}{ResNet18}}  & C10  & 16.35\%   & 19.43\%  & 17.21\% & 15.9\%    & 13.89\% & 16.0\%    & 0\%  & 0\%       \\
                                   & C100 & 44.68\%   & 49.33\%  & 45.16\% & 39.82\%   & 43.73\% & 44.0\%    & 0\%  & 0\%       \\
                                   & S10  & 24.45\%   & 27.61\%  & 24.51\% & 25.98\%   & 24.75\% & 24.0\%    & 0\%  & 0\%       \\
                                   & UTK  & 13.68\%   & 13.55\%  & 14.51\% & 12.59\%   & 7.58\%  & 13.0\%    & 0\%  & 0\%       \\ \hline
        \multirow{4}{*}{\rotatebox{90}{ResNet34}}  & C10  & 16.15\%   & 17.78\%  & 16.98\% & 15.67\%   & 3.43\%  & 15.0\%    & 0\%  & 0\%       \\
                                   & C100 & 41.44\%   & 45.36\%  & 41.98\% & 38.95\%   & 44.24\% & 41.0\%    & 0\%  & 0\%       \\
                                   & S10  & 22.13\%   & 25.3\%   & 23.71\% & 23.44\%   & 26.71\% & 22.0\%    & 0\%  & 0\%       \\
                                   & UTK  & 13.07\%   & 14.27\%  & 13.37\% & 12.28\%   & 10.1\%  & 13.0\%    & 0\%  & 0\%       \\ \hline
        \multirow{4}{*}{\rotatebox{90}{VGG16\_BN}} & C10  & 8.54\%    & 10.99\%  & 8.71\%  & 12.57\%   & 10.92\% & 8.0\%     & 0\%  & 0\%       \\
                                   & C100 & 36.08\%   & 41.98\%  & 36.16\% & 36.04\%   & 5.79\%  & 35.0\%    & 0\%  & 0\%       \\
                                   & S10  & 13.01\%   & 18.4\%   & 13.11\% & 15.89\%   & 18.3\%  & 13.0\%    & 0\%  & 0\%       \\
                                   & UTK  & 11.29\%   & 13.17\%  & 11.61\% & 12.87\%   & 9.63\%  & 12.0\%    & 0\%  & 0\%       \\ \hline
        \multirow{4}{*}{\rotatebox{90}{VGG19\_BN}} & C10  & 10.16\%   & 11.61\%  & 11.32\% & 11.48\%   & 13.03\% & 10.0\%    & 0\%  & 0\%       \\
                                   & C100 & 34.69\%   & 38.14\%  & 35.2\%  & 36.98\%   & 11.07\% & 34.0\%    & 0\%  & 0\%       \\
                                   & S10  & 15.06\%   & 19.28\%  & 16.11\% & 17.94\%   & 20.82\% & 15.0\%    & 0\%  & 0\%       \\
                                   & UTK  & 12.13\%   & 13.18\%  & 12.54\% & 13.3\%    & 13.08\% & 12.0\%    & 0\%  & 0\%       \\ \hline
        \end{tabular}

    \end{adjustbox}
    \end{table}

\begin{table}[!htbp]
    \caption{\textit{Gradient-base Attack Accuracy} regarding representative defense schemes.}
    \centering
    \label{tbl:evaluate_solution:grad_attack}
    \begin{adjustbox}{max width=\linewidth}

    \begin{tabular}{cccccccccc}
    \hline
                               &      & White-box & DarkneTZ & Serdab  & Magnitude & SOTER   & ShadowNet & Ours   & Black-box \\ \hline
    \multirow{4}{*}{\rotatebox{90}{AlexNet}}   & C10  & 69.73\%   & 56.63\%  & 65.49\% & 56.51\%   & 65.35\% & 72.0\%    & 50.0\% & 50.0\%    \\
                               & C100 & 62.47\%   & 53.42\%  & 56.95\% & 78.98\%   & 50.61\% & 64.0\%    & 50.0\% & 50.0\%    \\
                               & S10  & 71.2\%    & 58.71\%  & 63.89\% & 69.14\%   & 63.34\% & 72.0\%    & 50.0\% & 50.0\%    \\
                               & UTK  & 64.34\%   & 62.76\%  & 58.76\% & 59.76\%   & 58.35\% & 64.0\%    & 50.0\% & 50.0\%    \\ \hline
    \multirow{4}{*}{\rotatebox{90}{ResNet18}}  & C10  & 70.01\%   & 63.68\%  & 67.51\% & 60.13\%   & 53.29\% & 70.0\%    & 50.0\% & 50.0\%    \\
                               & C100 & 89.07\%   & 72.78\%  & 88.63\% & 72.15\%   & 76.78\% & 89.0\%    & 50.0\% & 50.0\%    \\
                               & S10  & 77.27\%   & 65.15\%  & 75.91\% & 64.38\%   & 61.51\% & 76.0\%    & 50.0\% & 50.0\%    \\
                               & UTK  & 63.91\%   & 58.0\%   & 61.35\% & 56.96\%   & 52.46\% & 64.0\%    & 50.0\% & 50.0\%    \\ \hline
    \multirow{4}{*}{\rotatebox{90}{ResNet34}}  & C10  & 68.98\%   & 55.44\%  & 57.56\% & 57.26\%   & 50.06\% & 69.0\%    & 50.0\% & 50.0\%    \\
                               & C100 & 85.02\%   & 75.2\%   & 84.17\% & 71.32\%   & 68.44\% & 85.0\%    & 50.0\% & 50.0\%    \\
                               & S10  & 72.23\%   & 64.8\%   & 70.25\% & 66.78\%   & 63.2\%  & 72.0\%    & 50.0\% & 50.0\%    \\
                               & UTK  & 63.36\%   & 55.39\%  & 57.81\% & 58.07\%   & 52.18\% & 64.0\%    & 50.0\% & 50.0\%    \\ \hline
    \multirow{4}{*}{\rotatebox{90}{VGG16\_BN}} & C10  & 68.86\%   & 52.52\%  & 67.35\% & 59.29\%   & 49.93\% & 71.0\%    & 50.0\% & 50.0\%    \\
                               & C100 & 87.8\%    & 70.66\%  & 87.88\% & 73.58\%   & 50.0\%  & 87.0\%    & 50.0\% & 50.0\%    \\
                               & S10  & 65.49\%   & 68.91\%  & 66.11\% & 66.91\%   & 52.62\% & 65.0\%    & 50.0\% & 50.0\%    \\
                               & UTK  & 63.17\%   & 61.94\%  & 63.01\% & 61.11\%   & 49.86\% & 63.0\%    & 50.0\% & 50.0\%    \\ \hline
    \multirow{4}{*}{\rotatebox{90}{VGG19\_BN}} & C10  & 64.66\%   & 55.12\%  & 63.28\% & 61.28\%   & 50.01\% & 65.0\%    & 50.0\% & 50.0\%    \\
                               & C100 & 68.75\%   & 50.0\%   & 87.53\% & 75.22\%   & 50.0\%  & 87.0\%    & 50.0\% & 50.0\%    \\
                               & S10  & 64.75\%   & 63.08\%  & 66.11\% & 67.62\%   & 64.02\% & 65.0\%    & 50.0\% & 50.0\%    \\
                               & UTK  & 63.24\%   & 62.07\%  & 62.3\%  & 57.99\%   & 55.32\% & 63.0\%    & 50.0\% & 50.0\%    \\ \hline
    \end{tabular}

    \end{adjustbox}
    \end{table}

\subsection{Qualitative Impacts of Configurations}
\label{append:sec:other_metrics:optimal_config_qualitative}

\begin{figure*}[!ht]
\centering
\includegraphics[width=\linewidth]{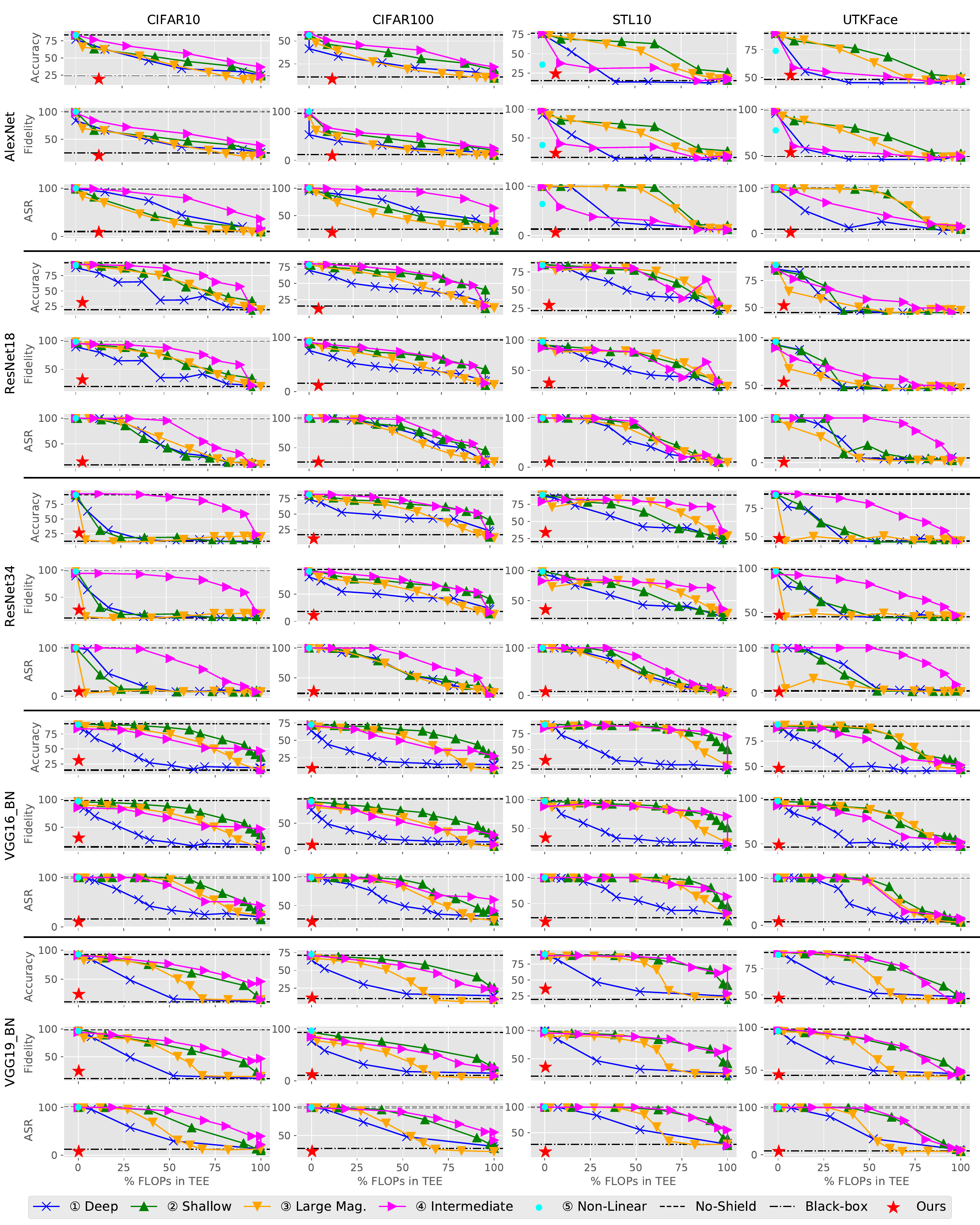}
\caption{Model stealing results in terms of accuracy, fidelity, and ASR.}
\label{fig:acc_flops_append_one_fig}
\end{figure*}

\begin{figure*}[!ht]
\centering
\includegraphics[width=\linewidth]{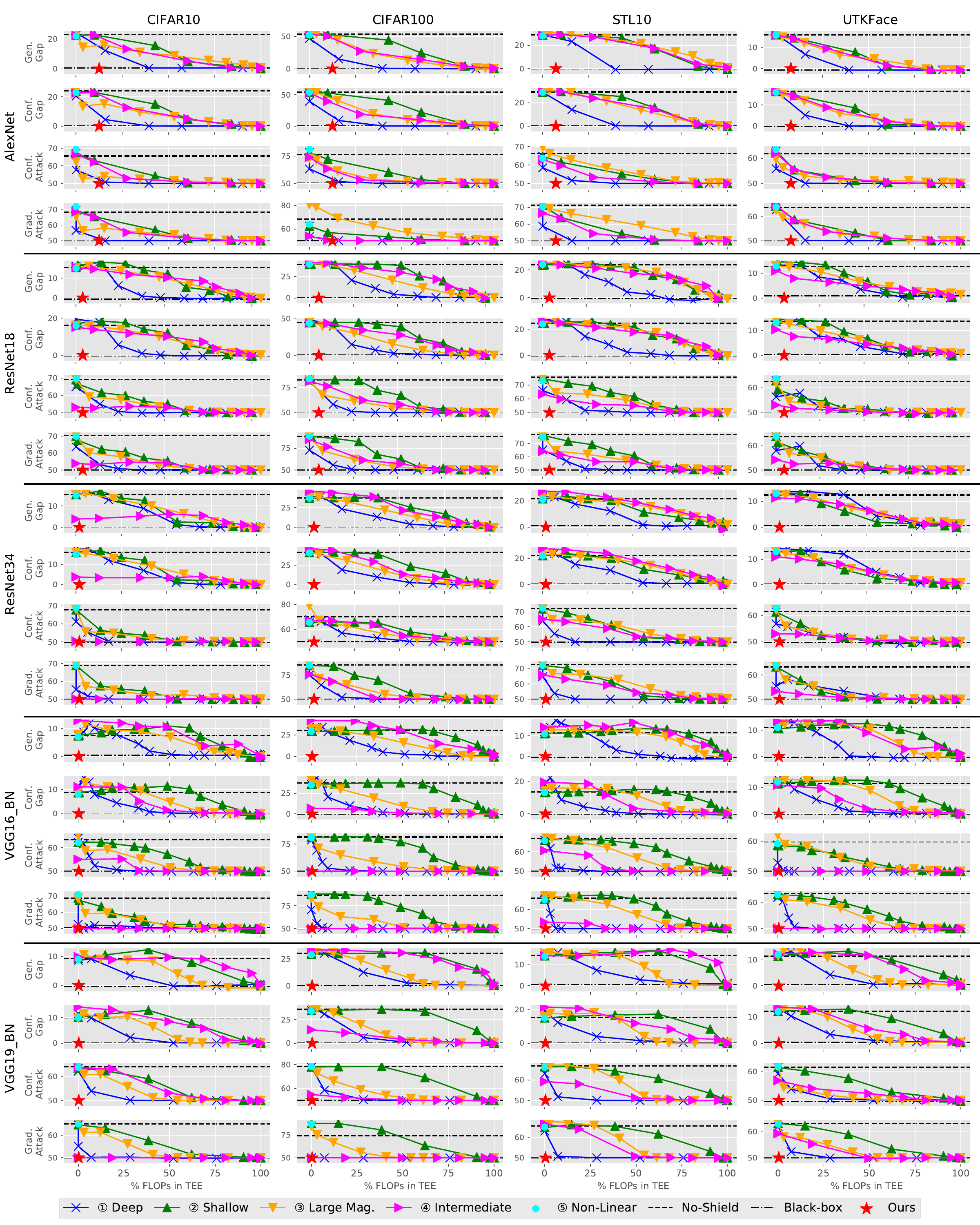}
\caption{Membership inference results of generalization gap, confidence gap, confidence-based membership inference attack accuracy, and gradient-based membership inference attack accuracy.}
\label{fig:mia_flops_append_one_fig}
\end{figure*}

We display the curves for all the \tsdp\ solutions in \F~\ref{fig:mia_flops_append_one_fig} and \F~\ref{fig:acc_flops_append_one_fig}. In \F~\ref{fig:mia_flops_append_one_fig}, we display the accuracy, fidelity, and ASR for all the models and datasets. In \F~\ref{fig:acc_flops_append_one_fig}, we displayed the confidence-based attack accuracy, gradient-based attack accuracy, generalization gap, and confidence gap. Note that the results of accuracy and confidence-based attack accuracy are consistent with the results in \S~\ref{sec:evaluation_dilemma_results} and \S~\ref{append:sec:resnet34_vgg19_results}. The tendencies of curves are generally consistent with the conclusions in the main paper. 

\subsection{Quantitative Results of ``Sweet Spot'' Configurations}
\label{append:sec:other_metrics:optimal_config_quantitative}
This part includes the evaluation results of representative defense schemes in \S~\ref{sec:evaluation_dilemma_results}. \T~\ref{tbl:optimal_config:fidelity} to \T~\ref{tbl:optimal_config:grad_attack} reports the results of the other five metrics that are not reported in the main paper. Note that the configuration of our approach (last column) is the same from \T~\ref{tbl:optimal_config:fidelity} to \T~\ref{tbl:optimal_config:grad_attack} because the size of the slices is fixed after the training phase and is not affected by the $Security$ metrics.

\begin{table}[!htbp]
    \caption{Different $Utility(C^*)$ ($\% FLOPs(C^*)$) values of ``sweet spot'' w.r.t. \textit{Fidelity}.}
    \centering
    \label{tbl:optimal_config:fidelity}
    \begin{adjustbox}{max width=\linewidth}

    \begin{tabular}{@{}ccccccc@{}}
    \toprule
                               &      & Deep     & Shallow  & Large Mag. & Intermediate & Ours    \\ \midrule
    \multirow{4}{*}{AlexNet}   & C10  & 100.00\% & 100.00\% & 81.18\%    & 100.00\%     & 12.48\% \\
                               & C100 & 100.00\% & 100.00\% & 90.58\%    & 100.00\%     & 12.48\% \\
                               & S10  & 39.44\%  & 100.00\% & 100.00\%   & 84.31\%      & 7.12\%  \\
                               & UTK  & 39.44\%  & 100.00\% & 71.82\%    & 60.69\%      & 8.01\%  \\ \midrule
    \multirow{4}{*}{ResNet18}  & C10  & 100.00\% & 100.00\% & 100.00\%   & 100.00\%     & 3.80\%  \\
                               & C100 & 100.00\% & 100.00\% & 94.71\%    & 100.00\%     & 5.33\%  \\
                               & S10  & 100.00\% & 100.00\% & 100.00\%   & 100.00\%     & 3.80\%  \\
                               & UTK  & 37.46\%  & 38.55\%  & 61.48\%    & 100.00\%     & 4.58\%  \\ \midrule
    \multirow{4}{*}{ResNet34}  & C10  & 55.49\%  & 87.03\%  & 20.35\%    & 100.00\%     & 1.83\%  \\
                               & C100 & 100.00\% & 100.00\% & 94.96\%    & 100.00\%     & 2.56\%  \\
                               & S10  & 100.00\% & 100.00\% & 100.00\%   & 100.00\%     & 1.73\%  \\
                               & UTK  & 37.47\%  & 56.05\%  & 5.22\%     & 100.00\%     & 1.92\%  \\ \midrule
    \multirow{4}{*}{VGG16\_BN} & C10  & 100.00\% & 100.00\% & 100.00\%   & 100.00\%     & 0.34\%  \\
                               & C100 & 100.00\% & 100.00\% & 87.43\%    & 100.00\%     & 0.47\%  \\
                               & S10  & 100.00\% & 100.00\% & 100.00\%   & 100.00\%     & 0.40\%  \\
                               & UTK  & 69.23\%  & 100.00\% & 100.00\%   & 100.00\%     & 0.60\%  \\ \midrule
    \multirow{4}{*}{VGG19\_BN} & C10  & 75.79\%  & 100.00\% & 100.00\%   & 100.00\%     & 0.27\%  \\
                               & C100 & 100.00\% & 100.00\% & 67.98\%    & 100.00\%     & 0.37\%  \\
                               & S10  & 100.00\% & 100.00\% & 100.00\%   & 100.00\%     & 0.35\%  \\
                               & UTK  & 75.79\%  & 100.00\% & 67.98\%    & 95.30\%      & 0.37\%  \\ \midrule
    \end{tabular}

    \end{adjustbox}
    \end{table}

\begin{table}[!htbp]
    \caption{Different $Utility(C^*)$ ($\% FLOPs(C^*)$) values of ``sweet spot'' w.r.t. \textit{ASR}.}
    \centering
    \label{tbl:optimal_config:asr}
    \begin{adjustbox}{max width=\linewidth}

    \begin{tabular}{@{}ccccccc@{}}
    \toprule
                               &      & Deep     & Shallow  & Large Mag. & Intermediate & Ours    \\ \midrule
    \multirow{4}{*}{AlexNet}   & C10  & 100.00\% & 100.00\% & 90.58\%    & 100.00\%     & 12.48\% \\
                               & C100 & 100.00\% & 100.00\% & 100.00\%   & 100.00\%     & 12.48\% \\
                               & S10  & 100.00\% & 100.00\% & 95.29\%    & 84.31\%      & 7.12\%  \\
                               & UTK  & 90.06\%  & 100.00\% & 100.00\%   & 100.00\%     & 8.01\%  \\ \midrule
    \multirow{4}{*}{ResNet18}  & C10  & 100.00\% & 100.00\% & 100.00\%   & 100.00\%     & 3.80\%  \\
                               & C100 & 100.00\% & 100.00\% & 94.71\%    & 100.00\%     & 5.33\%  \\
                               & S10  & 100.00\% & 100.00\% & 94.71\%    & 100.00\%     & 3.80\%  \\
                               & UTK  & 47.95\%  & 76.03\%  & 44.93\%    & 100.00\%     & 4.58\%  \\ \midrule
    \multirow{4}{*}{ResNet34}  & C10  & 55.49\%  & 56.05\%  & 5.22\%     & 100.00\%     & 1.83\%  \\
                               & C100 & 100.00\% & 100.00\% & 100.00\%   & 100.00\%     & 2.56\%  \\
                               & S10  & 99.52\%  & 100.00\% & 100.00\%   & 100.00\%     & 1.73\%  \\
                               & UTK  & 100.00\% & 56.05\%  & 74.82\%    & 100.00\%     & 1.92\%  \\ \midrule
    \multirow{4}{*}{VGG16\_BN} & C10  & 100.00\% & 100.00\% & 100.00\%   & 100.00\%     & 0.34\%  \\
                               & C100 & 100.00\% & 100.00\% & 100.00\%   & 100.00\%     & 0.47\%  \\
                               & S10  & 100.00\% & 100.00\% & 100.00\%   & 100.00\%     & 0.40\%  \\
                               & UTK  & 100.00\% & 100.00\% & 100.00\%   & 100.00\%     & 0.60\%  \\ \midrule
    \multirow{4}{*}{VGG19\_BN} & C10  & 100.00\% & 100.00\% & 67.98\%    & 100.00\%     & 0.27\%  \\
                               & C100 & 99.52\%  & 100.00\% & 67.98\%    & 100.00\%     & 0.37\%  \\
                               & S10  & 75.79\%  & 100.00\% & 82.18\%    & 100.00\%     & 0.35\%  \\
                               & UTK  & 100.00\% & 100.00\% & 67.98\%    & 100.00\%     & 0.37\%  \\ \bottomrule
    \end{tabular}

    \end{adjustbox}
    \end{table}

\begin{table}[!htbp]
    \caption{Different $Utility(C^*)$ ($\% FLOPs(C^*)$) values of ``sweet spot'' w.r.t. \textit{Generalization Gap}.}
    \centering
    \label{tbl:optimal_config:gen_gap}
    \begin{adjustbox}{max width=\linewidth}

    \begin{tabular}{@{}ccccccc@{}}
    \toprule
                               &      & Deep     & Shallow  & Large Mag. & Intermediate & Ours    \\ \midrule
    \multirow{4}{*}{AlexNet}   & C10  & 39.44\%  & 100.00\% & 100.00\%   & 84.31\%      & 12.48\% \\
                               & C100 & 90.06\%  & 100.00\% & 100.00\%   & 100.00\%     & 12.48\% \\
                               & S10  & 39.44\%  & 100.00\% & 100.00\%   & 100.00\%     & 7.12\%  \\
                               & UTK  & 100.00\% & 100.00\% & 100.00\%   & 84.31\%      & 8.01\%  \\ \midrule
    \multirow{4}{*}{ResNet18}  & C10  & 100.00\% & 100.00\% & 100.00\%   & 100.00\%     & 3.80\%  \\
                               & C100 & 100.00\% & 100.00\% & 100.00\%   & 100.00\%     & 5.33\%  \\
                               & S10  & 71.96\%  & 100.00\% & 100.00\%   & 100.00\%     & 3.80\%  \\
                               & UTK  & 71.96\%  & 100.00\% & 100.00\%   & 100.00\%     & 4.58\%  \\ \midrule
    \multirow{4}{*}{ResNet34}  & C10  & 100.00\% & 100.00\% & 94.96\%    & 100.00\%     & 1.83\%  \\
                               & C100 & 100.00\% & 100.00\% & 100.00\%   & 100.00\%     & 2.56\%  \\
                               & S10  & 68.47\%  & 100.00\% & 100.00\%   & 100.00\%     & 1.73\%  \\
                               & UTK  & 80.01\%  & 93.52\%  & 100.00\%   & 100.00\%     & 1.92\%  \\ \midrule
    \multirow{4}{*}{VGG16\_BN} & C10  & 63.21\%  & 93.98\%  & 100.00\%   & 100.00\%     & 0.34\%  \\
                               & C100 & 100.00\% & 100.00\% & 87.43\%    & 100.00\%     & 0.47\%  \\
                               & S10  & 69.23\%  & 100.00\% & 100.00\%   & 100.00\%     & 0.40\%  \\
                               & UTK  & 63.21\%  & 100.00\% & 100.00\%   & 100.00\%     & 0.60\%  \\ \midrule
    \multirow{4}{*}{VGG19\_BN} & C10  & 100.00\% & 100.00\% & 82.18\%    & 100.00\%     & 0.27\%  \\
                               & C100 & 99.52\%  & 100.00\% & 67.98\%    & 100.00\%     & 0.37\%  \\
                               & S10  & 100.00\% & 100.00\% & 82.18\%    & 100.00\%     & 0.35\%  \\
                               & UTK  & 99.52\%  & 100.00\% & 67.98\%    & 100.00\%     & 0.37\%  \\ \bottomrule
    \end{tabular}

    \end{adjustbox}
    \end{table}

\begin{table}[!htbp]
    \caption{Different $Utility(C^*)$ ($\% FLOPs(C^*)$) values of ``sweet spot'' w.r.t. \textit{Confidence Gap}.}
    \centering
    \label{tbl:optimal_config:conf_gap}
    \begin{adjustbox}{max width=\linewidth}

    \begin{tabular}{@{}ccccccc@{}}
    \toprule
                               &      & Deep     & Shallow  & Large Mag. & Intermediate & Ours    \\ \midrule
    \multirow{4}{*}{AlexNet}   & C10  & 57.19\%  & 100.00\% & 100.00\%   & 100.00\%     & 12.48\% \\
                               & C100 & 90.06\%  & 100.00\% & 100.00\%   & 100.00\%     & 12.48\% \\
                               & S10  & 100.00\% & 100.00\% & 100.00\%   & 100.00\%     & 7.12\%  \\
                               & UTK  & 100.00\% & 100.00\% & 95.29\%    & 100.00\%     & 8.01\%  \\ \midrule
    \multirow{4}{*}{ResNet18}  & C10  & 100.00\% & 100.00\% & 100.00\%   & 100.00\%     & 3.80\%  \\
                               & C100 & 100.00\% & 100.00\% & 100.00\%   & 100.00\%     & 5.33\%  \\
                               & S10  & 85.48\%  & 100.00\% & 100.00\%   & 100.00\%     & 3.80\%  \\
                               & UTK  & 100.00\% & 100.00\% & 100.00\%   & 100.00\%     & 4.58\%  \\ \midrule
    \multirow{4}{*}{ResNet34}  & C10  & 99.52\%  & 100.00\% & 100.00\%   & 100.00\%     & 1.83\%  \\
                               & C100 & 100.00\% & 100.00\% & 94.96\%    & 100.00\%     & 2.56\%  \\
                               & S10  & 100.00\% & 100.00\% & 100.00\%   & 100.00\%     & 1.73\%  \\
                               & UTK  & 100.00\% & 93.52\%  & 100.00\%   & 100.00\%     & 1.92\%  \\ \midrule
    \multirow{4}{*}{VGG16\_BN} & C10  & 100.00\% & 93.98\%  & 87.43\%    & 100.00\%     & 0.34\%  \\
                               & C100 & 100.00\% & 100.00\% & 87.43\%    & 100.00\%     & 0.47\%  \\
                               & S10  & 69.23\%  & 100.00\% & 100.00\%   & 100.00\%     & 0.40\%  \\
                               & UTK  & 100.00\% & 100.00\% & 100.00\%   & 100.00\%     & 0.60\%  \\ \midrule
    \multirow{4}{*}{VGG19\_BN} & C10  & 100.00\% & 100.00\% & 82.18\%    & 100.00\%     & 0.27\%  \\
                               & C100 & 99.52\%  & 100.00\% & 100.00\%   & 100.00\%     & 0.37\%  \\
                               & S10  & 100.00\% & 100.00\% & 82.18\%    & 100.00\%     & 0.35\%  \\
                               & UTK  & 99.52\%  & 100.00\% & 67.98\%    & 100.01\%     & 0.37\%  \\ \bottomrule
    \end{tabular}

    \end{adjustbox}
    \end{table}

\begin{table}[!htbp]
    \caption{Different $Utility(C^*)$ ($\% FLOPs(C^*)$) values of ``sweet spot'' w.r.t. \textit{Gradient-based Attack Accuracy}.}
    \centering
    \label{tbl:optimal_config:grad_attack}
    \begin{adjustbox}{max width=\linewidth}

    \begin{tabular}{@{}ccccccc@{}}
    \toprule
                               &      & Deep    & Shallow & Large Mag. & Intermediate & Ours    \\ \midrule
    \multirow{4}{*}{AlexNet}   & C10  & 15.78\% & 60.56\% & 71.82\%    & 60.69\%      & 12.48\% \\
                               & C100 & 15.78\% & 60.56\% & 90.58\%    & 10.28\%      & 12.48\% \\
                               & S10  & 15.78\% & 60.56\% & 81.18\%    & 60.69\%      & 7.12\%  \\
                               & UTK  & 15.78\% & 60.56\% & 53.17\%    & 60.69\%      & 8.01\%  \\ \midrule
    \multirow{4}{*}{ResNet18}  & C10  & 23.97\% & 62.54\% & 61.48\%    & 72.80\%      & 3.80\%  \\
                               & C100 & 23.97\% & 76.03\% & 61.48\%    & 72.80\%      & 5.33\%  \\
                               & S10  & 23.97\% & 76.03\% & 76.61\%    & 72.80\%      & 3.80\%  \\
                               & UTK  & 23.97\% & 38.55\% & 44.93\%    & 11.02\%      & 4.58\%  \\ \midrule
    \multirow{4}{*}{ResNet34}  & C10  & 6.48\%  & 56.05\% & 40.84\%    & 0.60\%       & 1.83\%  \\
                               & C100 & 18.01\% & 87.03\% & 58.31\%    & 36.36\%      & 2.56\%  \\
                               & S10  & 18.01\% & 56.05\% & 74.82\%    & 83.89\%      & 1.73\%  \\
                               & UTK  & 55.49\% & 38.02\% & 40.84\%    & 13.50\%      & 1.92\%  \\ \midrule
    \multirow{4}{*}{VGG16\_BN} & C10  & 0.00\%  & 48.83\% & 50.56\%    & 0.10\%       & 0.34\%  \\
                               & C100 & 6.02\%  & 90.97\% & 50.56\%    & 0.10\%       & 0.47\%  \\
                               & S10  & 6.02\%  & 90.97\% & 66.47\%    & 33.84\%      & 0.40\%  \\
                               & UTK  & 9.03\%  & 78.93\% & 66.47\%    & 0.10\%       & 0.60\%  \\ \midrule
    \multirow{4}{*}{VGG19\_BN} & C10  & 7.11\%  & 62.11\% & 40.90\%    & 0.11\%       & 0.27\%  \\
                               & C100 & 0.00\%  & 90.53\% & 40.90\%    & 0.11\%       & 0.37\%  \\
                               & S10  & 7.11\%  & 97.63\% & 54.49\%    & 49.82\%      & 0.35\%  \\
                               & UTK  & 7.11\%  & 90.53\% & 40.90\%    & 49.82\%      & 0.37\%  \\ \bottomrule
    \end{tabular}

    \end{adjustbox}
    \end{table}

\subsection{Comparison with Other Attack Assumptions}
\label{append:sec:assumption}
This part reports the comparison of fidelity and ASR between different attack assumptions in (\S~\ref{sec:experiment:other_assumption}). The fidelity is reported in \T~\ref{tbl:other_assumption:fidelity} and ASR is reported in \T~\ref{tbl:other_assumption:asr}.

\begin{table}[]
    \caption{Comparison of \textit{Fidelity} between different attack assumptions.}
    \label{tbl:other_assumption:fidelity}
    \centering
    \begin{adjustbox}{max width=0.8\linewidth}

    \begin{tabular}{@{}ccccc@{}}
    \toprule
                               &      & Hybrid  & Backbone & Victim  \\ \midrule
    \multirow{4}{*}{AlexNet}   & C10  & 19.84\% & 19.33\%  & 24.37\% \\
                               & C100 & 9.47\%  & 14.99\%  & 13.42\% \\
                               & S10  & 23.5\%  & 30.68\%  & 17.69\% \\
                               & UTK  & 53.45\% & 51.32\%  & 47.46\% \\ \midrule
    \multirow{4}{*}{ResNet18}  & C10  & 31.72\% & 25.6\%   & 17.64\% \\
                               & C100 & 11.21\% & 19.26\%  & 7.77\%  \\
                               & S10  & 29.27\% & 32.56\%  & 32.67\% \\
                               & UTK  & 53.45\% & 50.95\%  & 52.36\% \\ \midrule
    \multirow{4}{*}{ResNet34}  & C10  & 26.54\% & 23.72\%  & 26.05\% \\
                               & C100 & 10.97\% & 17.07\%  & 20.71\% \\
                               & S10  & 39.16\% & 31.82\%  & 41.99\% \\
                               & UTK  & 47.09\% & 50.5\%   & 50.95\% \\ \midrule
    \multirow{4}{*}{VGG16\_BN} & C10  & 31.11\% & 25.68\%  & 20.32\% \\
                               & C100 & 10.13\% & 18.73\%  & 6.41\%  \\
                               & S10  & 42.02\% & 31.99\%  & 32.25\% \\
                               & UTK  & 48.86\% & 53.36\%  & 52.0\%  \\ \midrule
    \multirow{4}{*}{VGG19\_BN} & C10  & 24.08\% & 25.52\%  & 22.1\%  \\
                               & C100 & 12.02\% & 18.5\%   & 15.01\% \\
                               & S10  & 41.91\% & 32.05\%  & 37.2\%  \\
                               & UTK  & 44.87\% & 50.95\%  & 49.18\% \\ \bottomrule
    \end{tabular}
\end{adjustbox}
\end{table}

\begin{table}[]
    \caption{Comparison of \textit{ASR} between different attack assumptions.}
    \label{tbl:other_assumption:asr}
    \centering
    \begin{adjustbox}{max width=0.8\linewidth}

    \begin{tabular}{@{}ccccc@{}}
    \toprule
                               &      & Hybrid  & Backbone & Victim  \\ \midrule
    \multirow{4}{*}{AlexNet}   & C10  & 8.62\%  & 5.85\%   & 10.15\% \\
                               & C100 & 18.94\% & 8.37\%   & 27.31\% \\
                               & S10  & 6.29\%  & 5.59\%   & 16.43\% \\
                               & UTK  & 2.62\%  & 3.79\%   & 14.29\% \\ \midrule
    \multirow{4}{*}{ResNet18}  & C10  & 14.99\% & 11.72\%  & 10.35\% \\
                               & C100 & 25.97\% & 29.87\%  & 25.32\% \\
                               & S10  & 10.75\% & 10.45\%  & 11.34\% \\
                               & UTK  & 1.49\%  & 5.07\%   & 4.18\%  \\ \midrule
    \multirow{4}{*}{ResNet34}  & C10  & 9.12\%  & 8.26\%   & 9.4\%   \\
                               & C100 & 26.73\% & 28.62\%  & 26.73\% \\
                               & S10  & 7.88\%  & 8.79\%   & 8.79\%  \\
                               & UTK  & 2.08\%  & 3.26\%   & 3.86\%  \\ \midrule
    \multirow{4}{*}{VGG16\_BN} & C10  & 10.48\% & 10.48\%  & 22.95\% \\
                               & C100 & 21.2\%  & 16.96\%  & 24.73\% \\
                               & S10  & 14.91\% & 15.79\%  & 39.18\% \\
                               & UTK  & 8.05\%  & 10.06\%  & 9.2\%   \\ \midrule
    \multirow{4}{*}{VGG19\_BN} & C10  & 7.78\%  & 10.83\%  & 15.56\% \\
                               & C100 & 21.79\% & 14.64\%  & 27.14\% \\
                               & S10  & 10.12\% & 11.85\%  & 36.13\% \\
                               & UTK  & 7.14\%  & 10.57\%  & 12.0\%  \\ \bottomrule
    \end{tabular}
\end{adjustbox}
\end{table}

\section{Security Under Other Assumptions of Data}
\label{append:sec:data_assumption}

\begin{figure*}[ht]
    \centering
    \includegraphics[width=\linewidth]{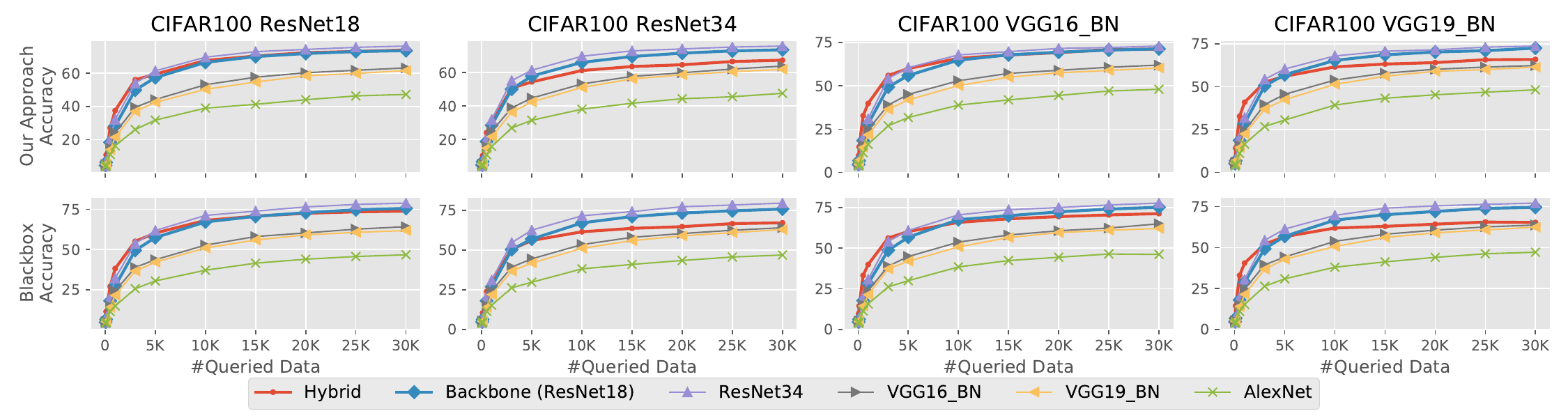}
    \caption{The comparison of \sys and black-box protection in terms of model \textit{accuracy}. }
    \label{fig:multi_arch_cifar100_accuracy}
    \end{figure*}

\begin{figure*}[ht]
\centering
\includegraphics[width=\linewidth]{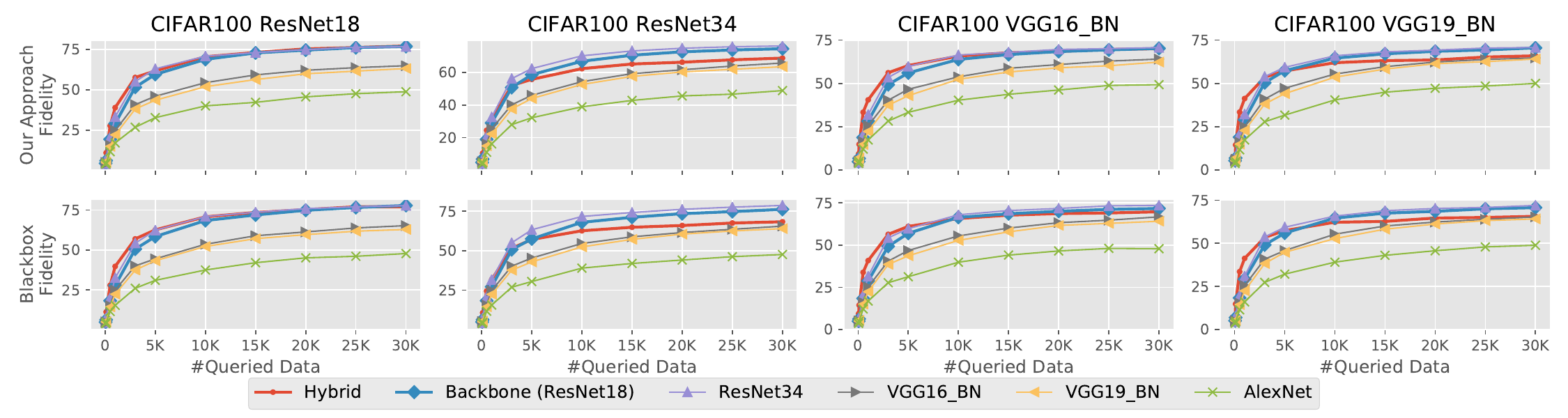}
\caption{The comparison of \sys and black-box protection in terms of model \textit{fidelity}. }
\label{fig:multi_arch_cifar100_fidelity}
\end{figure*}

\begin{figure*}[ht]
\centering
\includegraphics[width=\linewidth]{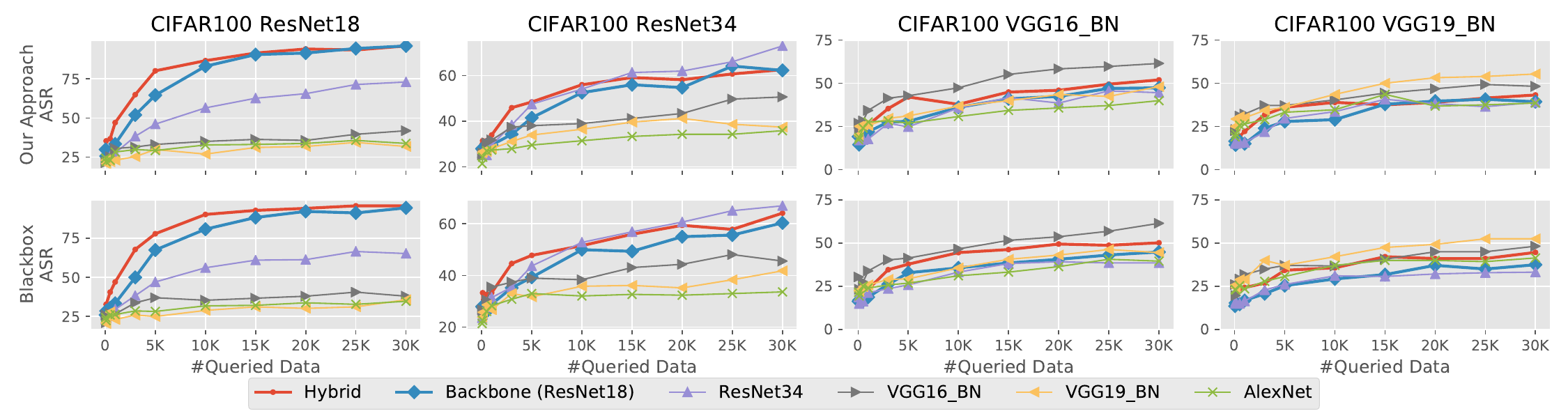}
\caption{The comparison of \sys and black-box protection in terms of \textit{ASR}. }
\label{fig:multi_arch_cifar100_asr}
\end{figure*}

The comparison of \sys and black-box protection under other assumptions of
queried data size w.r.t fidelity and ASR is displayed in
\F~\ref{fig:multi_arch_cifar100_accuracy},
\F~\ref{fig:multi_arch_cifar100_fidelity} and
\F~\ref{fig:multi_arch_cifar100_asr}. The fidelity results are similar to
accuracy, and the model performance is ordered by the model complexity. The
relative ratio of the backbone performance between \sys and the black-box the
setting is $1.0076$, which means the performance of \sys is only $0.76\%$ more
effective than the black-box scheme. For ASR, the model performance is not
ordered by model complexity. For each setting, the attack model with the highest
ASR is the one that has the same architecture as $M_{\rm vic}$. This result is
reasonable because adversarial examples transfer more easily between such
models. 

\end{document}